\documentclass[
twocolumn,
eqsecnum,
prd,
floatfix,
superscriptaddress,
showpacs
]{revtex4} 

\usepackage{graphicx}
\usepackage{dcolumn}
\usepackage{amsmath}
\usepackage{amssymb}
\usepackage{epstopdf}
\makeatletter
\def\btt#1{\texttt{\@backslashchar#1}}%
\DeclareRobustCommand\bblash{\btt{\@backslashchar}}%
\makeatother

\begin{document}
\title{Solar neutrino results in Super--Kamiokande--III}
\newcommand{\AFFicrr}{\affiliation{Kamioka Observatory, Institute for Cosmic Ray Research, University of Tokyo, Kamioka, Gifu 506-1205, Japan}}
\newcommand{\AFFkashiwa}{\affiliation{Research Center for Cosmic Neutrinos, Institute for Cosmic Ray Research, University of Tokyo, Kashiwa, Chiba 277-8582, Japan}}
\newcommand{\AFFipmu}{\affiliation{Institute for the Physics and
Mathematics of the Universe, University of Tokyo, Kashiwa, Chiba
277-8582, Japan}}
\newcommand{\AFFuam}{\affiliation{Department of Theoretical Physics, University Autonoma Madrid, 28049 Madrid, Spain}}
\newcommand{\AFFbu}{\affiliation{Department of Physics, Boston University, Boston, MA 02215, USA}}
\newcommand{\AFFbnl}{\affiliation{Physics Department, Brookhaven National Laboratory, Upton, NY 11973, USA}}
\newcommand{\AFFucd}{\affiliation{Department of Physics, University of California, Davis, Davis, CA 95616, USA}}
\newcommand{\AFFuci}{\affiliation{Department of Physics and Astronomy, University of California, Irvine, Irvine, CA 92697-4575, USA }}
\newcommand{\AFFcsu}{\affiliation{Department of Physics, California State University, Dominguez Hills, Carson, CA 90747, USA}}
\newcommand{\AFFcnm}{\affiliation{Department of Physics, Chonnam National University, Kwangju 500-757, Korea}}
\newcommand{\AFFduke}{\affiliation{Department of Physics, Duke University, Durham NC 27708, USA}}
\newcommand{\AFFfukuoka}{\affiliation{Junior College, Fukuoka Institute of Technology, Fukuoka, Fukuoka 811-0214, Japan}}
\newcommand{\AFFgmu}{\affiliation{Department of Physics, George Mason University, Fairfax, VA 22030, USA }}
\newcommand{\AFFgifu}{\affiliation{Department of Physics, Gifu University, Gifu, Gifu 501-1193, Japan}}
\newcommand{\AFFuh}{\affiliation{Department of Physics and Astronomy, University of Hawaii, Honolulu, HI 96822, USA}}
\newcommand{\AFFkanagawa}{\affiliation{Physics Division, Department of Engineering, Kanagawa University, Kanagawa, Yokohama 221-8686, Japan}}
\newcommand{\AFFkek}{\affiliation{High Energy Accelerator Research Organization (KEK), Tsukuba, Ibaraki 305-0801, Japan }}
\newcommand{\AFFkobe}{\affiliation{Department of Physics, Kobe University, Kobe, Hyogo 657-8501, Japan}}
\newcommand{\AFFkyoto}{\affiliation{Department of Physics, Kyoto University, Kyoto, Kyoto 606-8502, Japan}}
\newcommand{\AFFumd}{\affiliation{Department of Physics, University of Maryland, College Park, MD 20742, USA }}
\newcommand{\AFFmit}{\affiliation{Department of Physics, Massachusetts Institute of Technology, Cambridge, MA 02139, USA}}
\newcommand{\AFFmiyagi}{\affiliation{Department of Physics, Miyagi University of Education, Sendai, Miyagi 980-0845, Japan}}
\newcommand{\AFFnagoya}{\affiliation{Solar Terrestrial Environment
Laboratory, Nagoya University, Nagoya, Aichi 464-8602, Japan}}
\newcommand{\AFFsuny}{\affiliation{Department of Physics and Astronomy, State University of New York, Stony Brook, NY 11794-3800, USA}}
\newcommand{\AFFniigata}{\affiliation{Department of Physics, Niigata University, Niigata, Niigata 950-2181, Japan }}
\newcommand{\AFFokayama}{\affiliation{Department of Physics, Okayama University, Okayama, Okayama 700-8530, Japan }}
\newcommand{\AFFosaka}{\affiliation{Department of Physics, Osaka University, Toyonaka, Osaka 560-0043, Japan}}
\newcommand{\AFFseoul}{\affiliation{Department of Physics, Seoul National University, Seoul 151-742, Korea}}
\newcommand{\AFFshizuokasc}{\affiliation{Department of Informatics in
Social Welfare, Shizuoka University of Welfare, Yaizu, Shizuoka, 425-8611, Japan}}
\newcommand{\AFFshizuoka}{\affiliation{Department of Systems Engineering, Shizuoka University, Hamamatsu, Shizuoka 432-8561, Japan}}
\newcommand{\AFFskk}{\affiliation{Department of Physics, Sungkyunkwan University, Suwon 440-746, Korea}}
\newcommand{\AFFtohoku}{\affiliation{Research Center for Neutrino Science, Tohoku University, Sendai, Miyagi 980-8578, Japan}}
\newcommand{\AFFtokyo}{\affiliation{The University of Tokyo, Bunkyo, Tokyo 113-0033, Japan }}
\newcommand{\AFFtokai}{\affiliation{Department of Physics, Tokai University, Hiratsuka, Kanagawa 259-1292, Japan}}
\newcommand{\AFFtit}{\affiliation{Department of Physics, Tokyo Institute
for Technology, Meguro, Tokyo 152-8551, Japan }}
\newcommand{\AFFtsinghua}{\affiliation{Department of Engineering Physics, Tsinghua University, Beijing, 100084, China}}
\newcommand{\AFFwarsaw}{\affiliation{Institute of Experimental Physics, Warsaw University, 00-681 Warsaw, Poland }}
\newcommand{\AFFuw}{\affiliation{Department of Physics, University of Washington, Seattle, WA 98195-1560, USA}}

\AFFicrr
\AFFkashiwa
\AFFipmu
\AFFuam
\AFFbu
\AFFbnl
\AFFuci
\AFFcsu
\AFFcnm
\AFFduke
\AFFfukuoka
\AFFgifu
\AFFuh
\AFFkanagawa
\AFFkek
\AFFkobe
\AFFkyoto
\AFFmiyagi
\AFFnagoya
\AFFsuny
\AFFniigata
\AFFokayama
\AFFosaka
\AFFseoul
\AFFshizuoka
\AFFshizuokasc
\AFFskk
\AFFtokai
\AFFtokyo
\AFFtsinghua
\AFFwarsaw
\AFFuw
%

\AFFduke

\AFFkashiwa

\author{K.~Abe}
\AFFicrr
\author{Y.~Hayato}
\AFFicrr
\AFFipmu
\author{T.~Iida}
\author{M.~Ikeda}
\altaffiliation{Present address: Department of Physics, Kyoto University, Kyoto, Kyoto 606-8502, Japan}
\AFFicrr
\author{C.~Ishihara}
\author{K.~Iyogi} 
\author{J.~Kameda}
\author{K.~Kobayashi}
\author{Y.~Koshio}
\author{Y.~Kozuma} 
\author{M.~Miura} 
\AFFicrr
% ...

\collaboration{The Super-Kamiokande Collaboration}
\noaffiliation

\date{\today}

\begin{abstract}
The results of the third phase of the Super-Kamiokande solar neutrino measurement 
are presented and compared to the first and second phase results. 
With improved detector calibrations, 
a full detector simulation, and improved analysis methods, the systematic uncertainty on the total neutrino flux
is estimated to be $\pm 2.1\%$,
which is about two thirds of the systematic uncertainty for the first phase of Super-Kamiokande.
The observed $^8$B solar flux in the 5.0 to 20 MeV total electron energy region is 
2.32$\pm$ 0.04 (stat.) $\pm$ 0.05 (sys.) $\times 10^{6}~\text{cm$^{-2}$sec$^{-1}$}$ under the assumption of pure electron-flavor content,
in agreement with previous measurements. 
A combined oscillation analysis is carried out using 
 SK-I, II, and III data, and the results are also combined with the results of other solar neutrino experiments.
The best-fit oscillation parameters are obtained to be
$ \sin^2\theta_{12} = 0.30^{+0.02}_{-0.01}(\tan^2\theta_{12} = 0.42^{+0.04}_{-0.02})$ and 
$\Delta m^2_{21} = 6.2^{+1.1}_{-1.9}\times 10^{-5} \textrm{eV}^2$.
Combined with KamLAND results, the best-fit oscillation parameters are found to be
$ \sin^2\theta_{12} = 0.31 \pm 0.01 (\tan^2\theta_{12} = 0.44 \pm 0.03) $ and 
$\Delta m^2_{21} = 7.6 \pm 0.2 \times 10^{-5} \textrm{eV}^2$ .
The $^8$B neutrino flux obtained from global solar neutrino experiments is 
$5.3 \pm 0.2 $(stat.+sys.)$\times 10^{6}$cm$^{-2}$s$^{-1}$, while the $^8$B flux becomes 
$5.1 \pm 0.1 $(stat.+sys.)$\times 10^{6}$cm$^{-2}$s$^{-1}$ by adding KamLAND result.
In a three-flavor analysis combining all solar neutrino experiments, 
the upper limit of $\sin^2 \theta_{13}$ is 0.060 at 95\% C.L.. After combination with KamLAND results, 
the upper limit of $\sin^2 \theta_{13}$ is found to be 0.059 at 95\% C.L..

\end{abstract}


\maketitle

\section{Introduction}
The third phase of Super-Kamiokande (SK-III) began in October 2006 and ended in August 2008 when the
electronics were replaced. In SK-III, all 11129 PMTs have acrylic and
Fiber Reinforced Plastic (FRP) PMT covers (blast shields) which were
added at the start of SK-II in order to protect against propagating
shock waves from PMT implosions. 
In the inner detector, the active photodetector coverage is 40\% ( 40\% in SK-I and 19\% in SK-II).
Thanks to detector improvements and
superior analysis techniques, the SK-III's solar neutrino flux
measurement is more precise than either SK-I's \cite{sk_full_paper}(SK
before the accident) or SK-II's~\cite{sk_ii_paper} (SK with 46.5$\%$
of its PMTs) even with an exposure of only two years. In particular, the water purification system, event reconstruction and selection tools, as well as Monte Carlo detector simulation were improved.
They will be explained in Chapter II and III in detail.

In Chapter IV, the results of oscillation analyses are presented.
By adding SK-III data, it was found that 
the energy spectrum and the time variation of solar neutrinos obtained from our measurements
favor only the large mixing angle solution (LMA) by constraining the 
$^8$B  and hep neutrino flux to SNO NC flux~\cite{snoncdn,snoleta} and 
Standard Solar Model (SSM) prediction~\cite{ssm} respectively.
The first result of three-flavor neutrino oscillation analysis with the full SK data set 
will be shown as well as the two-flavor analysis result.
In the last section of Chapter IV, the $^8$B flux value obtained from the results of 
all solar neutrino experiments (global solar analysis) will be shown to compare with 
the prediction of the SSM.

\section{SK-III Performance}

\subsection{Water system}
 A major background for the solar neutrino observation at SK is the
 radioactivity from radon (Rn) from the U/Th decay chain in the
 water. The water in the detector is
 made from natural mine water using a very high performance water
 purification system.  Even though the water is extremely pure there is
 still some Rn remaining. The Rn background events are very similar to
 solar events, so it is very difficult to remove them using only
 analysis tools. To reduce it, we have upgraded the system since
 the end of SK-I, including the addition of a new heat exchanger and
 two reverse osmosis units during  the SK-II and III periods.
 
 In addition, we investigated the water flow in the detector by
 intentionally injecting radon-enriched water. Tracing the resulting
 background events in time from this injected Rn, we found stagnation
 of water in the top and bottom of the detector volume, which increased
 the background.  To counter this effect, we installed new pipes and
 changed the water flow. Previously, the water was supplied from the
 bottom of the inner detector (ID) and drained from the top of both the ID and outer
 detector (OD). Now, it is supplied from the ID bottom and drained from the top
and bottom in OD and the top in ID with a total flow of 60 tons/hr,
 which is two times faster than before. This final setting has been in
 effect since August, 2007. As a result of these improvements, we have
 a central region with half of SK-I's background, enabling a lowering of the energy threshold. 
 
 Note that the excessive background near the wall and bottom
 consisting of $\gamma$ rays due to the FRP cover also existed in SK-II. This background could not be reduced by improving the water system. 

\subsection{Event Reconstruction}
\subsubsection{Vertex}
The event vertex reconstruction for solar neutrino analysis performs a maximum likelihood fit
to the timing residuals of the Cherenkov signal as well as the dark noise background for
each testing vertex \cite{bonsai_smy}.  
The vertex reconstruction method in SK-III was initially installed in SK-II and further improved over SK-II.
It now has better resolution than SK-I.

\begin{figure}
 \begin{center}
 \includegraphics[width=7cm,clip]{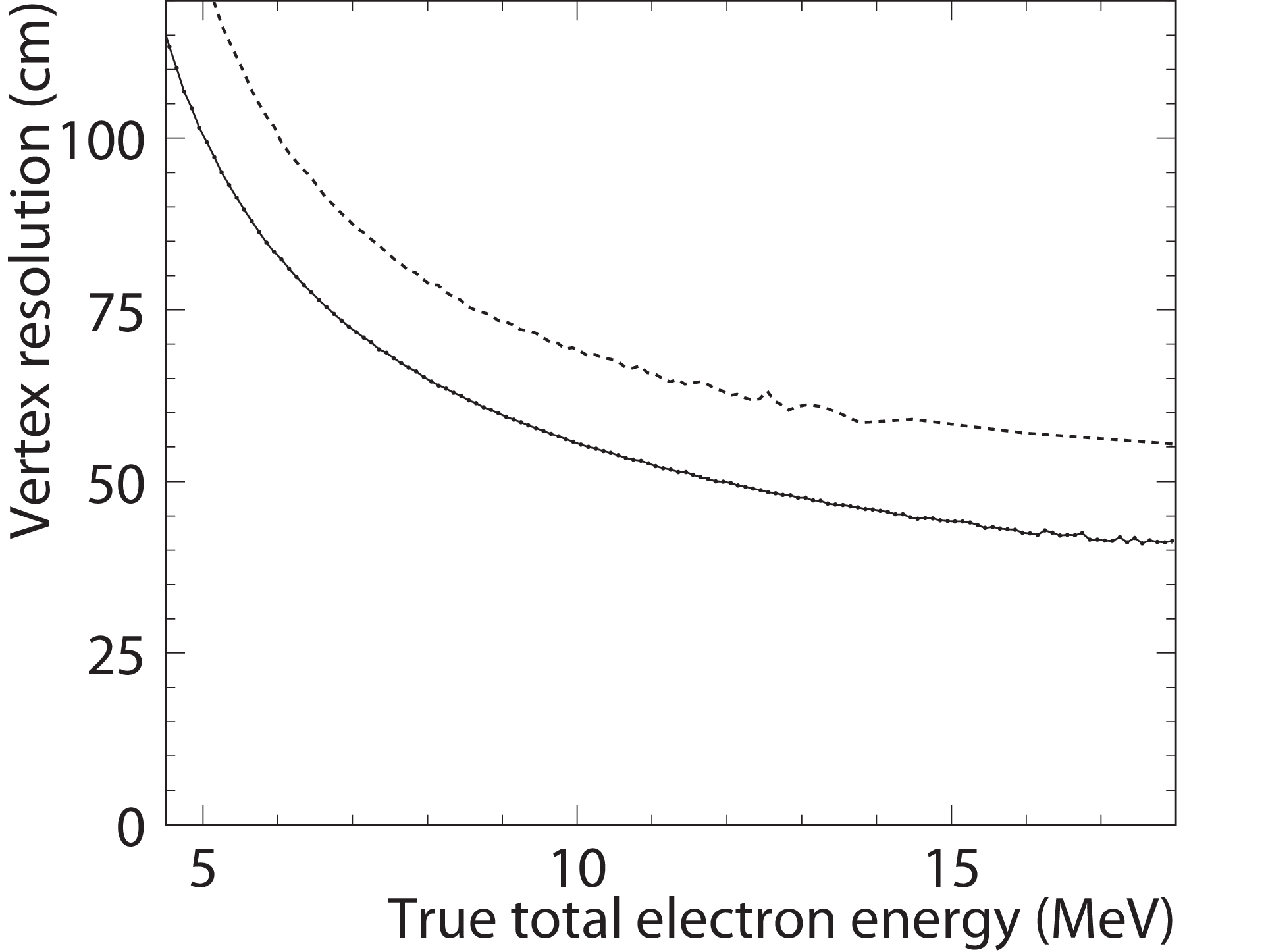} 
 \caption{Solid line shows the vertex resolution for SK-III as a function of the true total electron energy, while the dashed line shows that of SK-I.
  }\label{fig:vres}
 \end{center}
\end{figure}

Figure \ref{fig:vres} shows the vertex resolution for SK-III. The
vertex resolution is defined as the distance from the true vertex
position containing 68.3\% of all reconstructed vertices. The vertex
resolution in SK-I for 5~MeV electrons is 125~cm; here it is improved to 100~cm.

A bias in vertex reconstruction is called the ``vertex shift''.
The vertex shift is defined as the vector from the averaged vertex of
reconstructed events to that of the corresponding simulated Monte
Carlo (MC) events. Because the vertex shift results in events moving
in or out of the fiducial volume, it represents one of the main
systematic uncertainties for the solar neutrino flux measurement.

\begin{figure}
 \begin{center}
 \includegraphics[width=7cm,clip]{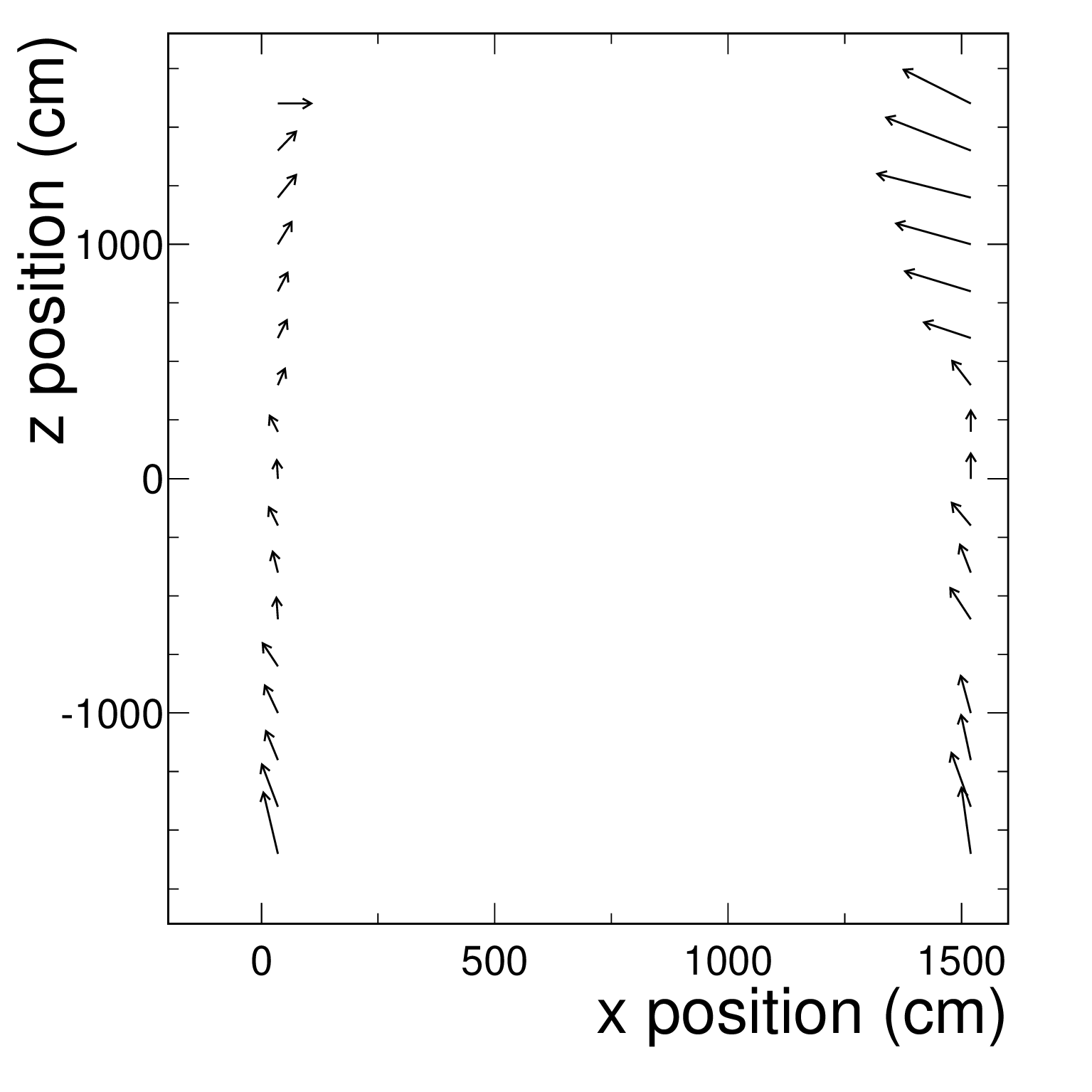} 
 \caption{Vertex shift of Ni calibration events. The origin of the
   arrows shows the true Ni source position and the direction
   indicates the averaged reconstructed position direction. The length
   of the arrow indicates the magnitude of the vertex shift. All
   vertex shifts are scaled by a factor of 20 to make them easier to see.
}\label{fig:versh}
\end{center}
\end{figure}

\begin{figure}
 \begin{center}
 \includegraphics[width=6cm,clip]{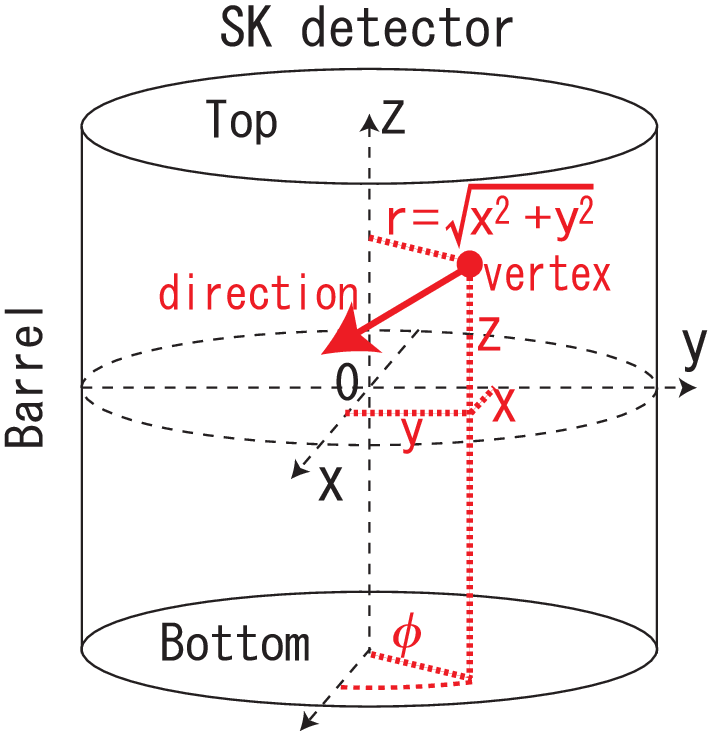} 
 \caption{ Definition of detector coordinate system.
  }\label{fig:def}
 \end{center}
\end{figure}

The vertex shift is measured by placing a Ni-Cf gamma ray
source~\cite{SKnim} at several positions inside the detector 
(Hereafter, the calibration using this Ni-Cf source is called ``Ni calibration'' or ``Ni events'').
 The reconstructed data vertices at the fiducial volume edge were shifted
more than 10~cm from the real source position inward toward the detector
center, while those of the MC simulation were shifted less than
3~cm.  It was found that this shift in data was due to an electronic
effect of the relative hit timing within a wide range ($\sim$100
nsec). We measured the timing linearity by artificially shifting the
common stop signal of individual TDCs for each hit channel.
We found that a correction of -0.7$\%$ to the hit
timing was required to restore linearity. After the correction was
applied, the vertex shift shortened
significantly. Figure~\ref{fig:versh} shows the vertex shift in
SK-III with timing correction. The definition of x and z in Figure
\ref{fig:versh} (and other variables) is explained in Figure
\ref{fig:def}.  

\subsubsection{Direction}

The direction reconstruction is based on the SK-I method: a likelihood
function is used to compare Cherenkov ring patterns between data and
MC simulation.  An energy dependence is now included in the
likelihood function for SK-III. The ring pattern distributions and
their energy dependences are simulated for several energy ranges using electron MC simulation events. 

\begin{figure}
 \begin{center}
 \includegraphics[width=7cm,clip]{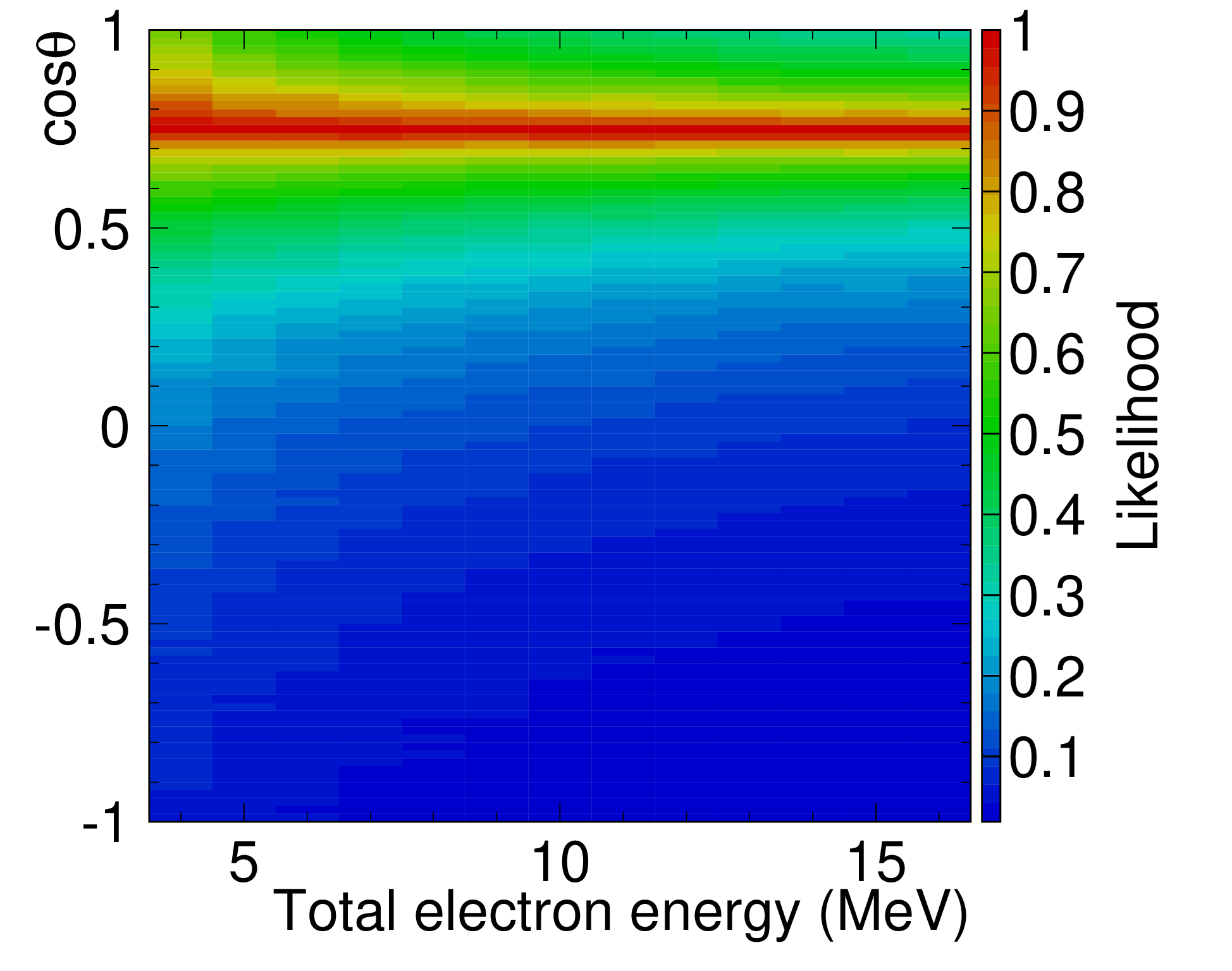} 
 \caption{ Likelihood value for reconstruction of event direction as a function of the reconstructed total electron energy and opening angle between the reconstructed direction and the direction from vertex to each hit PMT. 
  }\label{fig:dirlike}
 \end{center}
\end{figure}

\begin{figure}
 \begin{center}
 \includegraphics[width=7cm,clip]{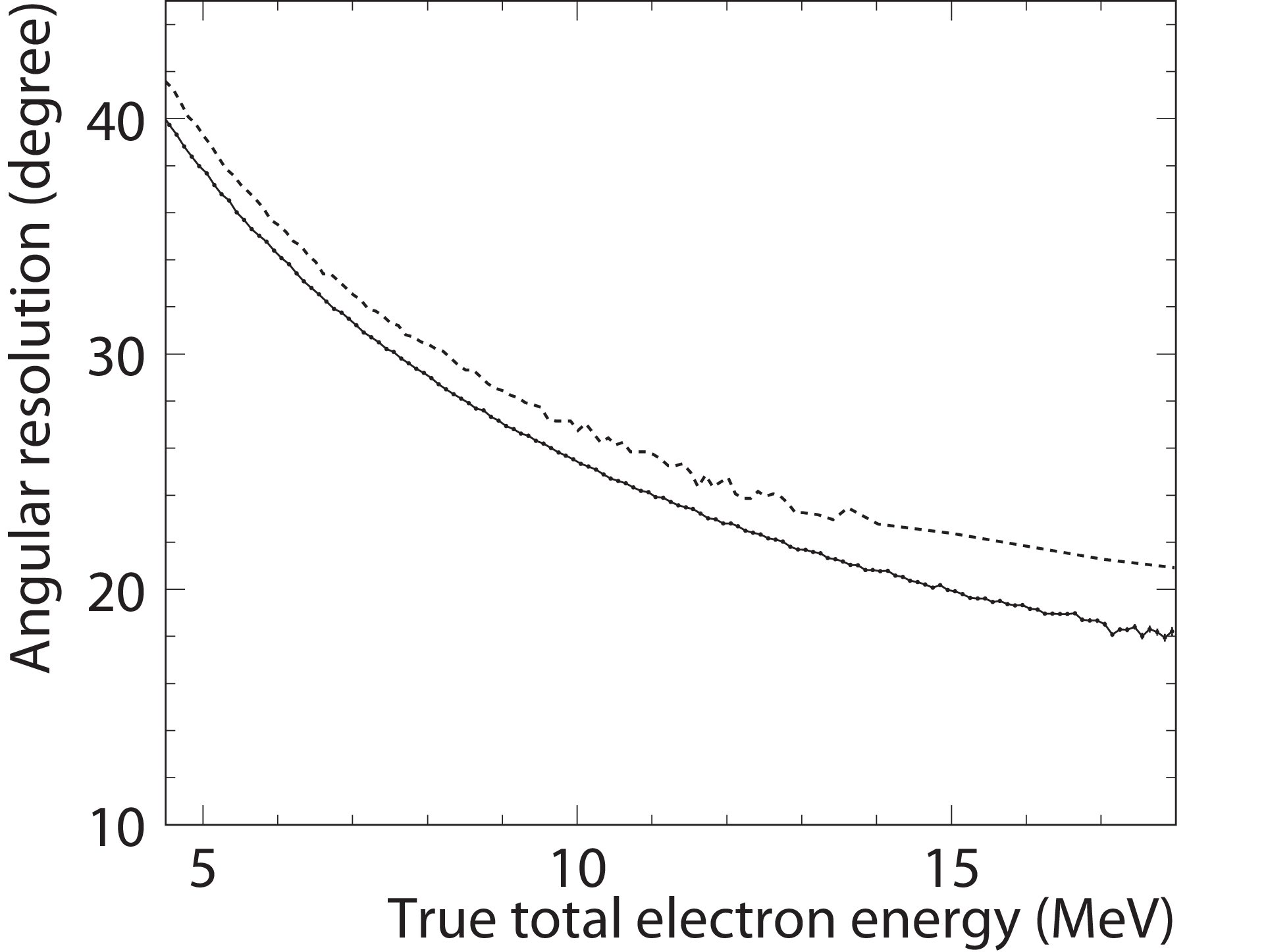} 
 \caption{ The solid line shows the angular resolution of SK-III as a function of the true total electron energy, while the dashed line shows that of SK-I.
  }\label{fig:dirres}
 \end{center}
\end{figure}

Figure \ref{fig:dirlike} shows the likelihood as a function
of the reconstructed event energy and the opening angle between the reconstructed
direction and the direction from the vertex to each hit PMT.
Figure \ref{fig:dirres} shows the absolute angular resolution, which is
defined as the angle of the cone around the true direction containing
68.3\% of the reconstructed directions. For SK-III, the angular
resolution is improved compared to SK-I by about 10\% at 10 MeV and is
close to the limit due to multiple Coulomb scattering of
electrons. Note that the improvement of vertex reconstruction also
contributes to the improvement of angular resolution, especially in
the low energy region below 6.5~MeV.

\subsubsection{Energy}
The reconstruction of event energy is similar to that for
SK-I~\cite{sk_full_paper}. The most important modifications with respect to
SK-I  are due to photo-cathode coverage and blast shields.  Starting with the
number of in-time hit PMTs ($N_{50}$ coincident within 50 ns after the
subtraction of time-of-flight (TOF) of Cherenkov photon from the
reconstructed vertex to the hit PMT position), several
corrections, described below, are made.  The resulting effective hit sum
$N_{\mbox{\tiny{\textit{eff}}}}$ has less position dependence than
$N_{50}$.  From $N_{\mbox{\tiny{\textit{eff}}}}$, we determine energy.
This procedure is further outlined in ~\cite{sk_full_paper}, and is also explained in \cite{sk_ii_paper}.
 
The definition of $N_{eff}$ is:
%
%
\begin{eqnarray}
 N_{eff} &=& \sum_{i}^{N_{50}} \{ (X_{i}+\epsilon_{tail}-\epsilon_{dark})\times \frac{N_{all}}{N_{alive}}\nonumber \\ 
  && \times \frac{1}{S(\theta_i,\phi_i) } \times  \exp(\frac{r_i}{ \lambda(t)}) \times G_i(t)  \} 
\end{eqnarray}

\noindent where the explanations for the factors are as follows:
\begin{description}

\item[$X_i$:]
This factor estimates the effect of multiple photo-electrons in the i-th hit
PMT.
If an event occurs close to a detector wall 
and is directed towards the same wall, the Cherenkov cone does not have
much distance to expand, and the observed number of hits is small.
The correction $X_i$ for this effect is defined as
%
%
\begin{equation}
X_i = \left\{
\begin{array}{ll}
 \frac{\log \frac{1}{1-x_i} }{x_i}, &\quad x_i < 1 \\
 3.0, &\quad x_i = 1
\end{array}
\right.
\end{equation}
where $x_i$ is the ratio of hit PMTs in a 3$\times$3 PMT region
surrounding the i-th PMT to the total number of live PMTs
in the same area. The $-\log(1-x_i)$ term is the estimated number of photons
per one PMT in that area and is determined from Poisson statistics.
When $x_i=1$, 3.0 is assigned to $X_i$.

\item[ $\epsilon_{tail}$:]
Some Cherenkov photons being scattered or reflected arrive late at the PMT,
and make late hits outside the 50~nsec time window.
To correct the effect of the late hits, the term 
%
%
\begin{equation}
\epsilon_{tail} = \frac{N_{100}-N_{50}-N_{alive}\times R_{dark}\times 50~n\text{sec}}{N_{50}}
\end{equation}
is added where $N_{100}$ is the maximum number of hits found by 
a 100~nsec sliding time window search.

\item[$\epsilon_{dark}$:]
This factor corrects for hits due to dark noise in the PMTs. 
%
%
\begin{equation}
\epsilon_{dark} = \frac{50n\text{sec} \times N_{alive} \times R_{dark}}{N_{50}}
\end{equation}
where $N_{alive}$ is the number of all live inner detector (ID) PMTs and $R_{dark}$
is the measured dark rate for a given data taking period.

\item[$\frac{N_{all}}{N_{alive}}$:]
This factor is for the time variation of the number of dead PMTs.
$N_{all}$ is total number of PMTs and for SK-III it is 11129.

\item[$\frac{1}{S(\theta_i,\phi_i)}$:]
This factor accounts for the direction-dependent photocathode coverage.
$S(\theta_i,\phi_i)$ is the effective photocathode area of the i-th hit PMT as 
viewed from the angles ($\theta_i, \phi_i$) to take into account the shadowing of PMTs for
glancing angles $\theta_i$.
$S$ is determined by MC simulation with the FRP PMT covers;
the resulting $N_{eff}$ is checked by an electron linear accelerator
(LINAC) and an in-situ deuterium-tritium neutron generator (DT) calibration data.

\item[$\exp(\frac{r_i}{\lambda(t)}$):]
The water transparency is accounted for by this factor, where $r_i$ is the
distance from the reconstructed vertex to the i-th hit PMT. $\lambda(t)$
is the measured water transparency for a given data taking period.

\item[$G_i (t)$: ]
This factor adjusts the relative quantum efficiency of the PMTs.
 The differences in the quantum efficiency depend on the fabrication date
of the PMTs.
\end{description}

\begin{figure}
 \begin{center}
 \includegraphics[width=7cm,clip]{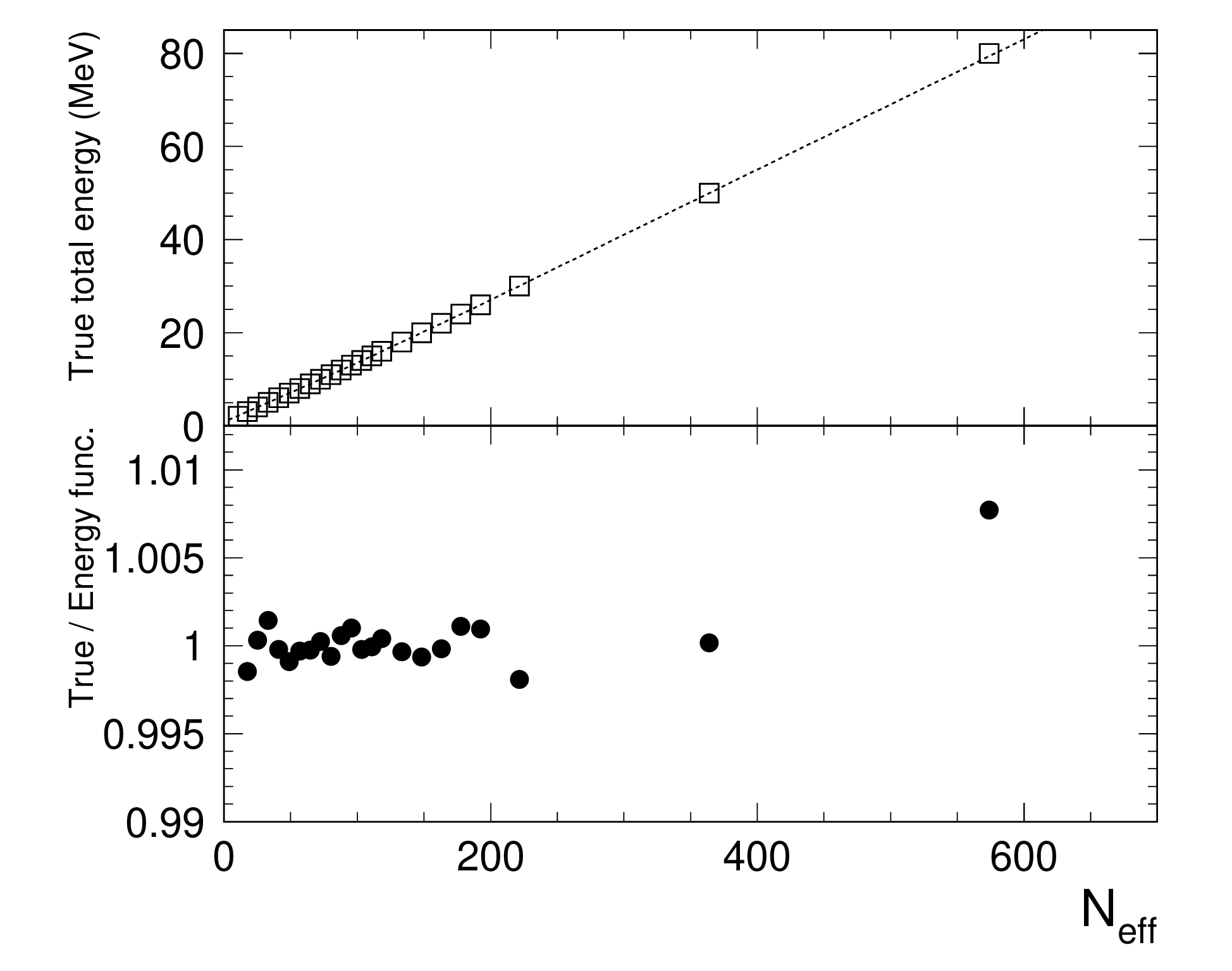}
 \caption{Relation between $N_{eff}$ and the true total electron energy (MeV) obtained from MC simulation.
The dotted line in the upper figure shows the fitted polynomial function.
The lower figure shows the deviation of the reconstructed energy from the polynomial function. \label{fig:neff_vs_energy}
}
 \end{center}
\end{figure}

After determining $N_{eff}$, an event's energy in MeV 
can be calculated as a function of $N_{eff}$. 
The relation between $N_{eff}$ and MeV is obtained using mono-energetic electron MC simulated events 
as shown in Figure \ref{fig:neff_vs_energy}.
The conversion function from $N_{eff}$ to MeV is determined by fitting the relation with
a fourth-order polynomial function for the lower energy region ($< 25$MeV)
and a first-order polynomial function for the  higher energy region ($\geq 25$ MeV).

The systematic uncertainty of the reconstructed energy is 
checked by LINAC and DT calibration. 

When calculating the energy for data events, the water transparency value,
as determined by decay electrons from cosmic-ray muons, measured for
six-day intervals, is used as an input parameter. 
For MC events, the change in water transparency and the relative quantum efficiency is simulated.

\begin{figure}
 \begin{center}
 \includegraphics[width=7cm,clip]{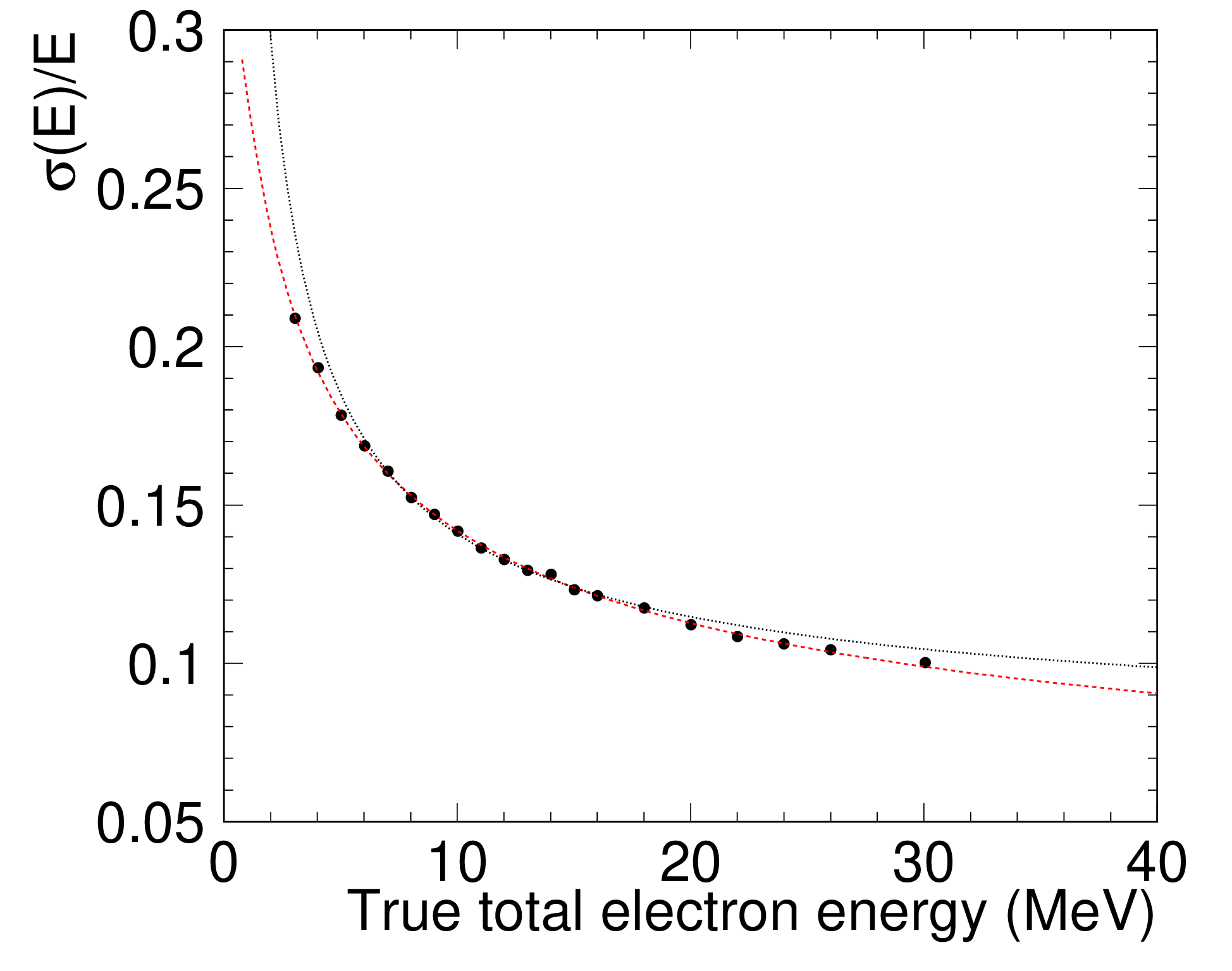} 
 \caption{Energy resolution function obtained by electron MC simulation\label{figeres}.
Black points show one standard deviation for a Gaussian fit of the MC simulation divided by the true total electron energy, 
while the red (dashed) line shows a fit to a polynomial function. The black (dotted) line shows SK-I energy resolution.
}
 \end{center}
\end{figure}

The detector's energy resolution is well described by a Gaussian function.
The energy resolution is described by
\begin{equation} \label{eq:eres2}
\sigma(E)=-0.123 + 0.376 \sqrt{E} + 0.0349 E,
\end{equation}
\noindent in units of MeV (see Figure \ref{figeres}). 
The SK-I resolution is $\sigma=0.2468 + 0.1492 \sqrt{E} + 0.0690 E$
which is shown also in Figure~\ref{figeres}. 
For low energy SK-III events, the energy resolution is
improved by 5\%, which is mainly due to the improved vertex reconstruction.

\subsection{Energy calibration}
As for SK-I, the primary instrument for energy calibration in SK-III is LINAC.  A detailed discussion of the LINAC calibration methods can be found elsewhere~\cite{linac}.  Single electrons are injected into the SK detector at various positions
and at energies between 4.4 and 18.9 MeV.  However, we could only take
data with 4.4 MeV electrons at two positions, because the tuning of the electron beam
is difficult for the lower energies. For this reason we did not include the 4.4 MeV
data, and the lowest energy we included in this analysis is 4.7 MeV. The reconstructed
energies of LINAC events are compared against those of the MC simulation to
determine the energy scale. The absolute correction factor for PMT
quantum efficiency was tuned to minimize any deviation between data
and MC. The effect of the water transparency change on the energy
scale was estimated as 0.22\% by averaging over all energies and
positions. The uncertainty of the electron beam energy, determined by a
Ge detector measurement, is 0.21\% (the same as for SK-I).

In addition to the LINAC calibration,  energy scale calibration is
done using $^{16}$N produced with DT~\cite{dtg}. The generated 14.2~MeV neutrons
exchange their charge with $^{16}$O. The produced $^{16}$N decays into
$\beta$ and $\gamma$  with a half-life of 7.13 seconds.  With a
Q-value of 10.4~MeV, $^{16}$N most probably decays into $^{16}$O, an
electron with maximum energy 4.3~MeV and a $\gamma$ ray of energy
6.1~MeV. The peak value of the energy distribution is taken to evaluate the energy scale. DT data-taking is faster than LINAC data taking, so more positions can be checked. 

The position dependence of the energy scale systematic uncertainty was
estimated  using only LINAC calibration data for SK-I and II, while for
SK-III, in addition DT calibration data are used to  take
into account the z-dependence and $\phi$-dependence of the energy scale.
The z-dependence is measured by LINAC, and the $\phi$-dependence is measured by DT.
The LINAC can only take data at $\phi= 180$ degrees, whereas the DT
generator can take data at five different positions at the 
same z position: center(r=4~m), $\phi=$ 0, 90, 180, and 270 degrees ($r=12$m).

\begin{figure}
 \begin{center}
 \includegraphics[width=7cm,clip]{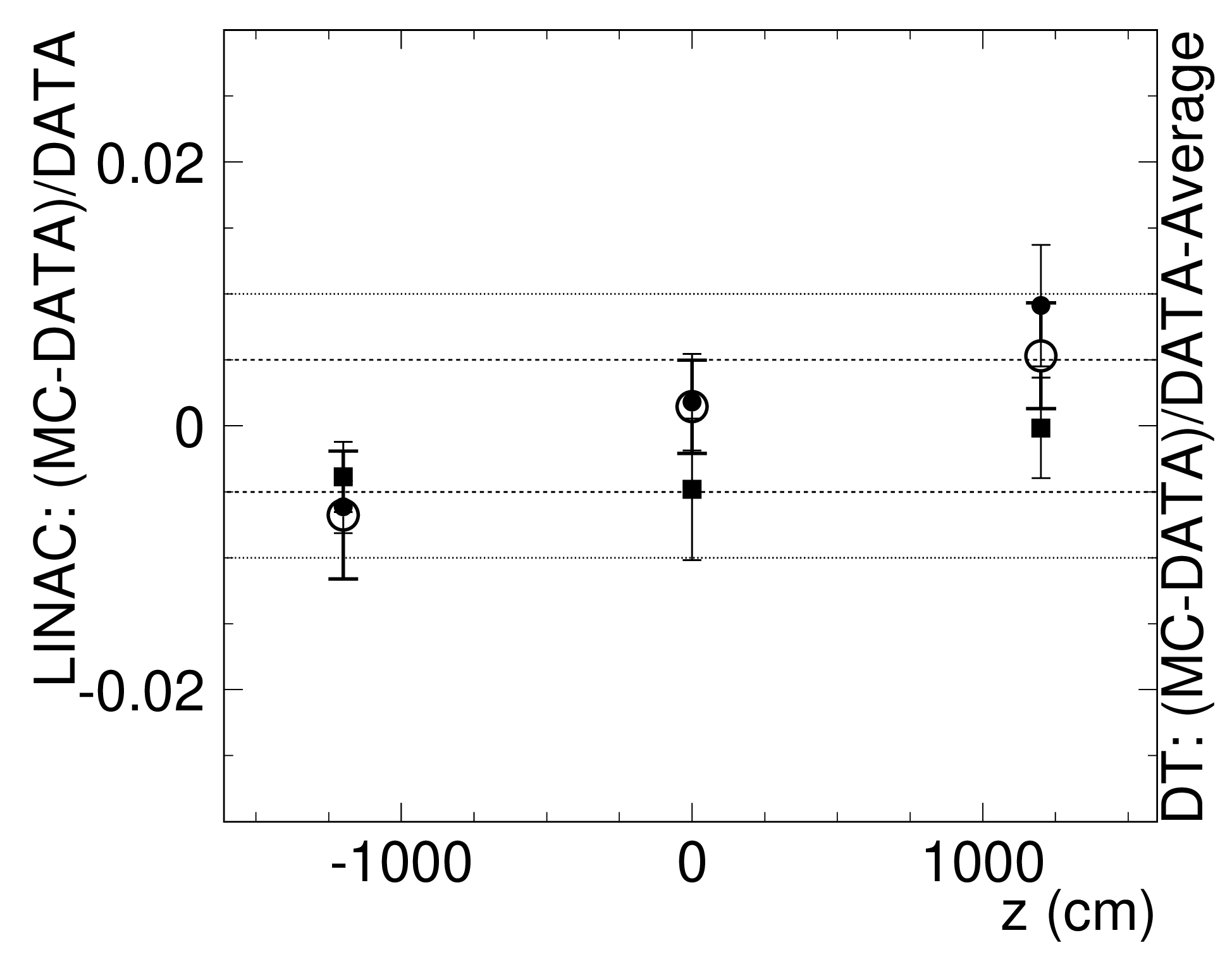} 
 \caption{ The z dependence of the energy scale measured by LINAC calibration. The marker shows an average of (MC-DATA)/DATA over four energies at each data-taking point. The error bars show the RMS of (MC-DATA)/DATA for four energies. 
 The  filled circle markers are for R=4~m (x=-4~m, y=0~m) and the square markers are for
r=12~m (x=-12~m, y=0~m). 
The open circle markers show the z dependence obtained by DT calibration.
The dashed and dotted lines show $\pm$1 and 0.5\%, respectively. The
  edge of the fiducial volume is the same as the edge of the plot window (from -1610~cm to 1610~cm).
  }\label{fig:e_zdep}
 \end{center}
\end{figure}

\begin{figure}
 \begin{center}
 \includegraphics[width=7cm,clip]{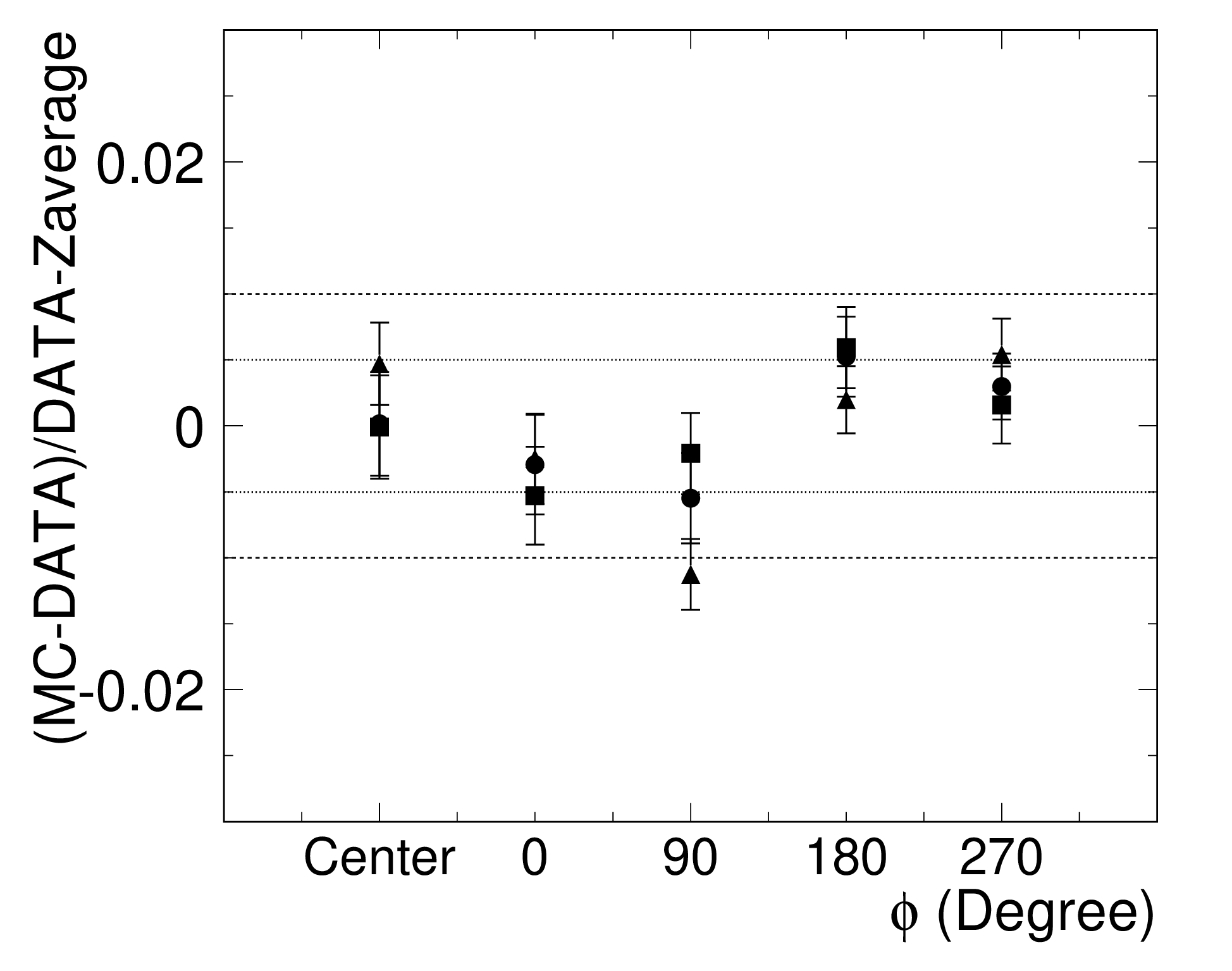} 
 \caption{ The $\phi$ dependence of the energy scale measured by DT calibration. The marker shows  (MC-DATA)/DATA for each point and the error bars show statistical uncertainty.
 The circle, square, and triangle markers are for z=+12~m, 0~m, and -12~m
 respectively. Zaverage is the average difference of energy scale between MC simulation and DT data obtained at the same z-position. The dashed and dotted lines show $\pm$1 and 0.5\%, respectively.
  }\label{fig:e_phidep}
 \end{center}
\end{figure}

Figure \ref{fig:e_zdep} shows the difference of the energy scale
between LINAC data and LINAC MC as a function of LINAC position.
By averaging over all positions, the z-dependence 
is estimated to be  0.06$\%$
(the difference between r=4~m and r=12~m is also included here.)

Figure \ref{fig:e_phidep} shows the difference of energy scale
between DT data and DT MC. The vertical axis is normalized
by the average of measurements at the same z positions. The mean value of the deviation
from the average is taken as the $\phi$-dependence of energy scale, 
which is 0.35$\%$.
A resulting uncertainty of $\pm$ 0.35\% for the overall position dependence is estimated.

\begin{figure}
 \begin{center}
 \includegraphics[width=7cm,clip]{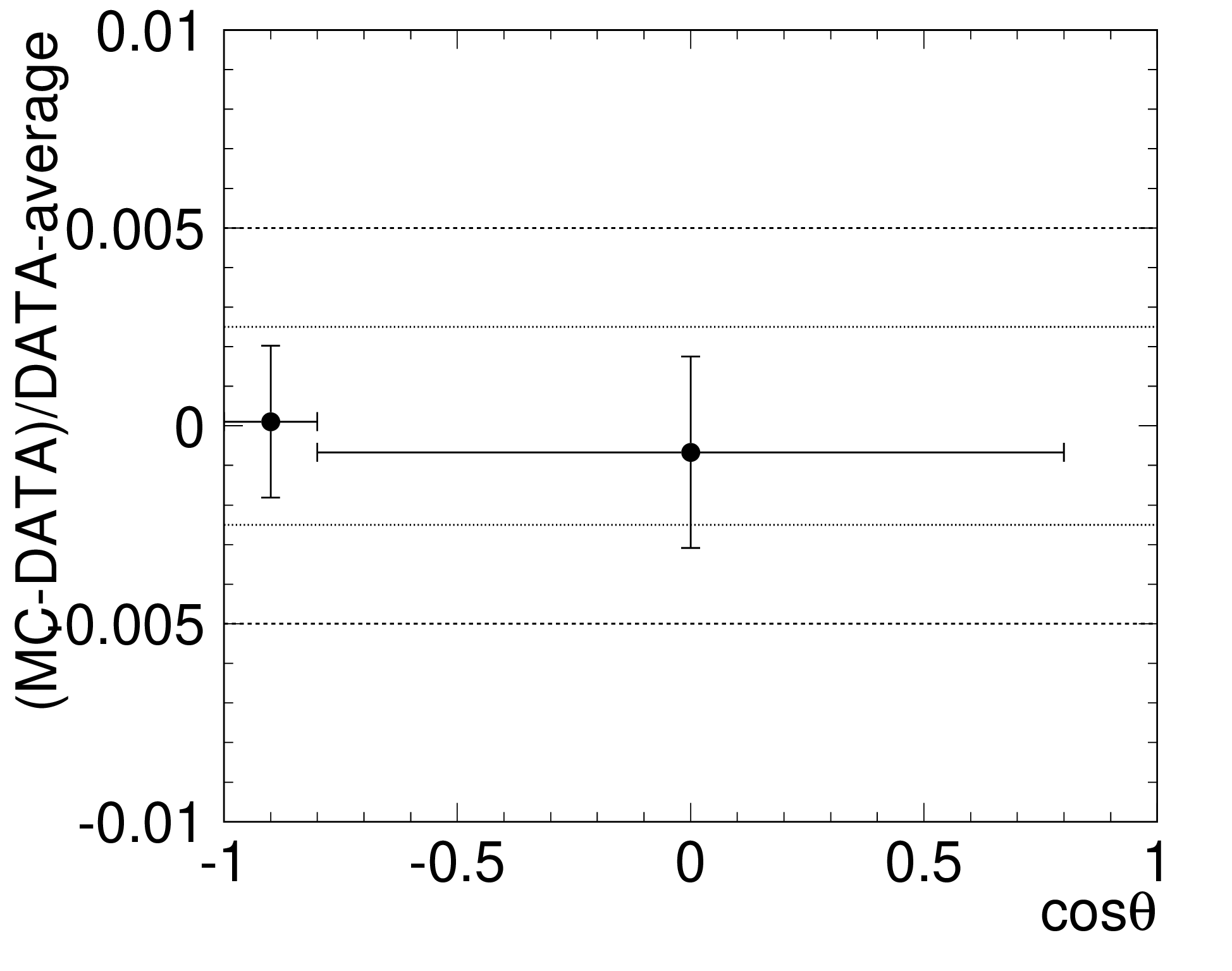} 
 \caption{ Energy scale difference between LINAC direction and other
   direction measured by DT calibration(LINAC direction range includes 76\% of 7 MeV electrons). The solid and dotted lines
   show $\pm$1 and 0.5\%, respectively.
  }\label{fig:e_dir}
 \end{center}
\end{figure}

$^{16}$N decays allow directional studies of the energy scale which
are not possible with the LINAC beam.  The observed energy at several
 positions in the detector is compared with the MC-simulated energy and the difference
is shown to agree with values obtained from LINAC data and MC.  The
$^{16}$N energy scale difference between LINAC direction (downward
direction) and the average of the other directions is estimated as $\pm$0.25\%, as shown in Figure \ref{fig:e_dir}. This difference is taken as the directional uncertainty of the energy scale.  

Finally, the energy scale uncertainty is calculated to be 0.53\% which is summarized in Table \ref{tabesys}.  This is slightly smaller than the SK-I estimated value of 0.64\%.  
\begin{table}
\begin{center}
\begin{tabular}{l l} \hline\hline
Position dep. & 0.35 $\%$\\ 
Direction dep. & 0.25 $\%$\\ 
Water transparency & 0.22 $\%$\\ \
LINAC energy sys. & 0.21 $\%$\\ \hline
Total  & 0.53 $\%$\\ \hline\hline
\end{tabular}
\end{center}
\caption{Systematic uncertainty of the energy scale\label{tabesys}.}
\end{table}

\begin{figure}[htbp]
\begin{center}
\includegraphics[width=7cm,clip]{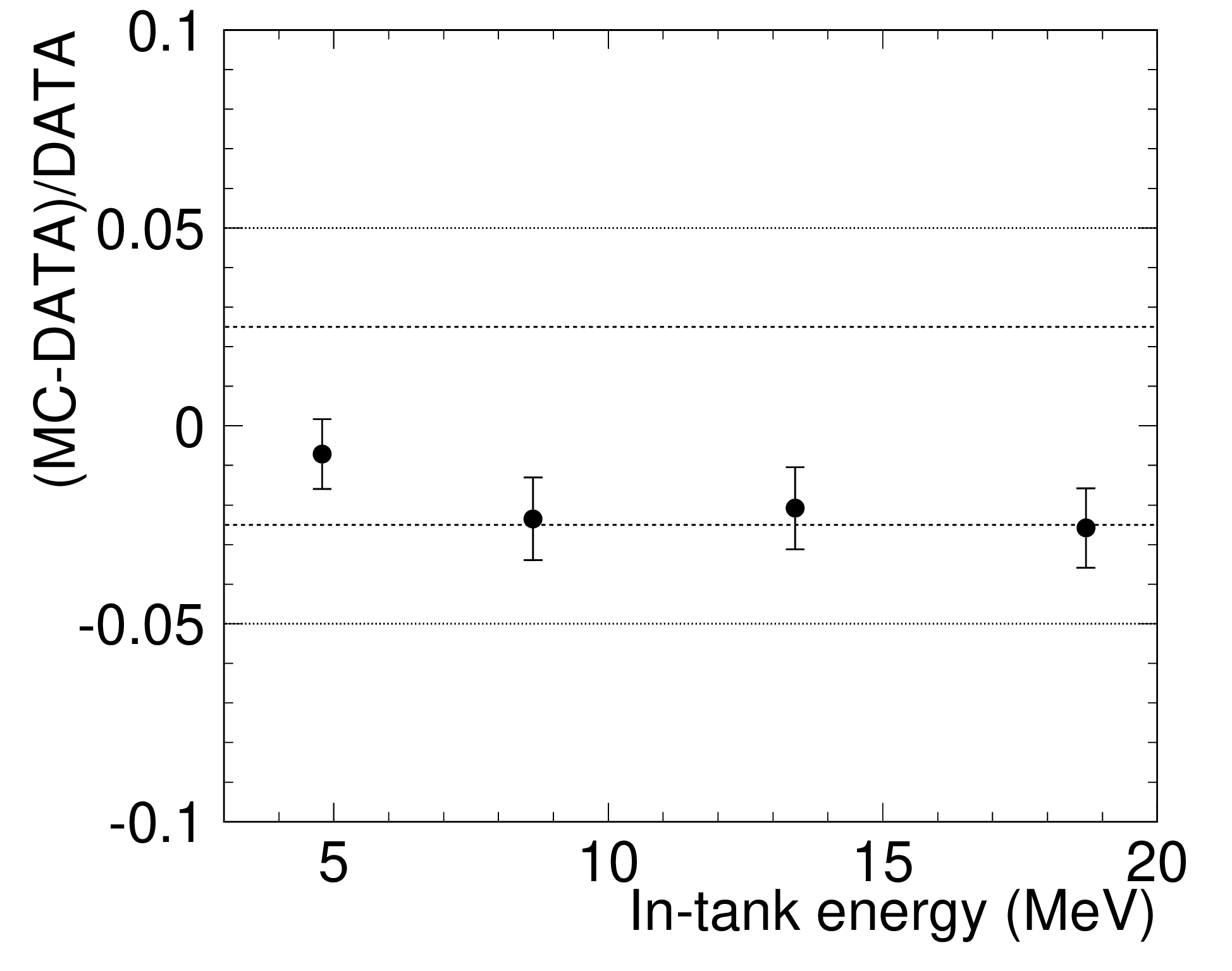}
\caption{Energy resolution difference between MC simulation and data as a function of energy, obtained by LINAC calibration.}
\label{fig:eres_sys}
\end{center}
\end{figure}

Energy resolutions of LINAC events are also compared for data and MC simulation.
Figure \ref{fig:eres_sys} shows the difference of the energy resolution between data and MC
as a function of the total electron energy. From Figure~\ref{fig:eres_sys}, 
$\pm$2.5\% systematic uncertainty is assigned to energy correlated systematic uncertainty 
for the spectrum measurement.

\begin{figure}[htbp]
\begin{center}
\includegraphics[width=7cm,clip]{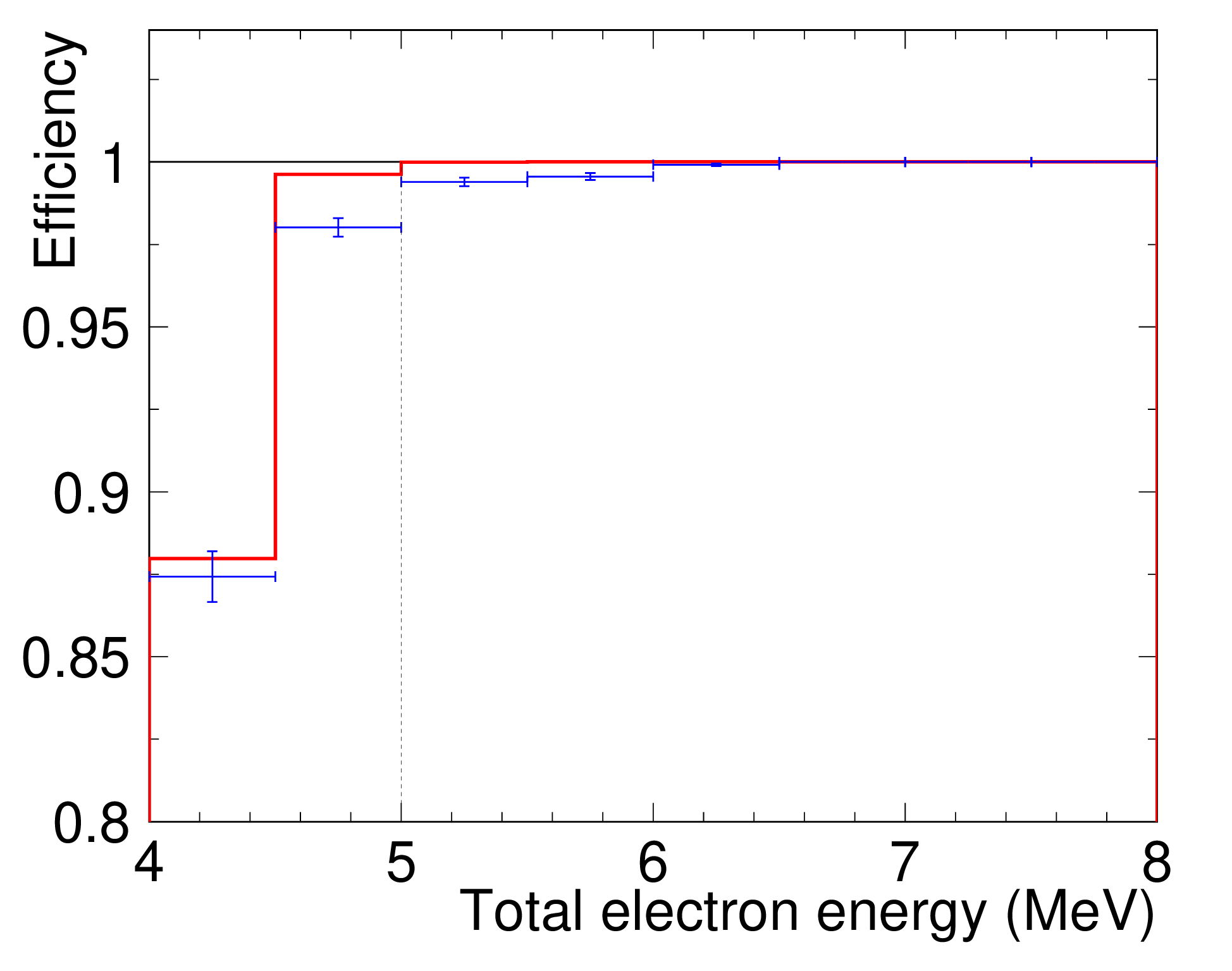}
\caption{Trigger efficiency as a function of energy.  Markers are $^{16}$N calibration data and the solid histogram is MC simulation. The vertical dashed line shows the analysis threshold, 5.0~MeV.}
\label{fig:trigger_efficiency}
\end{center}
\end{figure}

Quantitative estimates of trigger efficiencies are also obtained from
$^{16}$N data.  The lowest hardware threshold setting has been in effect
since April 2008.  At this setting, the SK-III trigger achieved more than 99\% efficiency at
4.5~MeV total electron energy. Before this time, the trigger efficiency was more than 99\% at 5.0~MeV total electron energy. Figure~\ref{fig:trigger_efficiency} shows the trigger efficiencies of the lowest threshold period in SK-III. 

\subsection{Optical calibration}
\subsubsection{Light Propagation in Water}

\begin{figure}[]
 \begin{center}
 \includegraphics[width=7cm,clip]{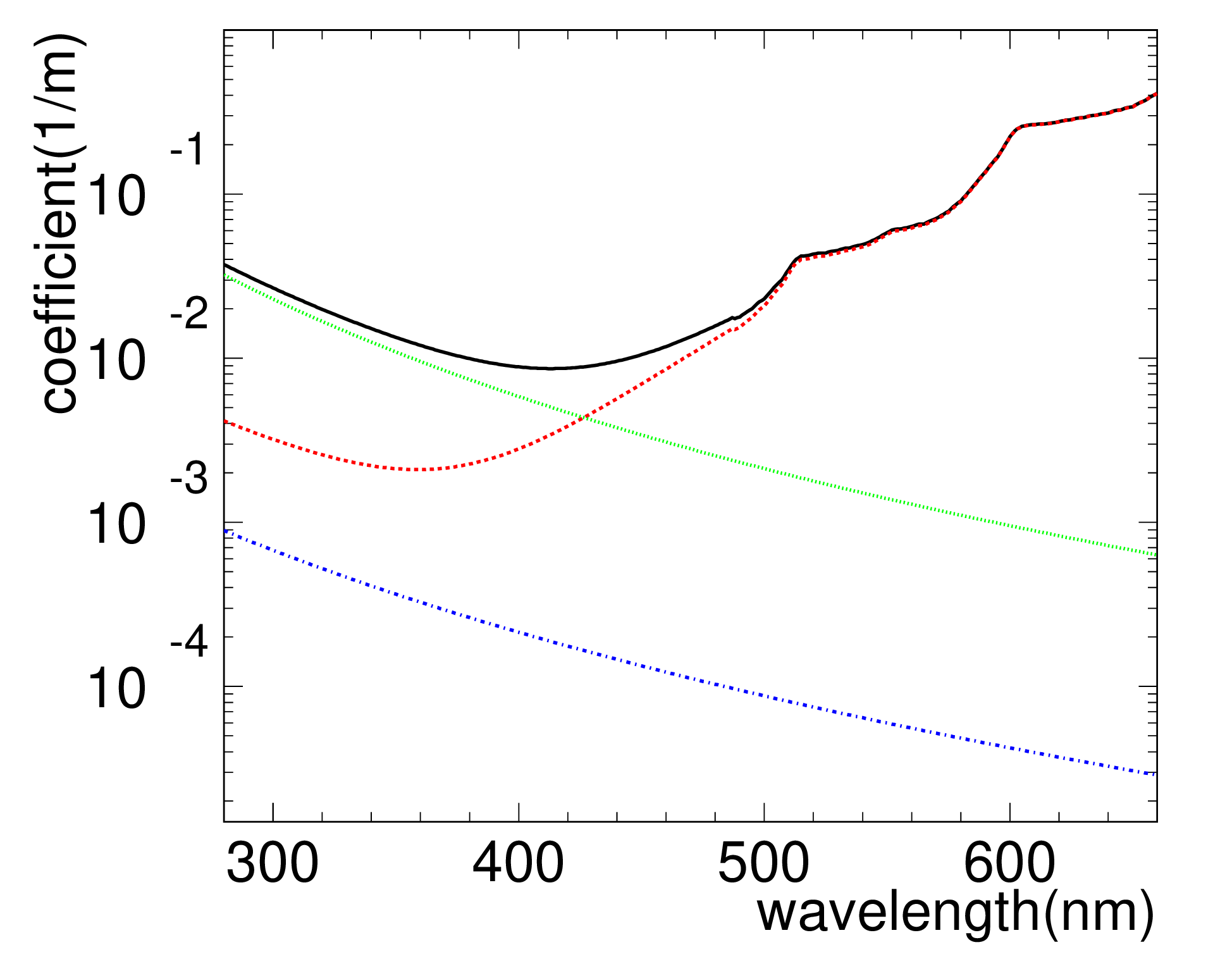} 
 \caption{Wavelength dependence of the water coefficients: scattering and absorption combined (black, solid), 
 absorption (red, dashed), Rayleigh scattering (green, dotted) and Mie scattering (blue, dashed-dotted).
The absorption coefficient is a function of water transparency and
the water transparency in this graph is 139 m.}
  \label{fig:water_par}
 \end{center}
\end{figure}

 For light propagation in water, a three-part model of light propagation consisting of
 absorption and two kinds of scattering is adopted for SK-III (as well
 as both SK-I and SK-II). In contrast to the previous phases, for
 SK-III the models are tuned using nitrogen/dye laser calibration. We
 measured the attenuation length of scattering and absorption for four
 wavelengths (337, 365, 400 and 420~nm) and tuned the water
 coefficients based on the measurement. The water coefficients are
 described by:
\begin{equation*}
\begin{split}
\alpha_{ray}(\lambda) & = \frac{r_1}{\lambda^4} \left(r_2 + \frac{r_3}{\lambda^{r_4}}  \right)\\
\alpha_{mie}(\lambda) & = \frac{m}{\lambda^4}\\
\alpha_{abs}(\lambda) & = \frac{a}{\lambda^4} + \alpha_{\mbox{\tiny long}}(\lambda,a)\\
\end{split}
\end{equation*}
Based on the data of Feb. 2007 and 2008, we determined $r_i$, $m$, and
$a$.  $\alpha_{\mbox{\tiny long}}(\lambda,a)$ was determined by a
third-party independent measurement introduced in
SK-II\cite{icam}\cite{sk_ii_paper}. Near the 400~nm region, we
modified the relation using nitrogen/dye laser calibration. The
crossing point of the SK and independent measurements\cite{pandf} is varied by
another parameter ($a$). The results of the determination of these parameters are shown in Figure \ref{fig:water_par}.
 
The time dependence and position dependence of the water quality are
described in the following.

\paragraph{Time Dependence}

The time variations of the water coefficients and the water
transparency are measured simultaneously by nitrogen/dye laser calibration and using
decay electrons from cosmic-ray muons. These
measurements confirmed that the change of the water transparency is
mainly caused by the change of the absorption coefficient. We obtained
the relation of the absorption coefficient and the water transparency
in SK-III using those data as well as water transparency as measured
by decay electrons from cosmic-ray muons, as for SK-I.

\begin{figure}[htbp]
\begin{center}
\includegraphics[width=8cm,clip]{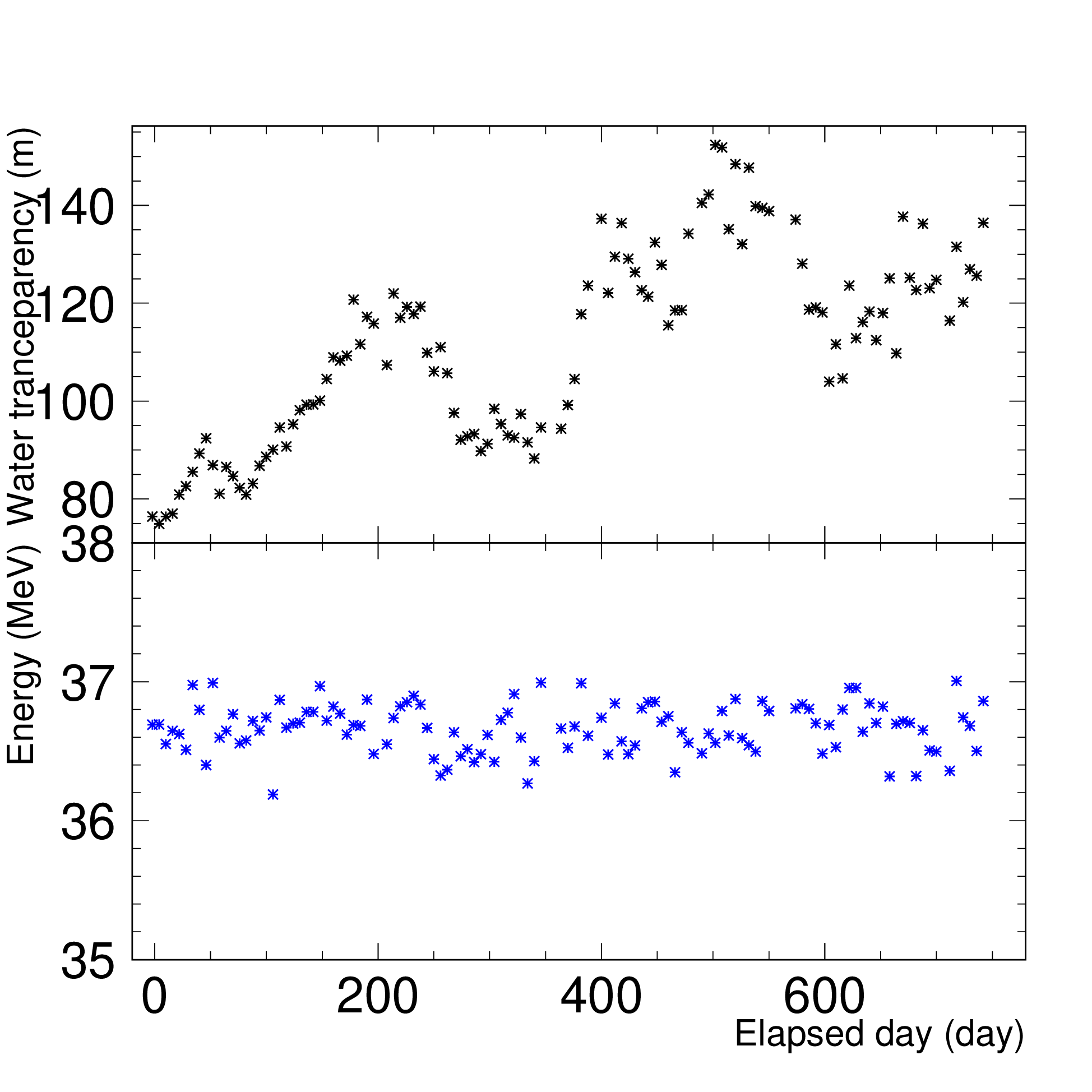}
\caption{The upper figure shows the time variation of the measured water 
  transparency (weighted by the Cherenkov spectrum) during SK-III.  The lower
 figure shows the stability of the peak energy of decay electrons in
 SK-III as a function of time.}
\label{fig:wt_energy}
\end{center}
\end{figure}

Figure \ref{fig:wt_energy} shows the time variation of the measured water transparency during SK-III
and the stability of the peak energy of the decay electrons in SK-III as a function of time.  
The stability of the energy scale has 0.47\% RMS.

\paragraph{Position Dependence}

It was found that the PMT hit rates measured for a Ni-Cf gamma-ray source
at the top region in the detector are systematically lower than those for
the bottom region by $3\sim5\%$.  This rate difference is denoted the
``top-bottom asymmetry''
(TBA). The MC simulation cannot reproduce the top-bottom asymmetry with a uniform attenuation length throughout the detector volume. In order to solve this problem, a simple model of the light absorption is introduced to take into account the dependence of the attenuation length on depth. In this modeling, depth-dependence of the $absorption$ parameter is considered, because the dominant contribution to the time variation of the water transparency is the absorption. The depth-dependence can be modeled as
\begin{equation}
\alpha_{abs}(\lambda,z) = \left\{
\begin{array}{ll}
\alpha_{abs}(\lambda)(1+\beta\cdot z) , &\quad  \text{for}\;z\geq  -1200\text{cm} \\
\alpha_{abs}(\lambda)(1-\beta\cdot 1200 ) , &\quad \text{for}\;z<-1200\text{cm}
\end{array}
\right.
\end{equation}
Below z=-1200~cm, the absorption coefficient is assumed to be uniform
due to convection of the water. $\beta$ is called the TBA parameter (unit is cm$^{-1}$)
and it varies in time,  because TBA varies. The $\beta$ for each period
was tuned to minimize the TBA for that period.

\begin{figure}
 \begin{center}
 \includegraphics[width=8cm,clip]{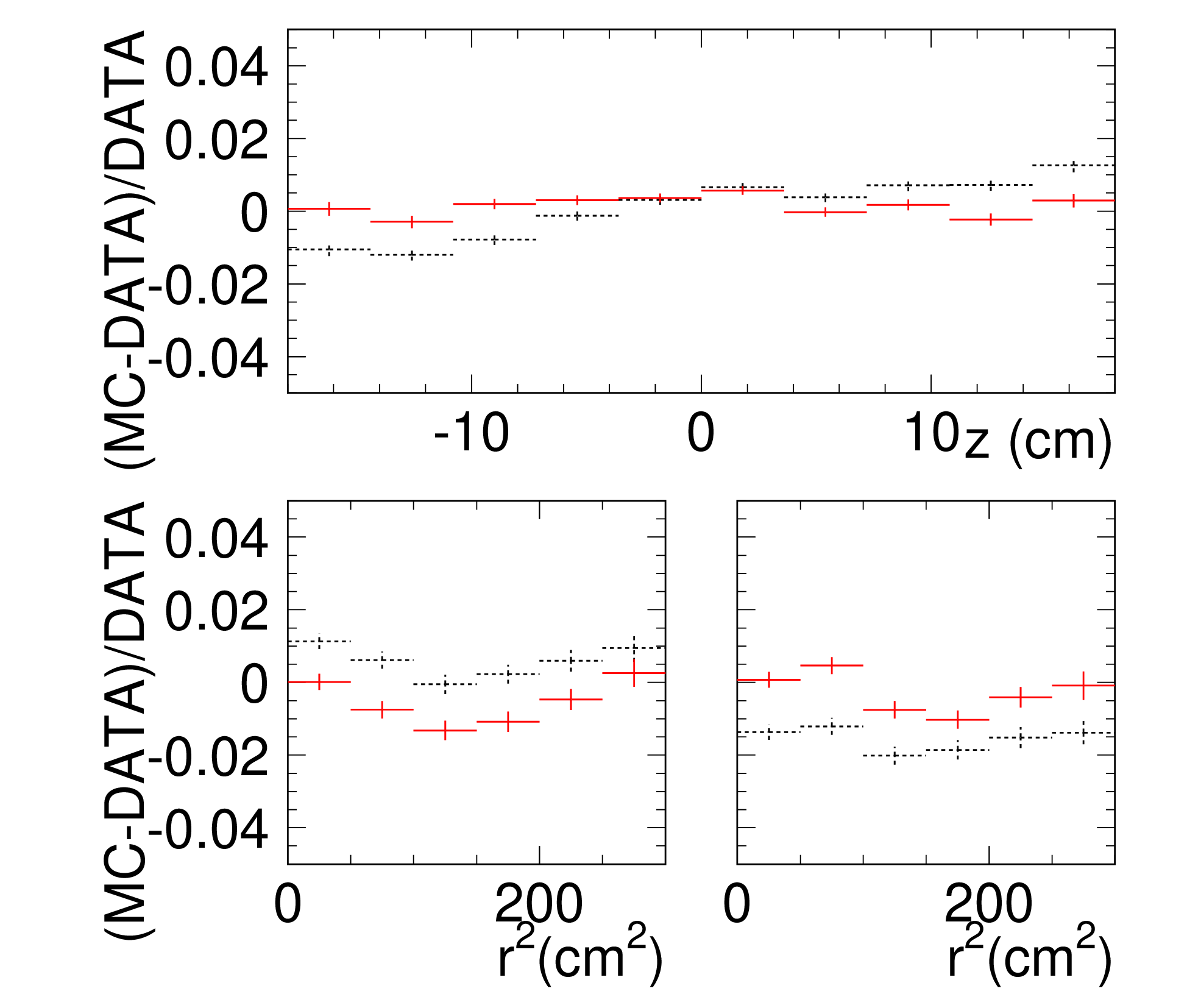} 
 \caption{ Hit rate difference, (MC-DATA)/DATA, for barrel (upper),
   top (left lower), and bottom (right lower). Black (dashed) shows
   $\beta$=0. Red (solid) shows the hit rate difference with tuned $\beta$.
   This calibration data was taken in Feb. 2008, the same period as the LINAC calibration.
 \label{fig:hitrate}}
\end{center}
\end{figure}

Figure \ref{fig:hitrate} shows a comparison of the hit-rate between
data and MC simulation with the tuned $\beta$ value, which is
8.85$\times 10^{-5}$ cm$^{-1}$, and without $\beta$ correction,
respectively. The hit-rate is defined as the number of hits in each
PMT in units of the averaged number of hits for all PMTs during the
calibration run. As shown by the figure, data and MC show better
agreement with the $\beta$ correction. 

\subsubsection{Reflectivity of Black Sheet}

In the detector simulation, the reflectivity of the black sheet which
covers the ID wall is calculated using the law of Fresnel, taking into account
the polarization of Cherenkov photons. To better model the real
situation,  we measure the reflectivity for three incident angles
using calibration data taken from a movable light injector and a black
sheet reflector.  We found that a value of half of  the SK-I reflectivity
gives better agreement and a wavelength-dependent correction is newly applied for SK-III.

 In addition, to better describe the diffuse and specular reflection on
 the black sheet,  we use two models for reflection: the Lambert model
 is used to describe diffuse reflection, and the Phong model is used for description of specular highlights\cite{sk_ii_paper}.

\subsubsection{PMT and Electronics}

\begin{figure}
 \begin{center}
 \includegraphics[width=7cm,clip]{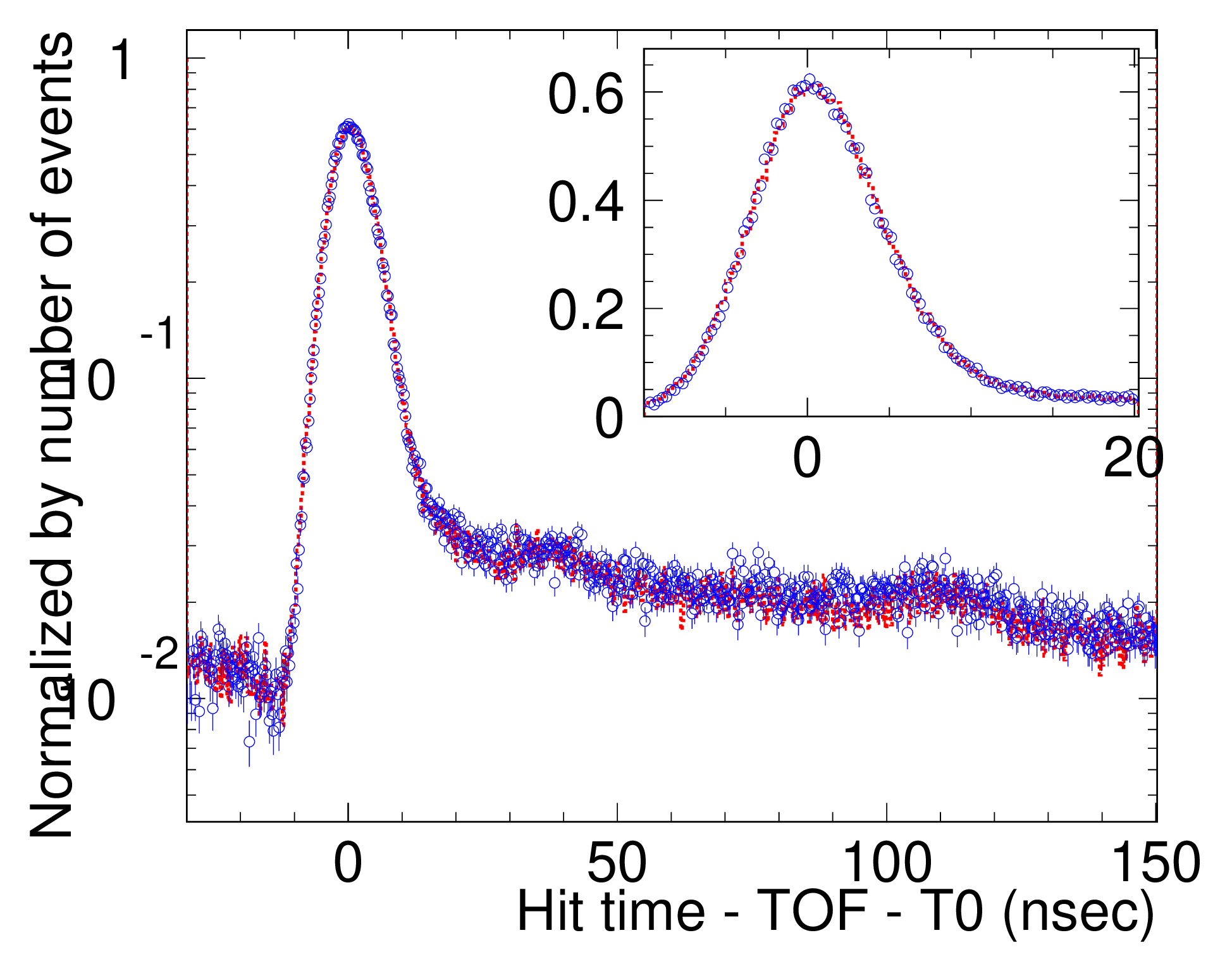} 
 \caption{Timing distribution of each hit PMT after tuning the MC
   simulation, for LINAC 5.1~MeV (x,z)=(-4~m,0~m) data. The blue histogram with open circles shows data and the red dashed histogram shows MC simulation. \label{fig:lin_t}}
 \end{center}
\end{figure}

 The PMT must be simulated as precisely as possible. Reflection and
 quantum efficiency are tuned using the nitrogen/dye laser calibration
 data.  In contrast to the description used for SK-I, we put a
 wavelength dependence into the reflection and an incident angle
 dependence into the quantum efficiency. The position and width of the
 PMT after-pulses and the timing resolution were tuned using data from single electrons injected into the SK detector by LINAC.
After these tunings, the timing distributions of LINAC data and MC simulation
agree very well, as shown in Figure \ref{fig:lin_t}.

\section{Data Analysis}
\subsection{Data set}

As for SK-I and II, SK-III has two trigger levels for solar neutrino analysis: low-energy (LE) and super-low-energy (SLE) thresholds. SLE triggered events are filtered online to reduce the amount of data written to limited storage space.  Events reconstructed outside the fiducial volume are rejected.  The data are reduced by a factor of approximately six.
From Aug. 4th, 2006 to Jan. 24th, 2007, only the LE trigger threshold
was applied.  The LE trigger is 100\% efficient above 6.5~MeV and the
livetime of this period is 121.7 days.  The threshold was lowered and
the first SLE data were taken with more than 99\% efficiency at 5.0 MeV from
Jan. 24th, 2007 to Apr. 17th, 2008. This period is called the SLE1 period, and the livetime of this period is 331.5 days. From Apr. 17th, 2008 to Aug. 18th, 2008, a lower SLE trigger threshold was applied and the SLE trigger had more than 98$\%$ efficiency at 4.5 MeV. This period is called SLE2 period, and the livetime of this period is 94.8 days.

Two neutrino samples are used for SK-III analysis. 
The first sample, for event energies between 6.5 and 20 MeV, has a total livetime of 547.9~days.
The second sample, for event energies between 5.0 and 6.5 MeV, has a total livetime of 298.2~days
after rejecting high background periods caused by radioactive impurities 
accidentally injected into the detector.

\subsection{Event selection}
Basic explanations of each selection step are as follows:
\begin{itemize}
\item Run selection\\
In this selection, short runs ($<$ 5 min.),
runs with hardware and/or software problems, or calibration runs are rejected.  

\item Cosmic-ray muon \\
Cosmic-ray muon events are rejected by a total charge cut ($<$ 2000 photo-electron (p.e.) in ID). 

\item Electronic noise reduction\\
Events due to electronic noise are rejected.

\item Fiducial volume cut \\
Events which have reconstructed vertex position within 2~m of the ID wall are also rejected. 

\item Spallation cut\\
This cut is to reject events caused by cosmic-ray muons. 

\item Event quality cut\\ \label{eqc}
In this step, results of event reconstructions are tested.
In particular,
(a) quality of vertex and direction reconstruction, and
(b) hit pattern, 
(c) result of the second vertex reconstruction used for SK-I~\cite{sk_full_paper}
are checked.
Events produced by  flasher PMTs are also rejected in this step.  

\item External event cut\\
This cut is to reject events induced by radioactivity from the 
PMTs or the detector wall structure.

\item Cosmogenic $^{16}$N cut\\
Events caused by decay of $^{16}$N are rejected. The $^{16}$N is produced 
when cosmic ray $\mu$ is captured by $^{16}$O in water.
 
\item Small clustered hit cut and tighter fiducial volume cut\\
These cuts reject events which have clustered hits (see \ref{sec:clik} and \ref{sec:tfv}).

\end{itemize}
For more detailed descriptions of the reduction steps,
please refer to \cite{sk_full_paper,sk_ii_paper}.  

\subsection{Small clustered hit cut for the lowest energy region}\label{sec:clik}

This cut is newly developed for SK-III to reduce the low energy
background in regions near the edge in the detector.
As described in the previous section, 
the target background  
is assumed to be triggered by a coincidence of dark 
hits and small clusters of hits due to radioactive sources 
in the FRP or the structure of the detector wall.  

We can separate this background from the solar neutrino signal
by searching for a small cluster in both space and time.  
A real neutrino signal at the edge region also has similar characteristics, 
but it causes a bigger cluster compared to the background events. 
Thus, the key is to evaluate the size of the hit cluster.

In order to check the size of the hit cluster for an event, a new cut variable was created.
This variable is a product of two factors. 
One factor is the radius of a circle which contains the clustered hits in a event,
and the other is the number of clustered hits  within  a 20~nsec time window. 
Figure \ref{fig:clik} shows the distribution of this cut variable
for  real events and for solar neutrino MC simulation events in a energy region 5.0 to 6.5~MeV. 
For this figure, a volume cut with $r>12$~m  and $z>-3$~m
is applied to study background events in the barrel, which are due to FRP cover radioactivity.

As expected, the real data sample which contains mostly FRP events shows smaller 
 values for this cut variable than simulated solar neutrino events.
A cut value of 75 is selected which gives maximum significance ($Signal/\sqrt{BG}$)
for the solar neutrino signal.
This cut is applied to events with energy $<$ 6.5 MeV 
whose vertex position is in an edge region.
The edge region is determined for 5.0-5.5 MeV, and 5.5-6.5 MeV
energy bins separately, to optimize the significance of the solar neutrino signal.
The criteria of the edge region are determined as follows:

\begin{figure}[]
 \begin{center}
 \includegraphics[width=7cm]{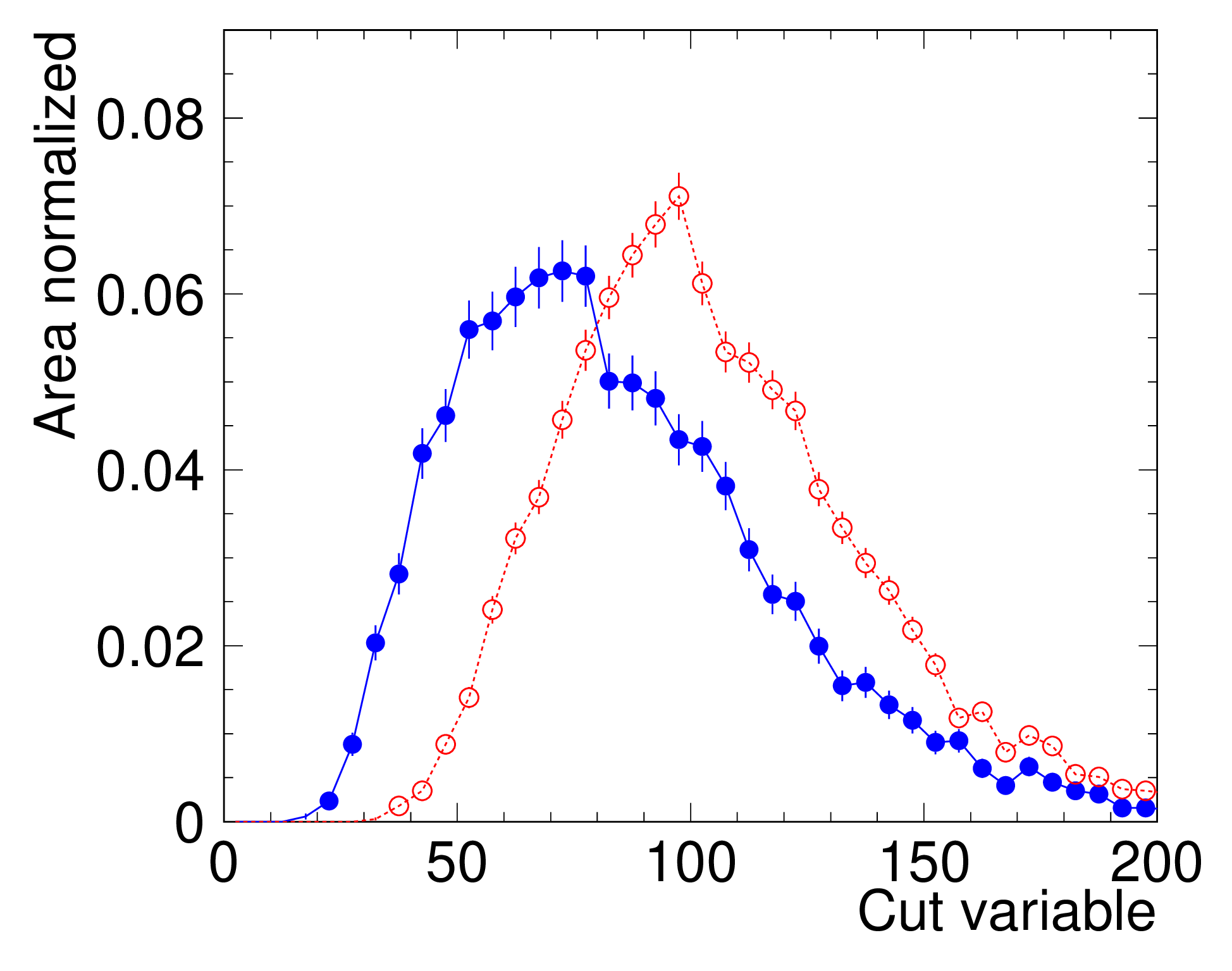} 
 \caption{Distribution of cut variable for the small clustered hit
   cut. Blue (filled circle) shows  background sample events which are selected from $r>12$~m and
   $z>-3$~m. Red (open circle) shows the solar neutrino MC simulation and the same volume cut is applied.
   For both data and MC events, event selection cuts up to the external event cut are applied.
   The reconstructed energies of these samples are in 5.0-6.5~MeV. 
  \label{fig:clik}}
 \end{center}
\end{figure}

\begin{equation}
\left\{
\begin{array}{ll}
  r^2 > 180~m^2 \text{ for 5.5 $\leq$E$<$ 6.5 MeV} \\
  r^2 > 155~m^2 \text{ for 5.0 $\leq$E$<$ 5.5 MeV} \\
\end{array}
\right.
\end{equation}
where $r$ is defined in Figure~\ref{fig:def}.
Figure \ref{fig:rr5065} shows the vertex distribution of the final data
sample for the 5.0-6.5~MeV energy region before and after the small clustered hits cut.
 
\begin{figure}[]
 \begin{center}
 \includegraphics[width=7cm]{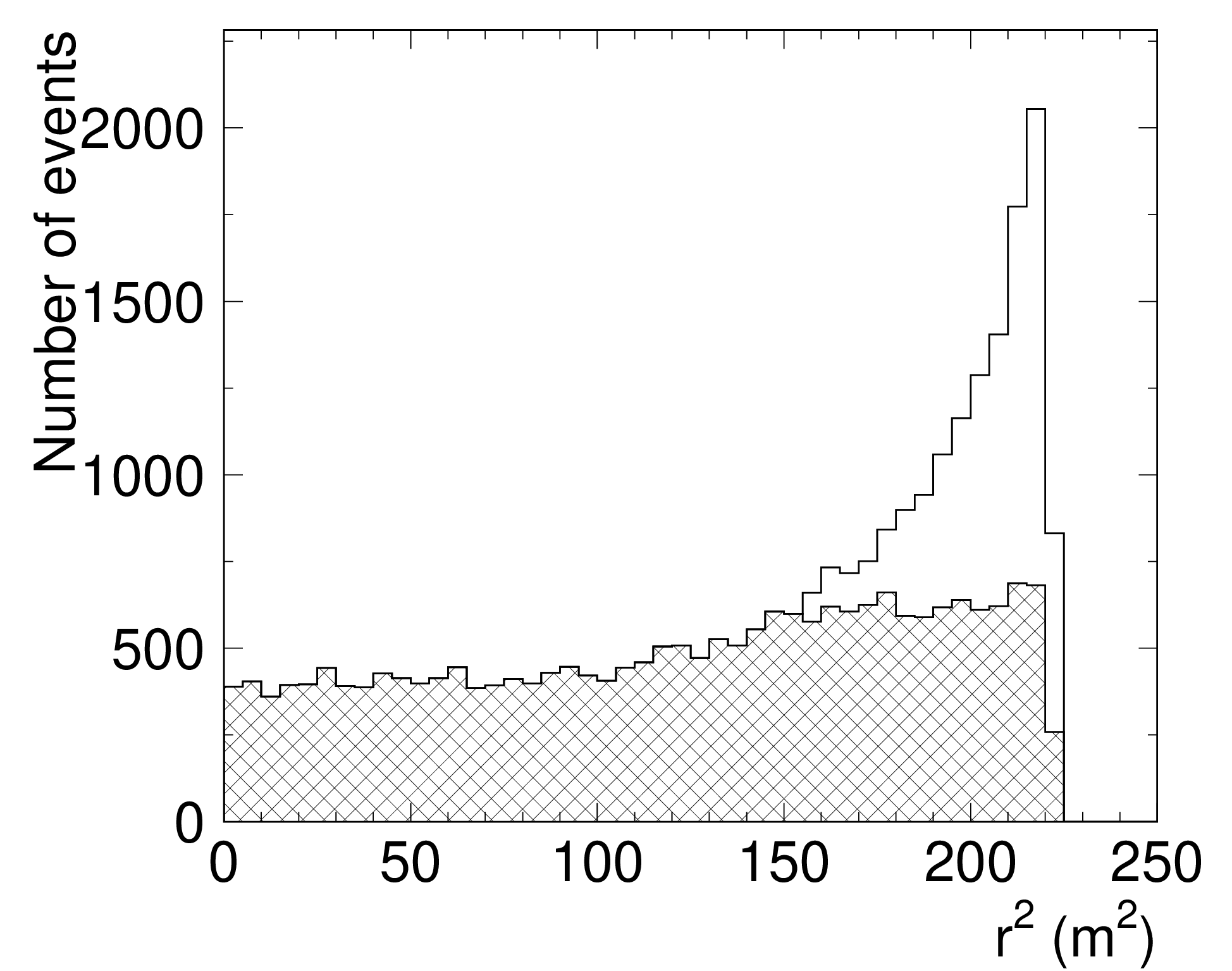} 
 \caption{ Vertex distribution of 5.0-6.5 MeV energy region before and after the small cluster hits cut.
 The horizontal axis shows $r^2$ and the vertical axis shows the number of events. Events with $z>-7$~m are selected to show  background events in the barrel.  \label{fig:rr5065}}
\end{center}
\end{figure}

\subsection{Tighter fiducial volume cut}\label{sec:tfv}
\begin{figure}[]
 \begin{center}
 \includegraphics[width=7cm,clip]{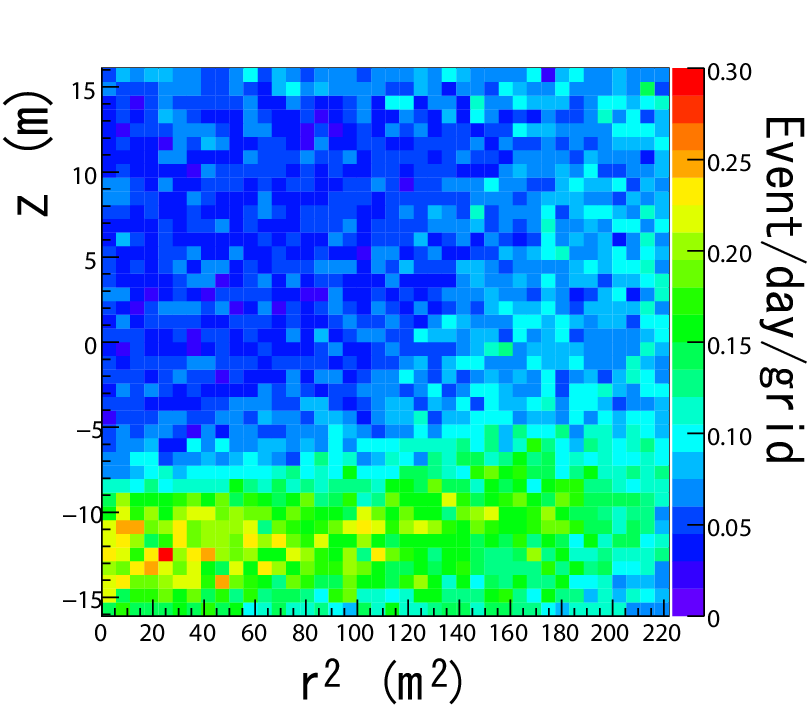} 
 \caption{ Vertex distribution of final data sample in the 5.0-6.5~MeV energy region (22.5~kton fiducial volume).
   \label{fig:v5065}}
\end{center}
\end{figure}

This cut is to reject the remaining background in the edge region.
As shown in  Figure \ref{fig:v5065}, background events in the
bottom region still remain in the final data sample.
This non uniformity of the vertex distribution of the background distorts
the angular distribution of the background, which 
causes a large systematic uncertainty for the day-night asymmetry of 
solar neutrino flux.

To set a tight fiducial volume, the significance as a function 
of detector radius is calculated.
Based on the significance calculation, 
the final value of fiducial volume is obtained for each energy region.
\begin{eqnarray}
 5.0-5.5~{\rm MeV}:& (r^2<180~\text{m}^2 \text{and } z>-7.5~\text{m})  = 13.3~\text{kton} \nonumber \\
 5.5-20~{\rm MeV}:& \text{no tight fiducial volume cut} = 22.5~\text{kton} \nonumber
\end{eqnarray}
Note that in the future, we hope to remove this tighter fiducial volume cut
by improving our signal extraction method, because solar neutrino events are
preferentially rejected in the low energy region by this cut.

\subsection{Summary of event selection}

\begin{figure} 
\begin{center}
 \includegraphics[width=7cm,clip]{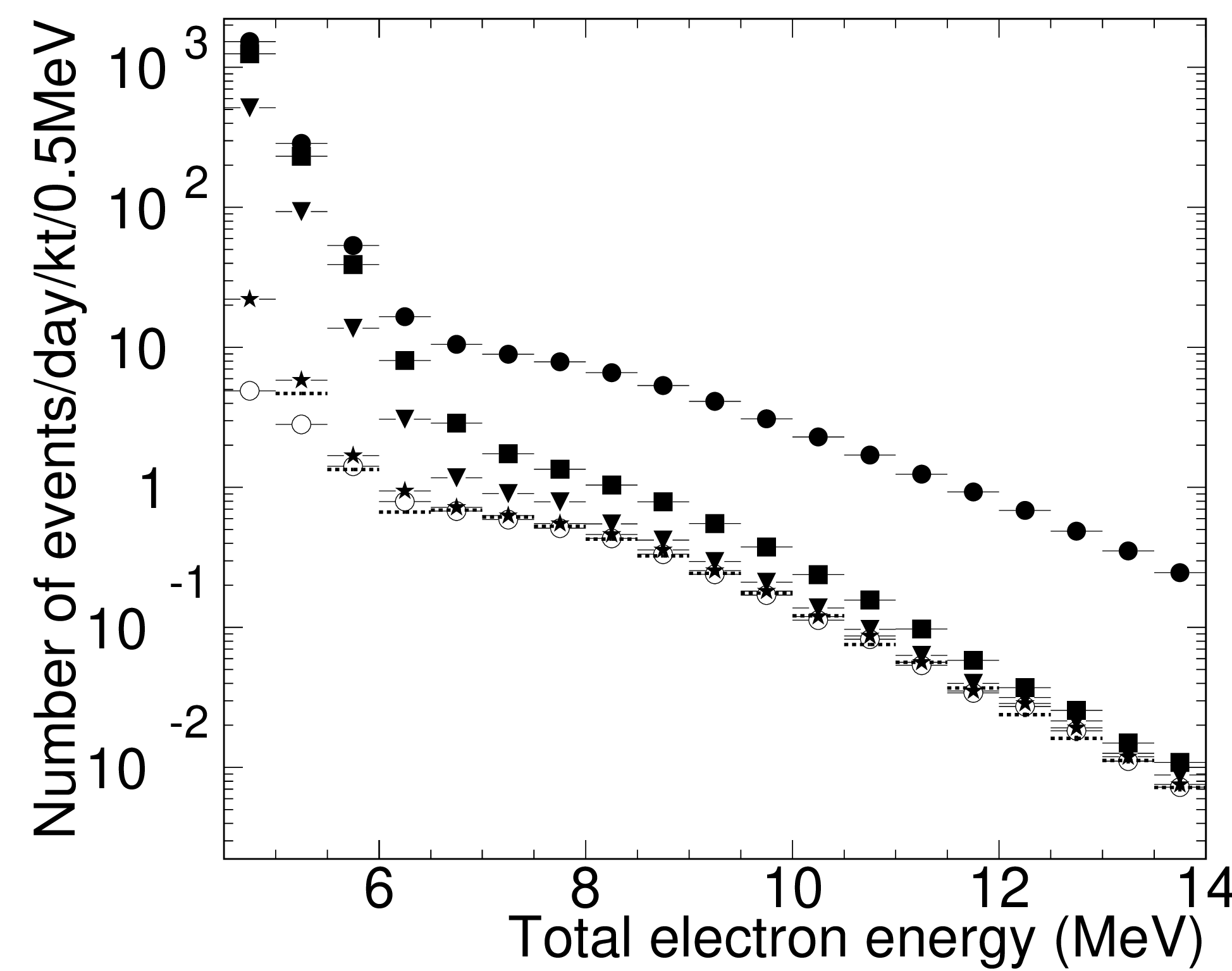} 
 \caption{ Number of events after each event reduction step as a
   function of energy. From the top, the markers show events after spallation cut, event quality cut, external event cut, $^{16}$N cut, and small clustered hit cut plus tight fiducial volume cut. The dashed histogram shows the number of events in the SK-I final sample.
  \label{fig:redstep}}
 \end{center}
\end{figure}

\begin{figure}
\begin{center}
 \includegraphics[width=7cm,clip]{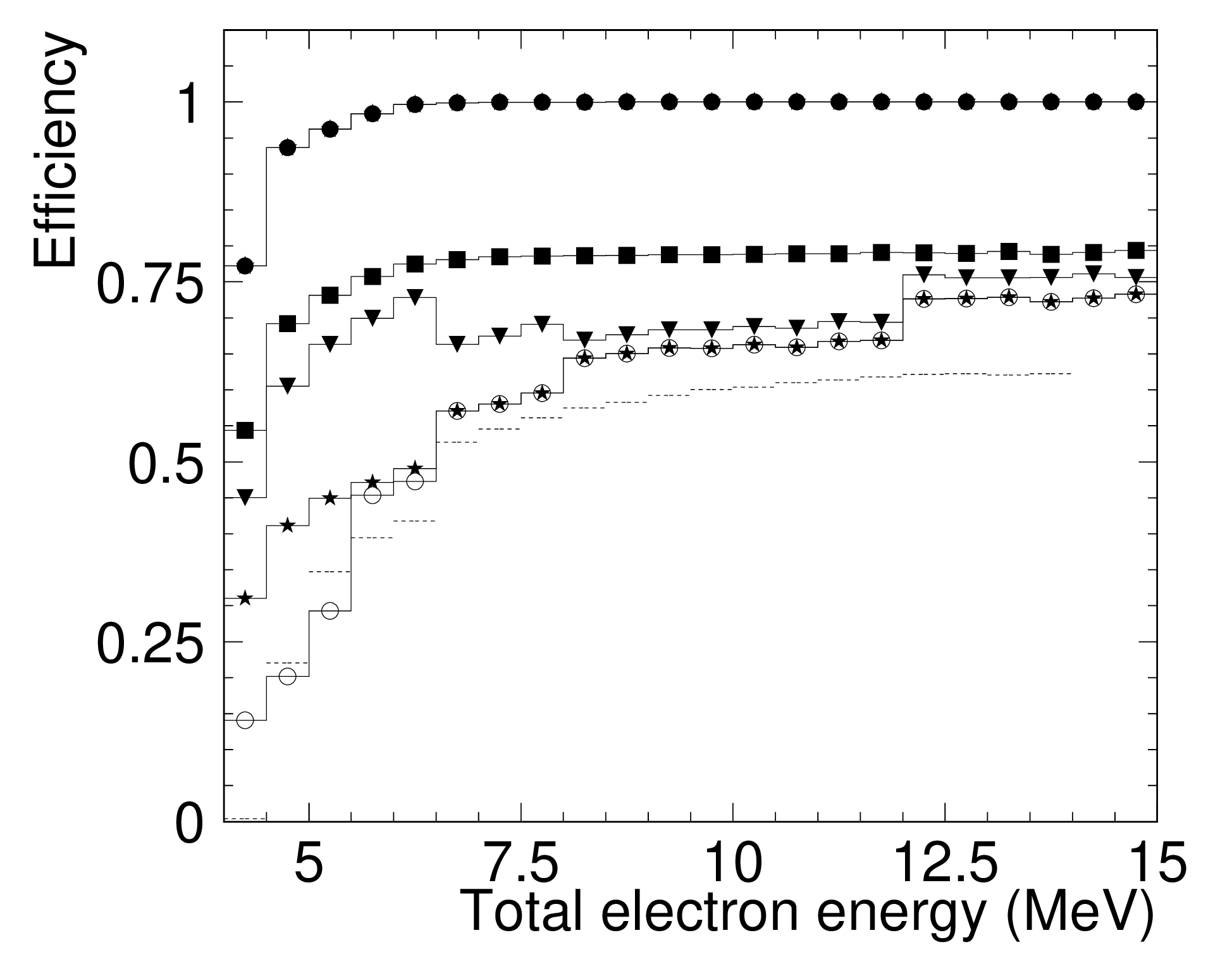} 
 \caption{ Reduction efficiency for the solar neutrino MC simulation.
  The definition of each histogram is the same as for Figure~\ref{fig:redstep}.
\label{fig:cuteff}}
\end{center}
\end{figure}

Figure \ref{fig:redstep}  shows the energy spectrum after
each step of the reduction and Figure \ref{fig:cuteff} 
shows the remaining efficiency of $^8$B solar neutrino MC
with respect to the reconstructed energy.
  The number of events after each reduction step is summarized in Table \ref{tab:redd} (real data) and \ref{tab:redm} (solar neutrino MC simulation).
While the event rate in the real data as a function of energy is the
same as for SK-I,
the cut efficiencies are improved by 10$\%$ in the final SK~III data sample.

\begin{table}
\centering
\begin{tabular}{ l r r }
\hline
Reduction step  & 5.0-6.5 MeV & 6.5-20.0 MeV \\
\hline
Spallation cut   & 1861770     & 114350 \\
Event quality cuts & 734843      & 58187  \\
External event cut    & 54820       & 43146 \\
$^{16}$N  cut    & 54351 & 39879 \\
Small cluster hits cut      & 42916 &  \\
Tight fidv. cut  & 24311 &  \\
\hline
Final           &  24311 & 39879 \\
\hline
\end{tabular}
\caption{Reduction results for data.}
\label{tab:redd}
\end{table}

\begin{table}
\centering
\begin{tabular}{ l r r r }
\hline
Reduction step  &  5.0-6.5 MeV & 6.5-20.0 MeV \\
\hline
Total            & 100 & 100 \\
Bad run cut      & 86 & 87 \\
Trigger condition & 72 & 86 \\
Flasher events cut   & 56 & 67 \\
Spallation+$^{16}$N cut  & 45 & 53 \\
Event quality cuts  & 45& 46 \\
External event cut   & 28 & 42 \\
Small cluster hits cut   & 27 &  \\
Tight fidv cuts & 24 &  \\
\hline
Final & 24 & 42 \\
\hline
\end{tabular}
\caption{Reduction results for solar neutrino MC events in \%. For the spallation cut and $^{16}$N cut, 
position-dependent dead time is considered.}
\label{tab:redm}
\end{table}

\subsection{Simulation of solar neutrinos}
The method to extract the energy spectrum and the flux of solar neutrinos in SK-III is based on that of SK-I and II. 
The  $^8$B and hep neutrino spectra are generated separately.
The total $^8$B and hep flux values are referred from the Standard Solar Model (SSM)~\cite{ssm}.
 We use the spectrum and its uncertainty of $^8$B and hep neutrinos 
 calculated by Winter~\cite{win06} and Bahcall~\cite{hep} respectively.
The event time of a solar neutrino event is simulated from 
the livetime of the SK-III full operation period, so that 
the expected zenith angle distribution of the solar neutrinos 
can be simulated correctly.

\subsection{Systematic uncertainty}
The following items are updated with respect to SK-I with the improved
calibration, detector simulation and analysis tools described above.
\begin{itemize}
\item Angular resolution\\
In order to estimate the systematic uncertainty on the total flux due to the angular resolution difference between data and MC simulation,
a predicted energy spectrum is made by artificially shifting the reconstructed direction of the solar neutrino MC events.
The shifted direction is calculated as 
the reconstructed direction $\pm$ the systematic uncertainty of angular resolution
with respect to the generated direction of the recoil electron.
After the solar angle fitting, the systematic uncertainty due to the 
uncertainty of angular resolution is estimated as $\pm 0.67\%$ 
on the total flux in 5.0-20 MeV region ($\pm 1.2\%$ for SK-I).
This method of estimation is the same as for SK-I, so the improvement is due to the 
new detector simulation and angular reconstruction.

\item Vertex shift and resolution\\
The vertex shift difference between data and MC is used to estimate
the systematic uncertainty of the fiducial volume. 
We measure this difference using Ni calibration rather than the LINAC, 
because LINAC electrons always point downwards while solar neutrino recoil electrons go in all directions. 
Also, very few positions near the edge of the fiducial volume can be probed by the LINAC. 
As an example, Figure~\ref{fig:vshift_ene} shows the observed vertex shift difference as a function of energy
at (15.2 m, -0.7 m, 12 m), where the vertex shift in the x direction is larger than other positions in Figure~\ref{fig:versh}. 
There is no evidence that the vertex shift difference is energy dependent.

To estimate the total flux uncertainty due to the vertex shift,
the reconstructed vertex positions of the solar neutrino MC events  
are artificially shifted outward.
Then, the fraction of events rejected by the fiducial volume cut due
to the shift is estimated.
Figure~\ref{fig:fidvol_sys} shows the systematic uncertainty on the fiducial volume due to the vertex shift
which is based on the results of the Ni calibration. 
Note that in SK-I, and II, the estimation was done by a calculation based on the
size of the vertex shift, whereas in SK-III, MC simulations after event selection
are used to take into account the cuts' efficiencies at the edge of the fiducial volume. 
This new method gives a more accurate estimation. 

This study shows that 0.54$\%$ of the MC events are rejected after shifting the vertex position, 
and this fraction is set to the total flux systematic uncertainty due to the vertex shift ($\pm 1.3\%$ for SK-I).
The improvement with respect to SK~I is mainly due to the reduction of the vertex shift.

In Figure~\ref{fig:fidvol_sys}, the step in the fiducial volume uncertainty between 5.0 and 6.0 MeV is
due to the tight fiducial volume cut for the lower energy region ($< 5.5$MeV).
This relative difference  of fiducial volume systematic uncertainty is taken into account in 
the uncertainty of the energy spectrum shape.  Between 5.5 and 20 MeV, since vertices are assumed to be shifted 
in the same direction for all energies, the relative differences are 0.1\%. 
Consequently, the systematic uncertainty of the spectrum shape is set to 0.1\% in that energy region.
In the 5.0-5.5 MeV region, the systematic error is set to 0.5\% because the uncertainty of the fiducial volume 
is relatively 0.5\% larger than for the other region.

\begin{figure}
\begin{center}
 \includegraphics[width=7cm,clip]{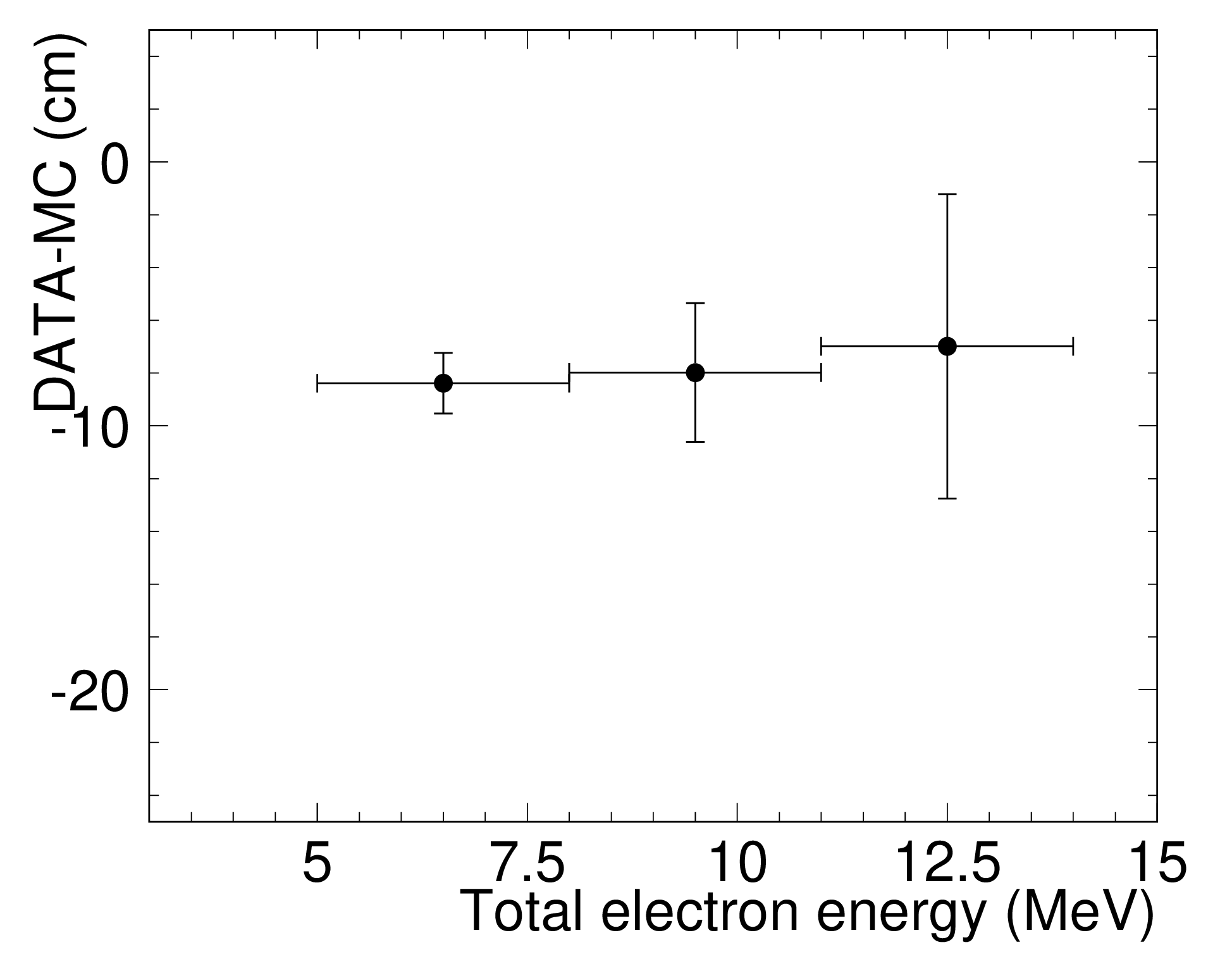} 
 \caption{ The vertex shift as a function of energy. The markers show the amount of vertex shift measured by Ni calibration
at x=15.2 m y=-0.7 m and z=12 m. For the vertical axis, negative values mean vertex shifts inward towards the center of the detector. 
\label{fig:vshift_ene}}
\end{center}
\end{figure}

\begin{figure}
\begin{center}
 \includegraphics[width=7cm,clip]{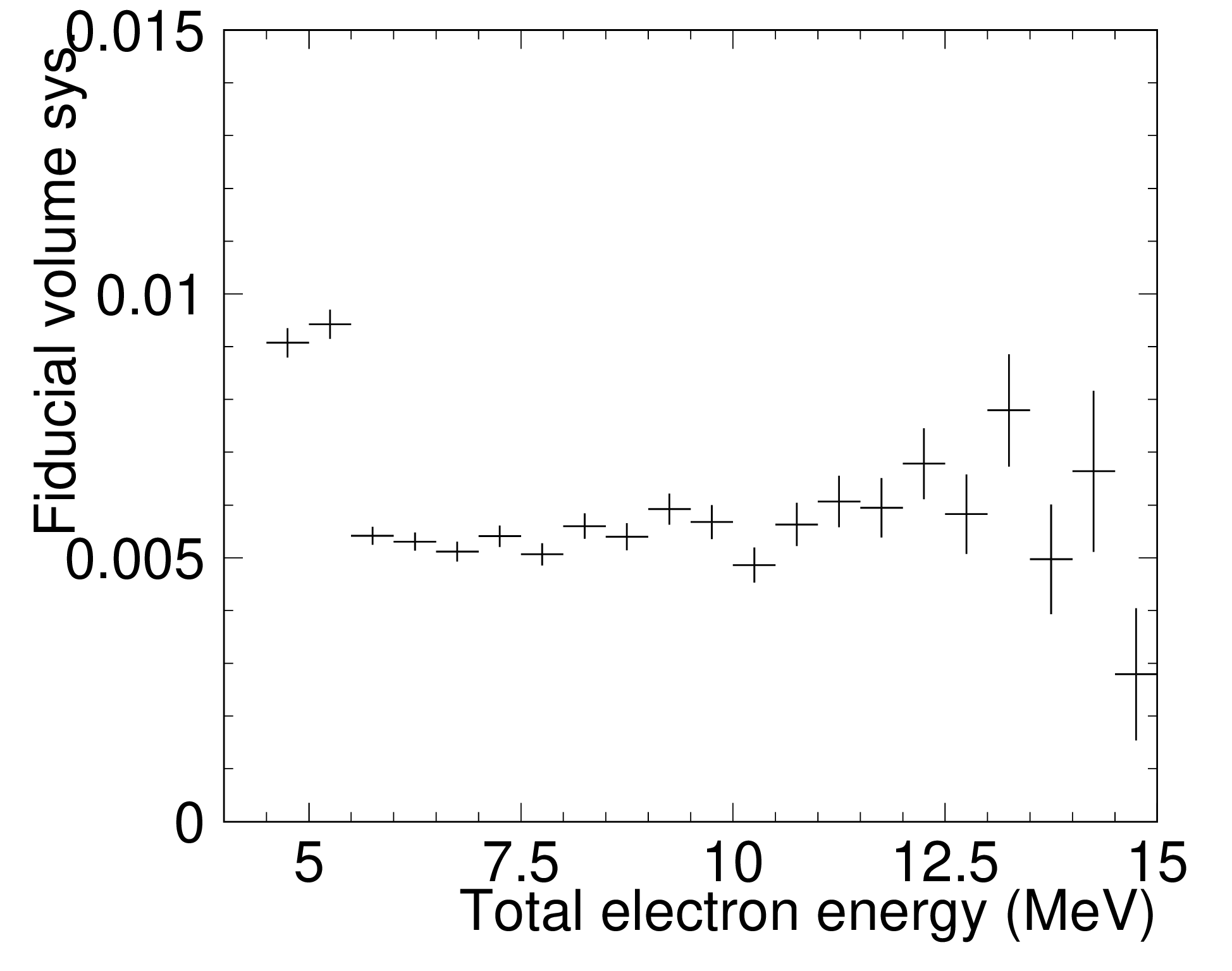} 
 \caption{ The systematic uncertainty of the fiducial volume as a function of energy. \label{fig:fidvol_sys}}
\end{center}
\end{figure}

The vertex resolution is also compared for data and MC simulation using LINAC events.
Figure~\ref{fig:vres_edep} shows the difference of vertex resolution as a function of energy.
This difference results in only a second order effect on the systematic uncertainty of the fiducial volume.
 \begin{figure}
\begin{center}
 \includegraphics[width=7cm,clip]{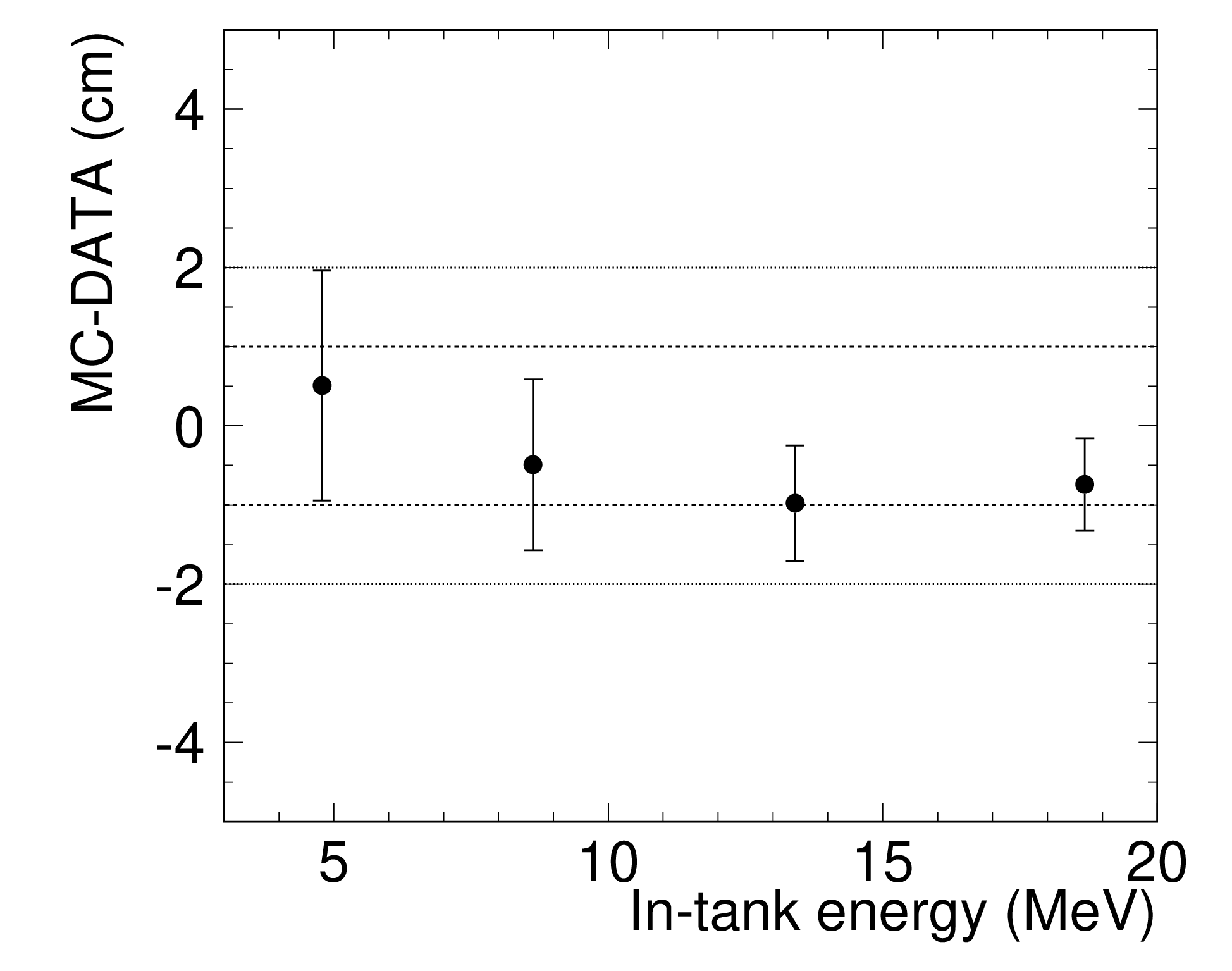} 
 \caption{ The difference of vertex resolution between MC simulation and data as a function of energy obtained by LINAC calibration. 
The dashed line and the dotted line show 1 cm and 2 cm difference respectively. 
\label{fig:vres_edep}}
\end{center}
\end{figure}

\item  Event quality cuts\\
 The systematic uncertainties associated with the event quality cuts 
are estimated  separately for (a), (b) and (c) of \ref{eqc}.
  For (a) and (b), the systematic uncertainties on the total flux are taken from  
  the differences of the efficiencies 
  between the LINAC data and MC simulation which are $\pm 0.4\%$ and $\pm0.25\%$, respectively.

  For (c), the second vertex cut,
  the estimation is done in the same way as for the first vertex fitter,
  resulting in a $0.45\%$ contribution to the total flux uncertainty.

The combined uncertainty is 0.65\% which is about one third of
SK-I's. The improvement is due to
the improved event selection methods and better-tuned MC simulation. 

\item Small clustered hits cut\\
Using DT data and MC simulation at the position (x=-12~m, y=0~m, z=0~m),
a difference in cut efficiency between DT data and MC events of $2\%$
is obtained.
Considering the entire fiducial volume and the energy region of the solar neutrino signal, 
this difference corresponds to a 0.5$\%$ uncertainty on the total flux
in the 5.0-20 MeV region.  A value of
2$\%$ is assigned to the spectral shape uncertainty for the 5.0 to 6.5 MeV bins.

\item Signal extraction method \\ 
The solar neutrino flux is obtained by fitting a solar angle distribution 
(see Figure~\ref{fig:solar_peak}). To check biases of the fitting method to the flux value,
the solar angle fit is applied to dummy data which have known 
numbers of signal and background events.
As a result, $\pm0.7\%$ difference is found between the input and the output number of signal events 
for the total flux, and $\pm2\%$ difference is found especially in the 
$ 5.0 < E < 5.5$ MeV region. 
\end{itemize}
\subsubsection{Summary of systematic uncertainties}

\begin{itemize} 
\item Flux, time variation, day-night asymmetry \\
The systematic uncertainties on total flux, time variation and day-night asymmetry are
summarized in Table \ref{tab:totalsys}.
The systematic uncertainty on the total flux is estimated to be 2.1$\%$.
This is about two thirds of the corresponding SK-I value. 
The main contributions to the improvement are
the vertex shift, angular resolution, and event selection uncertainties,
which are re-estimated for SK-III.

\begin{table}
\begin{center}
\begin{tabular}{l c  } \hline
Source            &  Total Flux       \\ \hline\hline
Energy scale      &$\pm 1.4   $ \\ 
Energy resolution &$\pm 0.2   $  \\ 
$^8$B spectrum      &$\pm 0.2   $            \\ 
Trigger efficiency        &$\pm 0.5  $   \\ 
Angular resolution   &$\pm 0.67   $ \\ 
Fiducial volume (vertex shift) &$\pm 0.54  $ \\ 
Event quality cuts         &  \\
- Quality cut &$\pm 0.4   $  \\ 
- Hit pattern cut   &$\pm 0.25  $  \\ 
- Second vertex     &$\pm 0.45  $  \\ 
Spallation        &$\pm 0.2   $ \\ 
External event cut     &$\pm 0.25  $ \\
Small cluster hits cut   &$\pm 0.5   $  \\ 
Background shape  &$\pm 0.1   $  \\ 
Signal extraction method&$\pm 0.7  $   \\ 
Livetime          &$\pm 0.1 $ \\
Cross section     &$\pm 0.5  $   \\ \hline
Total             &$\pm 2.1  $\\ \hline
\end{tabular}
\end{center}
\caption{ Summary of the systematic uncertainty of the total flux in $\%$.\label{tab:totalsys}}
\end{table}

\item Spectrum \\
The systematic uncertainty of the spectral shape consists of two components:
\begin{figure}[] 
\begin{center}
 \includegraphics[width=7cm]{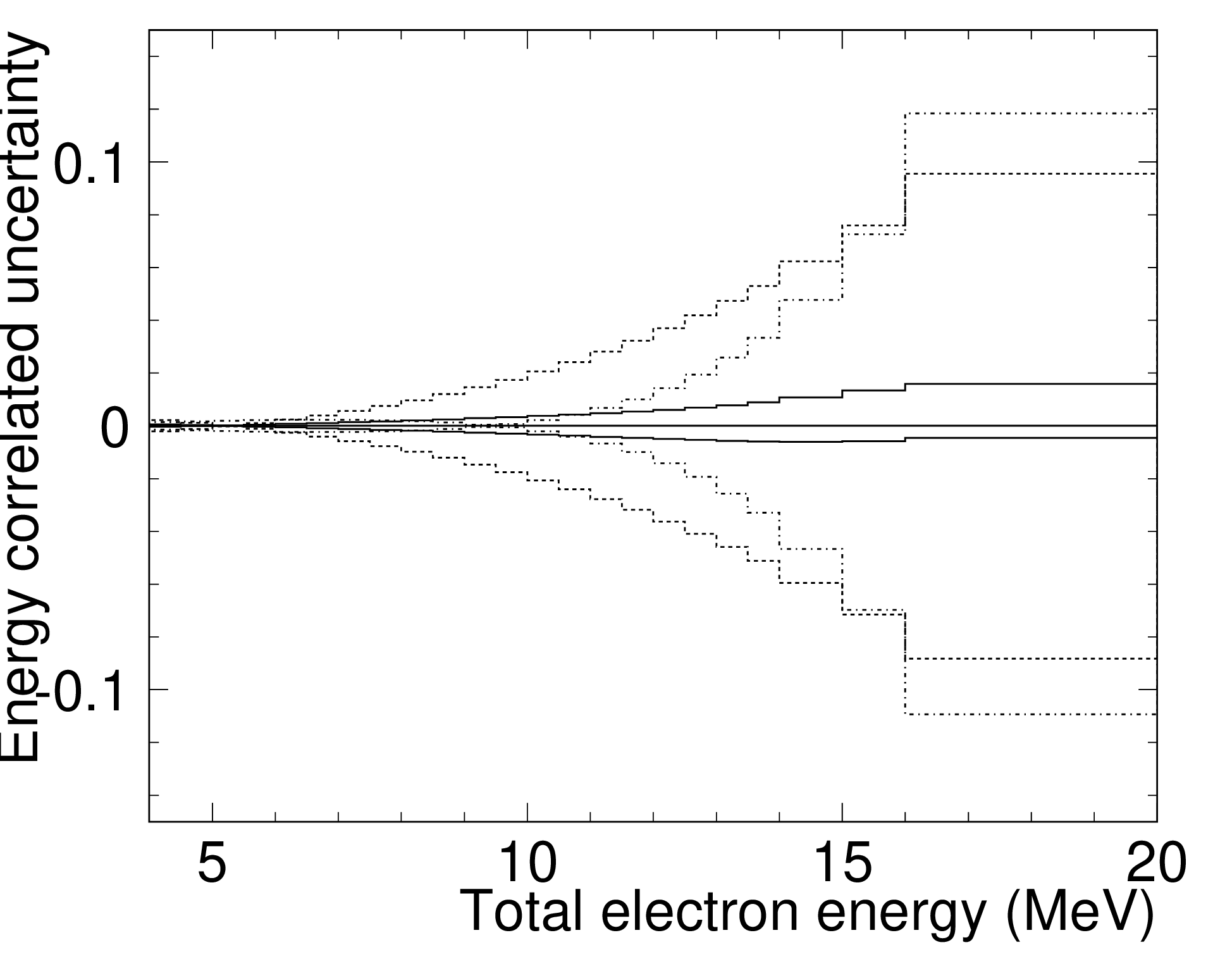} 
 \caption{ Energy-correlated systematic uncertainties. The solid,
   dotted, and dashed lines show the uncertainties of the $^8$B
   spectrum, the energy scale, and the energy resolution, respectively.}
   \label{fig:ecorr}
\end{center}
\end{figure}

\begin{itemize}
\item Energy-correlated: \\
The energy-correlated systematic uncertainties are obtained by counting the number of 
events in the solar neutrino MC simulation with  artificially shifted energy scale,
energy resolution and $^8$B $\nu$ energy spectrum.
The results of the calculations are shown in Figure \ref{fig:ecorr}.
These correlations are taken into account in the oscillation analysis.

\item Energy-uncorrelated:\\
The energy uncorrelated spectral  uncertainties are listed in Table \ref{tab:euncor}.
To obtain the effect on the energy spectral shape,
relative differences of each uncertainty between each energy bin are studied. 
For example, the uncertainty of the fiducial volume due to the vertex shift is 0.54\% in total, 
and the relative differences are obtained as 0.1\% for 5.5-20 MeV region,
but 0.5\% for 5.0-5.5 MeV region with the tight fiducial volume. 
These uncertainties are taken into account in the oscillation analysis without correlations.

\begingroup
\squeezetable
\begin{table}
\begin{center}
\begin{tabular}{l c c c c c c } \hline
Energy(MeV)  & 5-5.5     &5.5-6      &6-6.5      &6.5-7      &7-7.5      & 7.5-20   \\ \hline \hline
Trig eff     &$\pm 2.4  $&$\pm 0.9  $&$\pm 0.1  $& -         & -         &   -      \\ 
(a)         &$\pm 2.   $&$\pm 1.75 $&$\pm 1.5  $&$\pm1.25  $&$\pm 1.0  $&   -      \\ 
(b)     &    -      &     -     &    -      &$\pm0.25  $&$\pm0.25  $&$\pm0.25 $\\        
Small cluster hits cut  &$\pm 2.   $&$\pm 2.  $&$\pm 2.    $&     -     &    -      &    -     \\ 
External event cut  &$\pm 0.1  $&$\pm 0.1 $&$\pm 0.1   $&$\pm 0.1 $&$\pm 0.1   $&$\pm 0.1 $\\ 
Fiducial volume (vertex shift) &$\pm 0.5  $&$\pm 0.1 $&$\pm 0.1   $&$\pm 0.1 $&$\pm 0.1   $&$\pm 0.1 $\\ 
BG shape     &$\pm 0.2  $&$\pm 0.8 $&$\pm 0.2   $&$\pm 0.2 $&$\pm 0.2   $&$\pm 0.2 $\\ 
Sig.Ext.     &$\pm 2.1  $&$\pm 0.7 $&$\pm 0.7   $&$\pm 0.7 $&$\pm 0.7   $&$\pm 0.7 $\\ 
Cross section&$\pm 0.2  $&$\pm 0.2 $&$\pm 0.2   $&$\pm 0.2 $&$\pm 0.2   $&$\pm 0.2 $\\ \hline
Total        &$\pm 4.3  $&$\pm 3.0 $&$\pm 2.6   $&$\pm 1.5 $&$\pm 1.3   $&$\pm 0.8 $\\ \hline
\end{tabular}
\end{center}
\caption{Energy-uncorrelated systematic uncertainty on the observed 
spectrum shape in $\%$.\label{tab:euncor}}
\end{table}
\endgroup

\end{itemize}
\end{itemize}

\subsection{Total Flux Result}
Recoil electrons from elastic solar $\nu$-electron scattering are strongly forward-biased.
SK-III statistically separates solar $\nu$'s from background with an 
unbinned likelihood fit to the directional distribution with respect to the Sun.
For a live time of 548 days of SK-III data, from 5.0 to 20.0 MeV, the
extracted number of signal events is 8132$^{+133}_{-131}$(stat.) $\pm
186$(sys.).  The corresponding $^8$B flux is obtained using the $^8$B
spectrum of \cite{win06} to be: 

\begin{eqnarray*}
(2.32 \pm 0.04 (\textrm{stat.}) \pm 0.05 (\textrm{sys.})) \times 10^ 6~\textrm{cm}^{-2} \textrm{sec}^{-1}.
\end{eqnarray*}

This result is consistent with  SK-I $(2.38 \pm 0.02 (\textrm{stat.}) \pm 0.08 (\textrm{sys.}) \times 10^ 6~\textrm{cm}^{-2} \textrm{sec}^{-1})$ and  SK-II $(2.41 \pm 0.05 (\textrm{stat.}) ^{+0.16}_{-0.15} (\textrm{sys.}) \times 10^ 6~\textrm{cm}^{-2} \textrm{sec}^{-1})$.
The SK-I and II values are recalculated using the $^8$B spectrum of \cite{win06}.

\begin{figure}[htbp]
\begin{center}
\includegraphics[width=7cm]{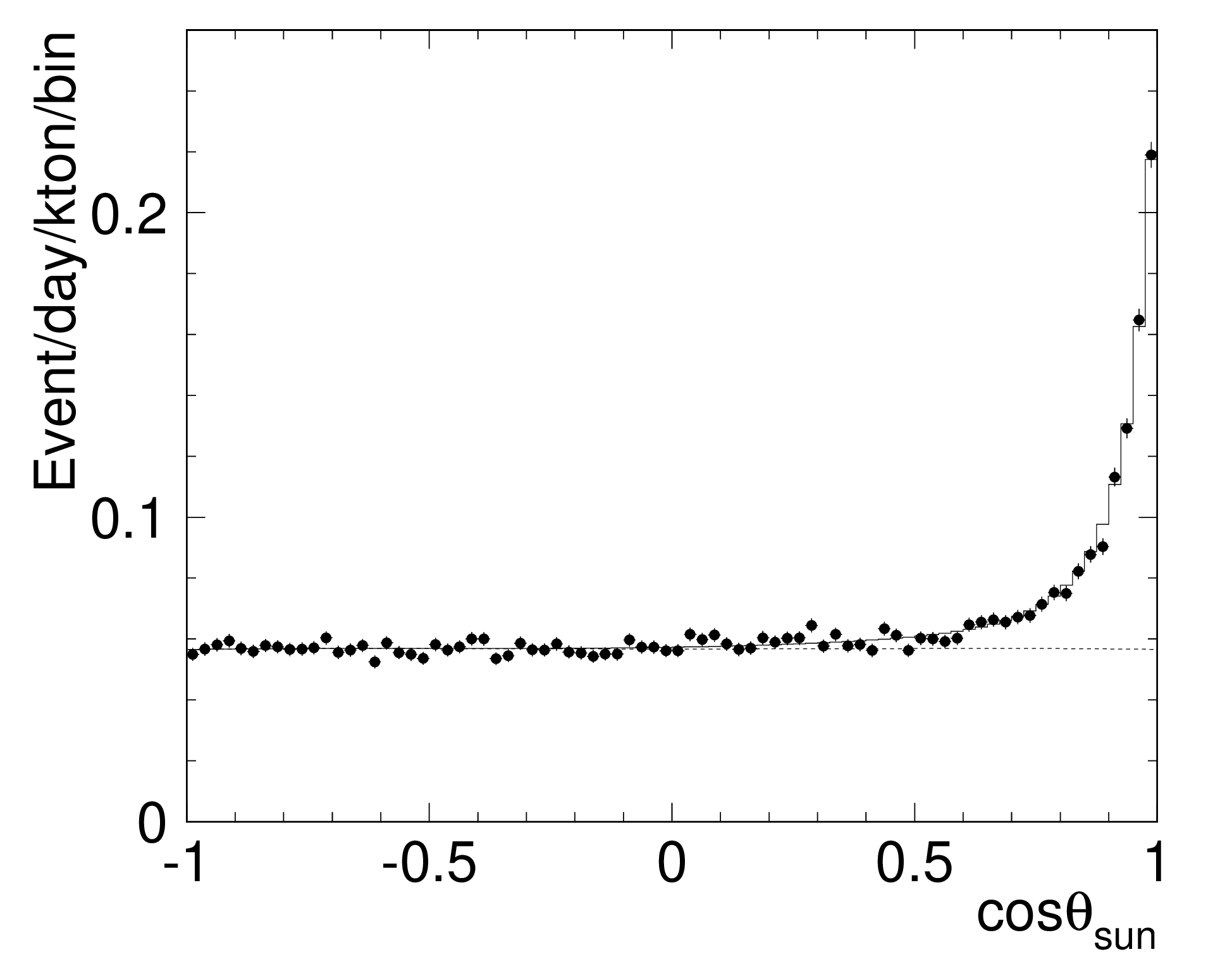}
\caption{The angular distribution of the solar neutrino final sample events.  The dotted line seen under the peak in the solar direction represents background contributions.}
\label{fig:solar_peak}
\end{center}
\end{figure}

 Figure~\ref{fig:solar_peak} shows the angular distribution of extracted solar neutrino events. 

\subsection{Energy Spectrum}

\begin{figure}[htbp]
\begin{center}
\includegraphics[width=7cm]{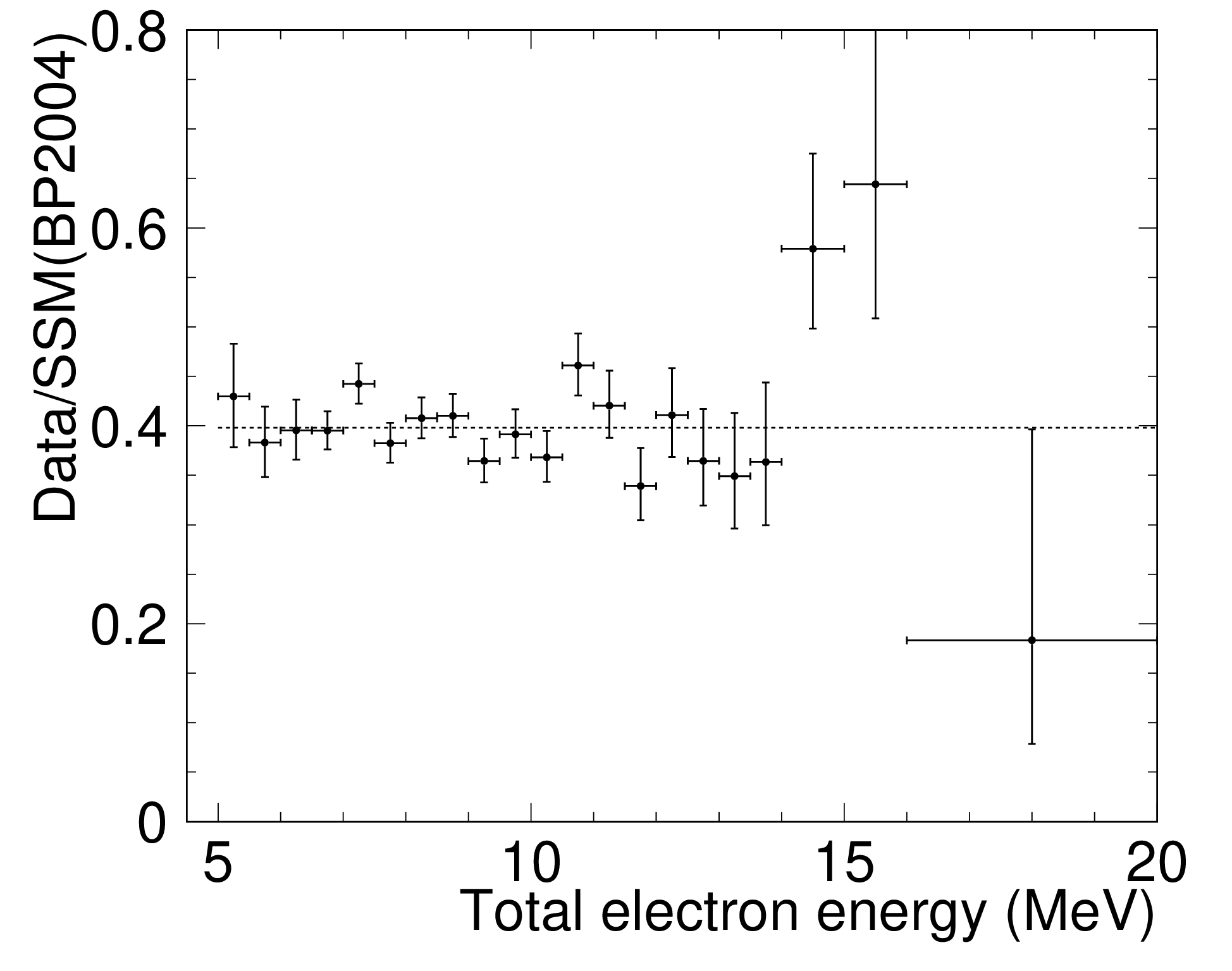}
\caption{Ratio of observed and expected energy spectra.  The dashed line represents the SK-III average.}
\label{fig:energy_spectrum}
\end{center}
\end{figure}

The recoil electron energy spectrum is obtained  from 5.0 to 20.0 MeV in 21 bins.
The definition of the energy bins is given in
Table VI, which shows the observed and expected event rates.
Figure~\ref{fig:energy_spectrum}
 shows the observed energy spectrum divided by the SSM(BP2004) \cite{ssm} without oscillation.  
The line in the Figure~\ref{fig:energy_spectrum} represents the total
SK-III average (flat data/SSM prediction without neutrino oscillation effect). 
The $\chi^2$ value is 27.1/20 dof for the flat prediction and 26.8/20 dof for the prediction with
the best fit neutrino oscillation parameters obtained by the global solar analysis (see Section \ref{sec:global}).
This result indicates no significant spectral distortion.

\begin{table*}[htbp] \label{tab:rates}
 \begin{center}

  \begin{tabular}{c|c|c|c|c|c}
   \hline
   \hline
   Energy    & \multicolumn{3}{c}{Observed rate} & \multicolumn{2}{c}{Expected rate}\\
   (MeV)     &  ALL & DAY & NIGHT & $^8$B & hep\\
             & $ -1 \leq \cos\theta_{\rm z} \leq 1 $ 
             & $ -1 \leq \cos\theta_{\rm z} \leq 0 $ 
             & $  0 <    \cos\theta_{\rm z} \leq 1 $  & &\\ 
   \hline
  \hline
 $ 5.0- 5.5$ & $ 83.3^{+ 10.3}_{- 10.0}$ & $ 94.6^{+ 15.8}_{- 15.0}$ & $ 73.5^{+ 13.7}_{- 13.1}$ & 193.4 & 0.334 \\ 
 $ 5.5- 6.0$ & $ 67.9^{+  6.4}_{-  6.2}$ & $ 75.2^{+  9.8}_{-  9.4}$ & $ 61.5^{+  8.5}_{-  8.0}$ & 177.0 & 0.321 \\ 
 $ 6.0- 6.5$ & $ 63.5^{+  5.0}_{-  4.8}$ & $ 55.9^{+  7.0}_{-  6.6}$ & $ 71.0^{+  7.1}_{-  6.7}$ & 160.4 & 0.310 \\ 
 $ 6.5- 7.0$ & $ 55.3^{+  2.7}_{-  2.6}$ & $ 51.3^{+  3.9}_{-  3.7}$ & $ 59.1^{+  3.9}_{-  3.7}$ & 139.7 & 0.289 \\ 
 $ 7.0- 7.5$ & $ 54.0^{+  2.5}_{-  2.4}$ & $ 55.9^{+  3.7}_{-  3.5}$ & $ 52.3^{+  3.5}_{-  3.4}$ & 121.9 & 0.271 \\ 
 $ 7.5- 8.0$ & $ 40.6^{+  2.2}_{-  2.1}$ & $ 39.9^{+  3.2}_{-  3.0}$ & $ 41.2^{+  3.1}_{-  2.9}$ & 105.8 & 0.257 \\ 
 $ 8.0- 8.5$ & $ 36.7^{+  1.9}_{-  1.8}$ & $ 37.5^{+  2.8}_{-  2.6}$ & $ 35.9^{+  2.6}_{-  2.5}$ &  89.8 & 0.240 \\ 
 $ 8.5- 9.0$ & $ 30.9^{+  1.7}_{-  1.6}$ & $ 28.7^{+  2.4}_{-  2.2}$ & $ 32.9^{+  2.4}_{-  2.3}$ &  75.0 & 0.223 \\ 
 $ 9.0- 9.5$ & $ 22.6^{+  1.4}_{-  1.3}$ & $ 20.0^{+  1.9}_{-  1.8}$ & $ 25.2^{+  2.1}_{-  1.9}$ &  61.8 & 0.205 \\ 
 $ 9.5-10.0$ & $ 19.5^{+  1.3}_{-  1.2}$ & $ 18.0^{+  1.8}_{-  1.6}$ & $ 20.8^{+  1.8}_{-  1.7}$ &  49.5 & 0.186 \\ 
 $10.0-10.5$ & $ 14.5^{+  1.0}_{-  1.0}$ & $ 15.2^{+  1.5}_{-  1.4}$ & $ 13.8^{+  1.5}_{-  1.3}$ &  39.2 & 0.169 \\ 
 $10.5-11.0$ & $ 14.0^{+  1.0}_{-  0.9}$ & $ 15.2^{+  1.5}_{-  1.3}$ & $ 13.0^{+  1.3}_{-  1.2}$ &  30.3 & 0.151 \\ 
 $11.0-11.5$ & $9.62^{+ 0.81}_{- 0.74}$ & $ 9.67^{+ 1.21}_{- 1.07}$ & $ 9.56^{+ 1.12}_{- 1.00}$ & 22.76 & 0.134 \\ 
 $11.5-12.0$ & $5.74^{+ 0.65}_{- 0.58}$ & $ 5.33^{+ 0.92}_{- 0.78}$ & $ 6.17^{+ 0.96}_{- 0.83}$ & 16.81 & 0.118 \\ 
 $12.0-12.5$ & $5.01^{+ 0.58}_{- 0.52}$ & $ 4.20^{+ 0.81}_{- 0.68}$ & $ 5.77^{+ 0.86}_{- 0.74}$ & 12.09 & 0.102 \\ 
 $12.5-13.0$ & $3.11^{+ 0.45}_{- 0.39}$ & $ 2.74^{+ 0.63}_{- 0.50}$ & $ 3.47^{+ 0.67}_{- 0.55}$ &  8.44 & 0.088 \\ 
 $13.0-13.5$ & $1.97^{+ 0.36}_{- 0.30}$ & $ 1.63^{+ 0.49}_{- 0.36}$ & $ 2.30^{+ 0.56}_{- 0.44}$ &  5.56 & 0.074 \\ 
 $13.5-14.0$ & $1.37^{+ 0.30}_{- 0.24}$ & $ 1.17^{+ 0.40}_{- 0.28}$ & $ 1.53^{+ 0.48}_{- 0.36}$ &  3.70 & 0.062 \\ 
 $14.0-15.0$ & $2.22^{+ 0.37}_{- 0.31}$ & $ 2.08^{+ 0.53}_{- 0.41}$ & $ 2.35^{+ 0.54}_{- 0.43}$ &  3.74 & 0.092 \\ 
 $15.0-16.0$ & $0.866^{+0.243}_{-0.182}$ & $0.394^{+0.298}_{-0.164}$ & $1.266^{+0.404}_{-0.288}$ & 1.285 & 0.059 \\ 
 $16.0-20.0$ & $0.117^{+0.136}_{-0.067}$ & $0.252^{+0.245}_{-0.121}$ & $0.000^{+0.130}_{-0.422}$ & 0.570 & 0.068 \\ 
 \hline
  \hline
 \end{tabular}
\caption{SK-III observed energy spectra expressed in units of event/kton/year.  The errors in the observed rates are statistical only.  The expected rates neglecting oscillation are for the BP2004 SSM flux values.  $\theta_{z}$ is the angle between the z-axis of the detector and the vector from the Sun to the detector.}

 \end{center}
\end{table*}

\section{Oscillation Analysis with SK-III result}
Oscillations of solar neutrinos have been studied by numerous experiments.
The results from such experiments have placed increasingly stringent constraints on the
mixing angle between neutrino mass and flavor eigenstates as well as
on the neutrino mass difference.  

In section \ref{sec:sk_osc}, the result of the two-flavor 
neutrino oscillation analysis using SK-I,II,III data is presented.  
A conventional two flavor analysis is done in order to compare directly with previous results.
The results from all other solar neutrino experiments and KamLAND are combined in section~\ref{sec:global}.
The two-flavor analysis  in SK-III is accomplished  in basically the same way as the previous SK-I and SK-II analyses.
Updates to the experimental data which are used in this analysis are explained in each section.

In section~\ref{sec:3f}, the results of the three flavor analysis are shown.
The experimental data have been improved statistically and systematically and now have a few percent uncertainty.
In this paper, we present the first result of a three-flavor analysis using full SK data information.

\subsection{$\chi^2$ definition}

The oscillation analysis of SK uses the spectrum,
 time variation (zenith angle dependence), and total flux
in determination of the solar neutrino oscillation parameters ($\theta_{12}, \Delta m_{12}$). 
For each set of oscillation parameters, the total $^8$B and hep neutrino fluxes are fit to the data.  
The entire SK-III observed spectrum is utilized for a 5.0 MeV threshold.  
The MSW~\cite{msw} $\nu_e$ survival probabilities are numerically calculated from the solar matter distribution provided by  SSM(BP2004). The absolute $^8$B and hep neutrino flux predictions of the SSM are only used as a normalization. The uncertainty of the $^8$B neutrino flux of the SSM is not used in the calculation of the $\chi^2$. We use the uncertainty of the SNO neutral current (NC) measurement instead (see the next section); hence, the absolute $^8$B flux predicted by the SSM does not affect the fitted $^8$B neutrino flux value. For the hep neutrino flux, we use the uncertainty of the SSM, since experimental uncertainty is still large; hence the fitted hep neutrino flux is constrained by the SSM prediction. The predicted neutrino spectrum is then converted to an expected SK-III rate spectrum by using the $\nu - e$ elastic scattering cross section and the SK-III detector energy resolution.  To account for the systematic uncertainties in energy resolution as well as the energy scale and the $^8$B neutrino spectrum model shape, the combined rate predictions are modified by energy shape factors, $f(E_i,\delta_B, \delta_S, \delta_R)$.  The quantities $\delta_B$, $\delta_S$, and $\delta_R$ represent uncertainty in the $^8$B neutrino spectrum, SK-III energy scale, and SK-III energy resolution respectively.  The function $f$ serves to shift the rate predictions corresponding to a given uncertainty $\delta$ in the data rate.  The following equation describes the SK-III spectrum $\chi^{2}$ along with energy-correlated systematic error shape factors applied to the expected rate:

\begin{equation}\label{eq:chi}
\begin{split}
\chi^2_{\textrm{SK-III}} =& \sum_{i=1}^{21} \frac{(d_i - (\beta b_i + \eta h_i) \times f(E_i,\delta_B,\delta_S,\delta_R))^2}{\sigma_i^2}\\
&+\left(\frac{\delta_B}{\sigma_B} \right)^2
+ \left( \frac{\delta_S}{\sigma_S} \right)^2
+ \left( \frac{\delta_R}{\sigma_R} \right)^2
+2 \Delta \log(\cal{L}) \\
&+ \frac{(\phi_{\text{SNO}}-\beta)^2}{\sigma_{\text{SNO}}^2}
+ \frac{(1-\eta)^2}{\sigma_{\text{hep}}^2},
\end{split}
\end{equation}
where $d_i$ is the observed rate divided by the expected, unoscillated rate for the $i^{th}$ energy bin.  Similarly, $b_i$ and $h_i$ are the predicted MSW oscillated rates divided by the unoscillated rate for $^8$B and hep neutrinos respectively.  $\beta$ ($\eta$) scales the $^8$B (hep) neutrino flux.  $\cal{L}$ is the unbinned time-variation likelihood for the SK-III solar zenith angle flux variation above a 5.0 MeV threshold.  This likelihood is analogous to the one used in SK-I and SK-II. The last two terms in the Equation~\ref{eq:chi} are the constraining terms of $^8$B and hep fluxes respectively. The value of $\phi_{\text{SNO}}$ is the NC value coming from the
measurement of SNO divided by the SSM flux value, and $\sigma_{\text{SNO}}$ is the accompanying uncertainty of their measurements ~\cite{snoncdn,snoleta}. The numerical values are    
$\phi_{\text{SNO}} = 0.899$ and $\sigma_{\text{SNO}} = 0.032$. $\sigma_{\text{hep}}$ is the uncertainty of SSM 
prediction on the hep neutrino flux, which is  16\% \cite{ssm}.

\subsection{Oscillation Results - SK with constrained flux}\label{sec:sk_osc}
\begin{figure}[htbp]
\begin{center}
\includegraphics[height=7cm,width=7cm,clip]{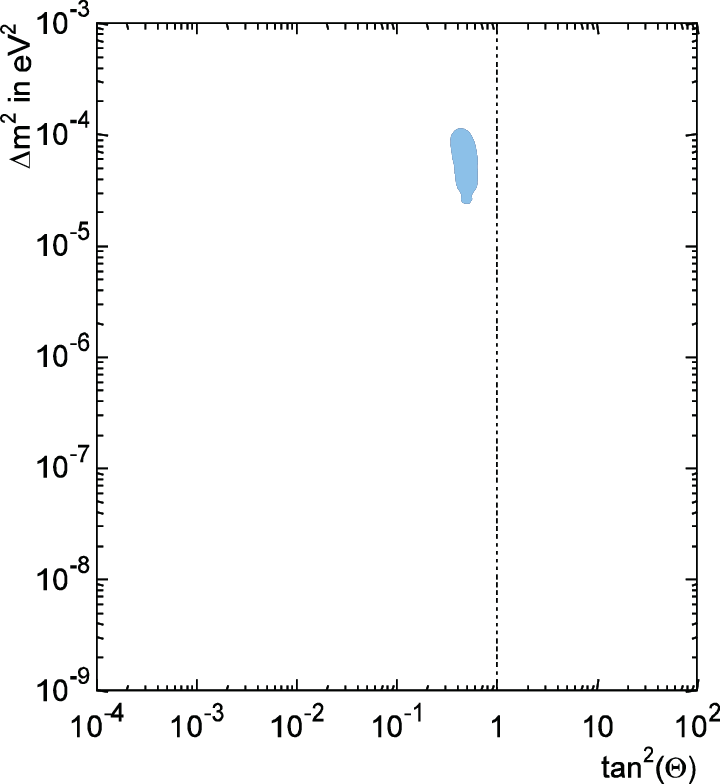}
\caption{95\% C.L. allowed region from SK-I,II,III combined
 analysis. The $^8$B flux is constrained by the SNO NC rate (LETA and phase-III).}
\label{fig:contours}
\end{center}
\end{figure}

 The oscillation analysis is performed by including $\chi^2$ terms corresponding to the SK-I and SK-II values (namely, the spectrum and unbinned time variation for SK-I and SK-II). 
 By constraining the $^8$B flux to the total NC flux value from SNO, allowed parameter regions can be obtained.  Figure~\ref{fig:contours} shows allowed regions at 95\% confidence level.
 The result is consistent with previous SK-I and II results.
This result is the first to show that the energy spectrum and the time variation of the solar neutrino flux measured by SK favor only LMA solution at 95\% C.L., by constraining the $^8$B neutrino flux to the SNO NC flux and the hep neutrino flux to the SSM prediction.

\subsection{Combined Oscillation Results from several experiments} \label{sec:global}

\begin{figure}[htbp]
\begin{center}
\includegraphics[height=7cm,width=7cm,clip]{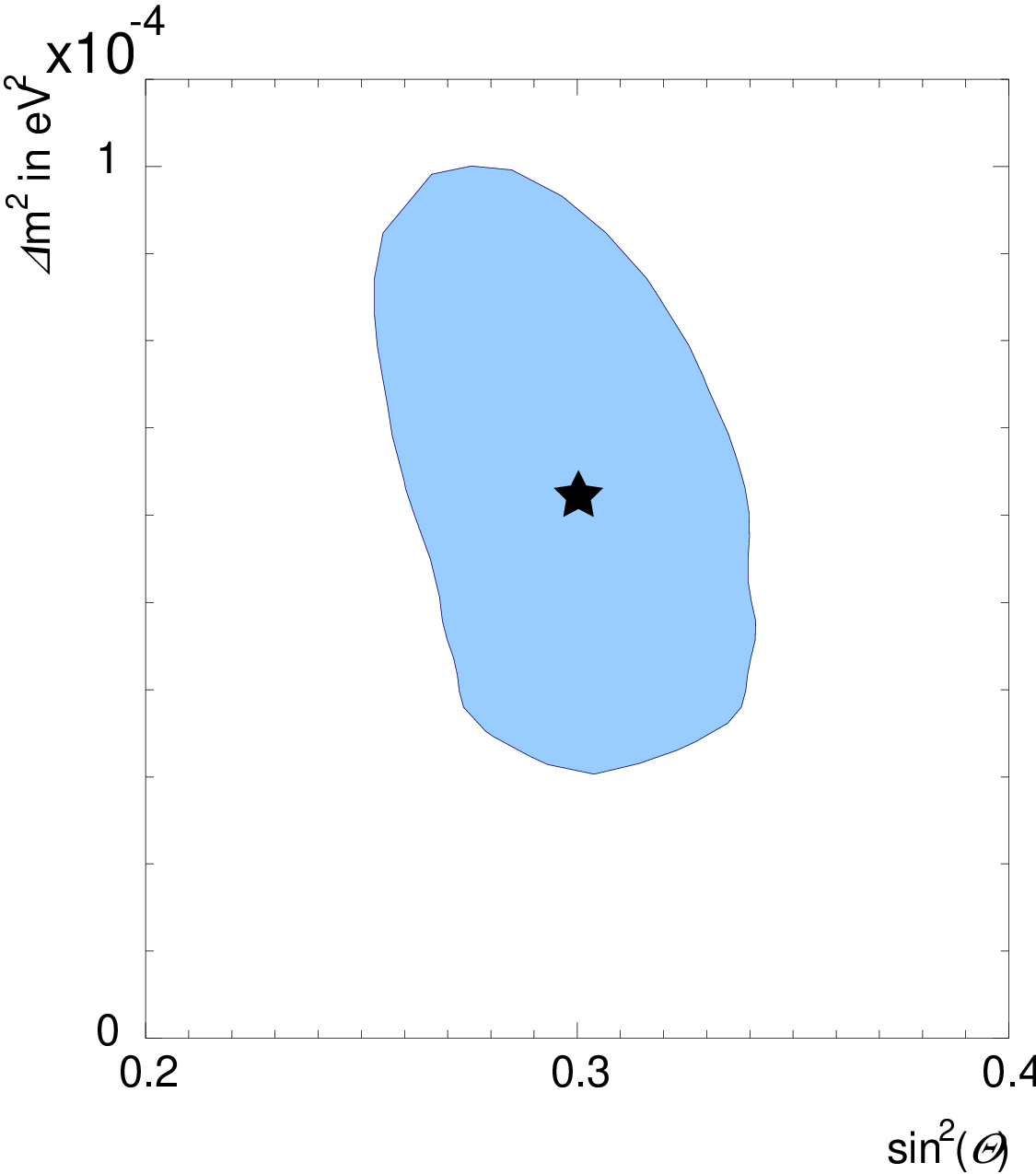}
\caption{Allowed region for all solar experiments  at 95\% C.L..}
\label{fig:global}
\end{center}
\end{figure}

\begin{figure}[htbp]
\begin{center}
\includegraphics[height=7cm,width=7cm,clip]{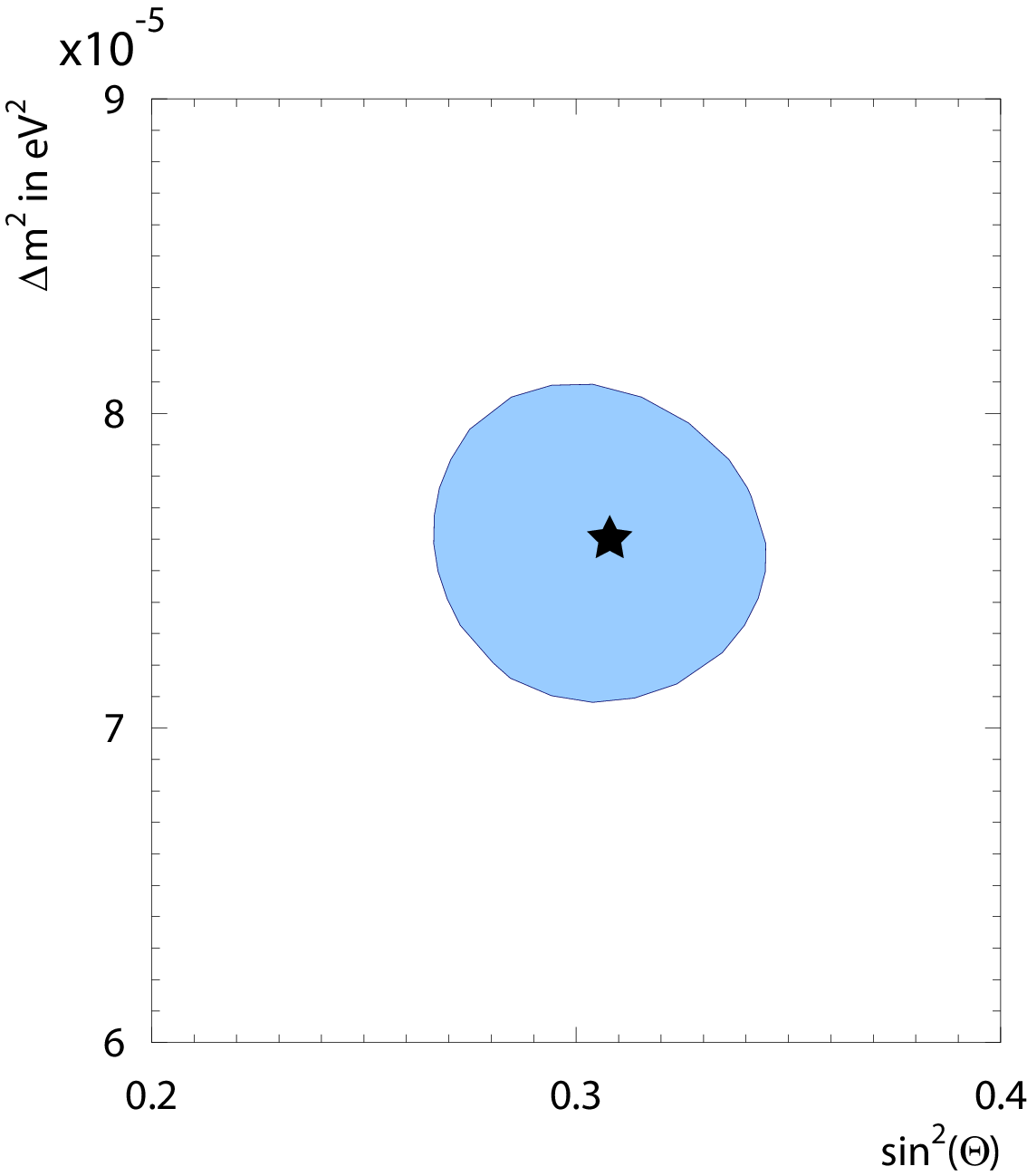}
\caption{Allowed region for all solar experiments and KamLAND for two-flavor analysis at 95\% C.L..}
\label{fig:global_kamland}
\end{center}
\end{figure}

The combination of other solar neutrino experimental results 
such as the SNO, Borexino and radiochemical results with the SK 
combined analysis is accomplished with a two-flavor neutrino
oscillation analysis by constructing a global $\chi^2$.
For the SNO results, the total CC rates observed
in the 306-day pure D$_2$O phase (SNO-I)~\cite{snod2o}, 391-day salt phases~(SNO-II)\cite{snosalt}, and 385-day NCD phase (SNO-III)~\cite{snoncdn},  the combined NC rates of LETA~\cite{snoleta} and SNO-III, and the predicted day-night asymmetry for SNO-I and II are used. 
In this analysis, the correlations between the SNO CC and NC rates are not taken into account.
The $^7\mathrm{Be}$ solar neutrino flux of Borexino's 192-day~\cite{borexino} and the radiochemical experiments of Homestake~\cite{cl}, GALLEX-GNO~\cite{gallex_gno}, and SAGE~\cite{sage} are then included into the global $\chi^2$ with the fluxes and their correlations calculated by SSM in a way shown by \cite{fogli}. Figure~\ref{fig:global} shows the combined solar allowed region. The best fit parameter set is 
$ \sin^2\theta_{12} = 0.30^{+0.02}_{-0.01}(\tan^2\theta_{12} = 0.42^{+0.04}_{-0.02})$ and 
$\Delta m^2_{21} = 6.2^{+1.1}_{-1.9}\times 10^{-5} \textrm{eV}^2$, consistent with the SK-I global analysis.
In addition, combining the above and KamLAND data, the best fit parameter set is 
$ \sin^2\theta_{12} = 0.31 \pm 0.01(\tan^2\theta_{12} = 0.44 \pm 0.03)$ and 
$\Delta m^2_{21} = 7.6 \pm 0.2 \times 10^{-5} \textrm{eV}^2$ as shown in Figure \ref{fig:global_kamland}.

\begin{figure}[htbp]
\begin{center}
\includegraphics[height=7cm,width=7cm,clip]{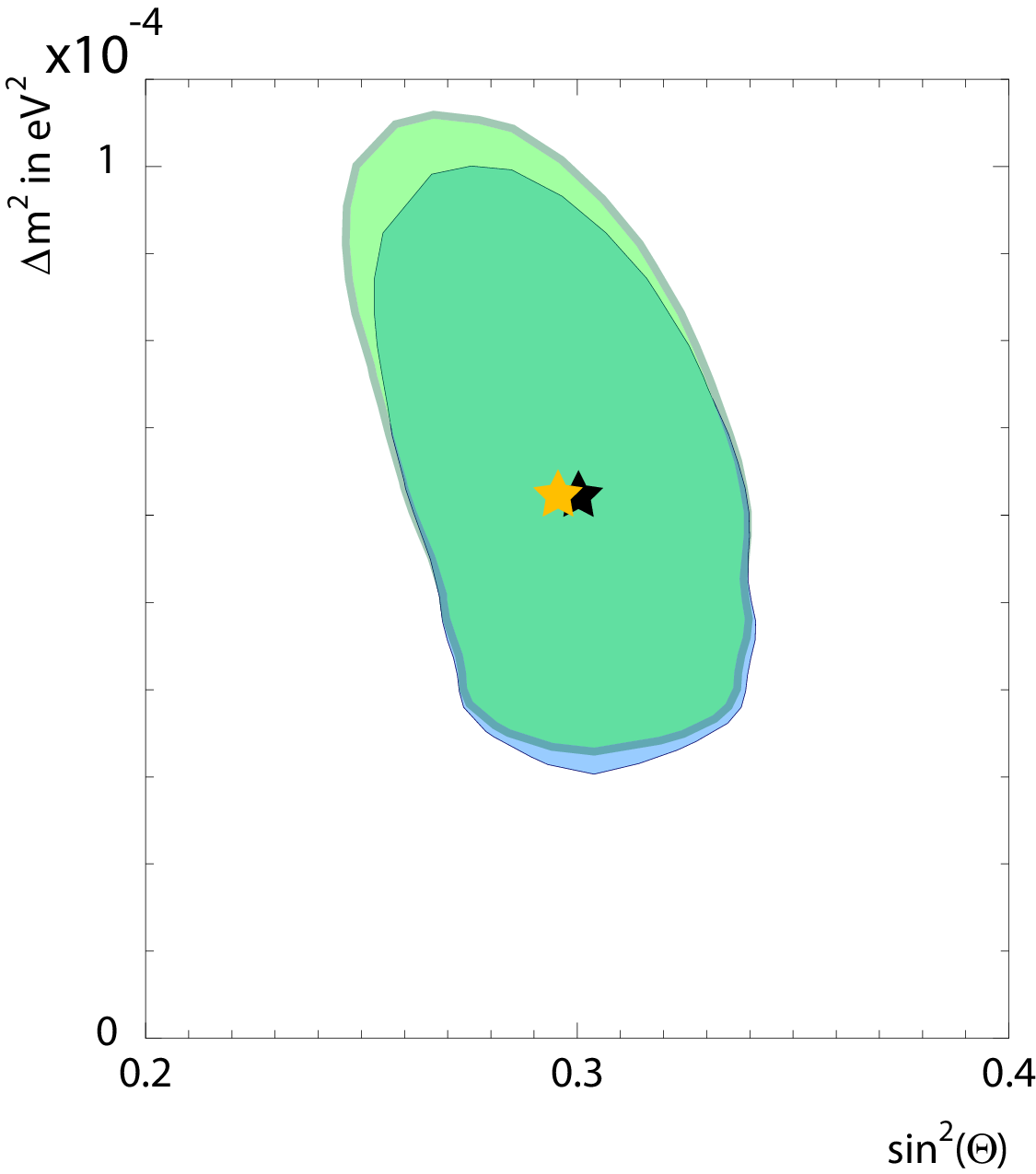}
\caption{Allowed region for all solar experiments for two-flavor analysis at 95\% C.L..
The light green contour and the yellow star show the result without SK-III, and the light blue contour and the black star show 
the result with SK-III, which is same as Figure~\ref{fig:global}.
}
\label{fig:w_wo_sk3}
\end{center}
\end{figure}

Figure~\ref{fig:w_wo_sk3} shows the 95\% allowed region for all solar experiments before and after the SK-III result 
is included. As shown in the figure, the SK-III result contributes about
5\% improvement to the uncertainty of 
$\Delta m_{12}^2$.

\subsection{Three flavor analysis} \label{sec:3f}
 In a three-flavor analysis, the calculation of oscillation probability is based on \cite{barger}.  The probability can be calculated with three parameters: $\theta_{12}$, $\theta_{13}$, and $\Delta m^2_{12}$, assuming $\Delta m^2_{12} \ll \Delta m^2_{23} \sim \Delta m^2_{13}$. We fixed $\Delta m^2_{23} = 2.4 \times 10^{-3}$~eV$^2$ and the normal hierarchy is assumed. For the solar neutrino oscillation, the other mixing parameters are irrelevant, but we set $\theta_{23}=\pi/4$ and $\delta_{CP}=0$ in our calculation.

As done for the two-flavor analysis, the oscillation probabilities
depending on different zenith angles are calculated, and then the rate of the radiochemical experiments, Borexino ($^7$Be neutrino flux), and SNO (CC for all phases, NC for LETA and phase-III) are calculated. SNO day-night asymmetries for phase I and II are also predicted. 

The KamLAND spectra and rates from surrounding reactors are calculated
based on the information from the KamLAND official database~\cite{kldb} and
the published paper~\cite{kamland}. Because the information in \cite{kldb}
is used for their second result~\cite{kam2nd}, we
normalized the calculated neutrino spectrum without oscillation to
that of Figure 1 of \cite{kamland}. The oscillation probability is
calculated using three-flavor vacuum oscillation. The amount of
background and systematic uncertainties for each energy bin are read
from Figure~1 of \cite{kamland}. The background is fixed in our
calculation of $\chi^2$. The systematic uncertainty for each energy
bin is treated as an energy-uncorrelated systematic uncertainty. The
systematic uncertainty of event rate on Table 1 of \cite{kamland} is
also taken into account.  We checked first that the contour of the two-flavor
analysis in Figure 2 of \cite{kamland} was reproduced with
$\theta_{13} = 0$.  We also checked that our three-flavor contour 
was very close to the contour presented in \cite{ichep08kl},
and consistent with their latest result \cite{kl3f}. 
 
\begin{figure}[htbp]
\begin{center}
\includegraphics[height=7cm,width=7cm,clip]{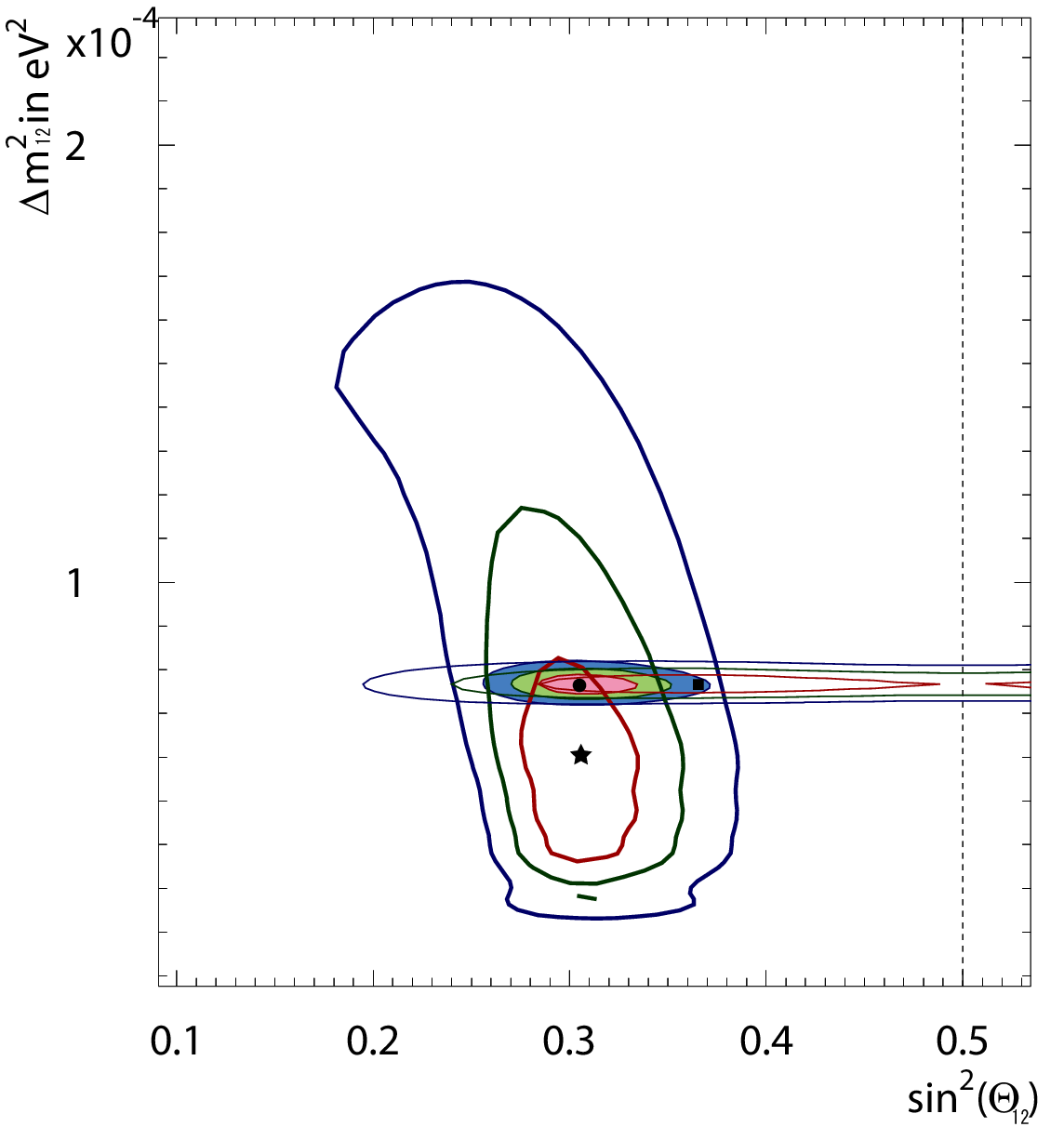}
\caption{Allowed region in solar parameter space $(\theta_{12}, \Delta
  m^2)$ obtained by the three-flavor analysis. The thick lines and the
  star mark show the allowed regions and the best fit point of
  the global solar analysis. The thin lines and the square mark
  show the allowed regions and the best fit point of our KamLAND
  analysis. The filled areas and the filled circle mark show the
  allowed regions and the best fit point of the combined analysis. For all regions, the innermost area (red), the middle area (green) and the outermost area (blue) show 68.3, 95, 99.7 \% C.L. respectively.}
\label{fig:3f_gk}
\end{center}
\end{figure}

 The oscillation parameters are scanned in the following regions:
 $10^{-5}~$eV$^2 <\Delta m^2_{12} < 2 \times 10^{-4}~ $eV$^2$, $0.1< \tan^2
 \theta_{12}< 1.0 $, and $0< \sin^2\theta_{13}< 0.25$. Figure
 \ref{fig:3f_gk} shows the allowed region of the solar neutrino
 parameters, ($\theta_{12}, \Delta m^2_{12}$), obtained by the
 three-flavor analysis of the global solar results. The allowed region
 for KamLAND obtained by our three-flavor analysis and the allowed
 region for global solar and KamLAND combined analysis are also shown
 in Figure \ref{fig:3f_gk}.  Inclusion of another oscillation parameter $\theta_{13}$ results
 in a weaker constraint on the solar parameter space.
  
\begin{figure}[htbp]
\begin{center}
\includegraphics[height=7cm,width=7cm,clip]{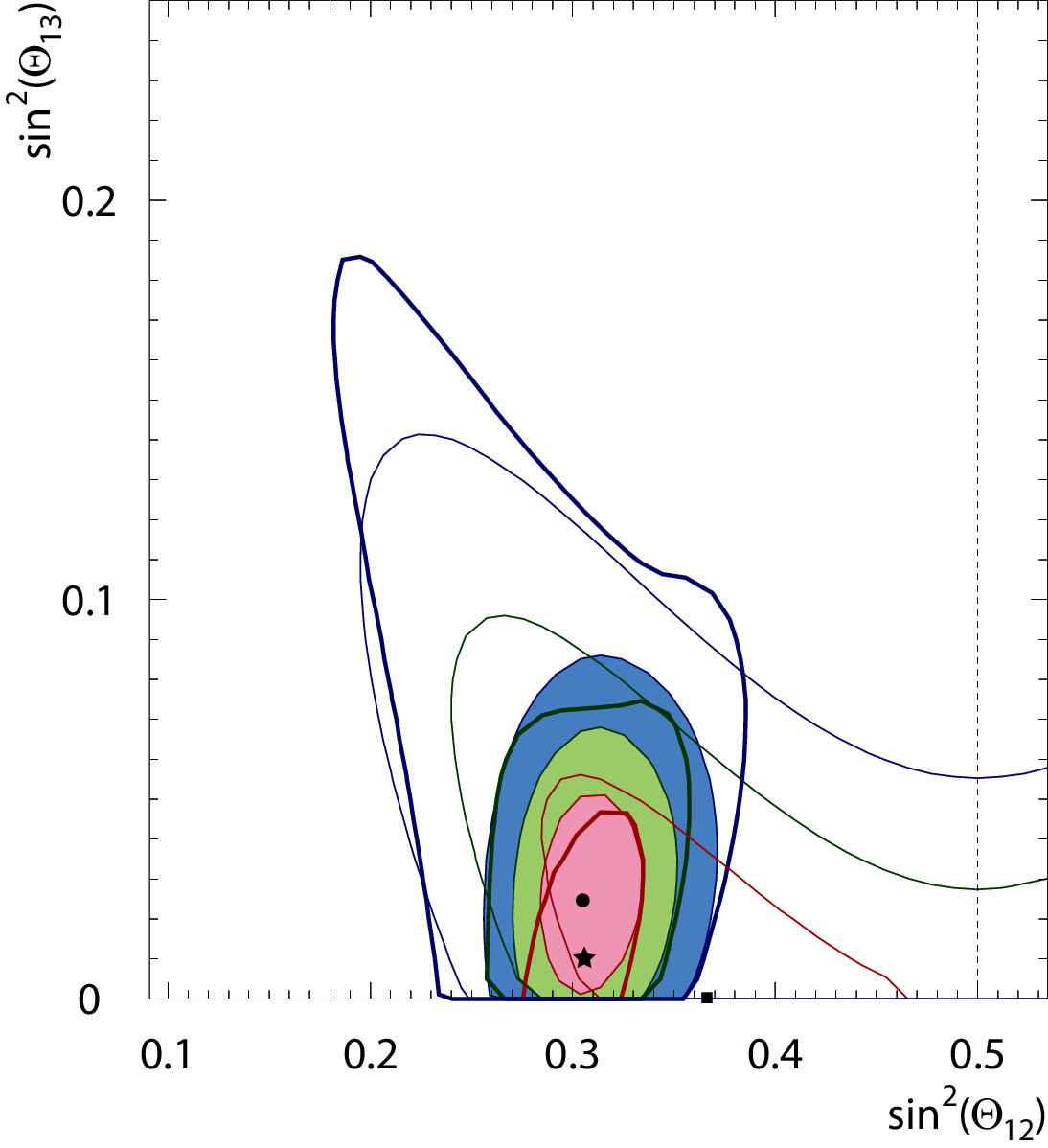}
\caption{Allowed region in $(\theta_{12}, \theta_{13})$ space obtained by the three-flavor analysis. 
The definitions of marks and lines are same as in Figure~\ref{fig:3f_gk}.}
\label{fig:3f_13_12}
\end{center}
\end{figure}

  Figure \ref{fig:3f_13_12} shows the allowed region in ($
  \theta_{12}, \theta_{13}$) space obtained from the global solar
  analysis and our KamLAND analysis.  As shown in the figure,  in the
  global solar contour, the larger value of $\theta_{13}$ prefers  the
  larger value of $\theta_{12}$, while in the KamLAND contour the
  larger value of $\theta_{13}$ prefers the smaller value of
  $\theta_{12}$.  The global solar analysis finds that the best fit
  values at 
$ \sin^2\theta_{12} = 0.31\pm0.03 ~(\tan^2\theta_{12} = 0.44\pm0.06)$ and 
$\Delta m^2_{21} = 6.0^{+2.2}_{-2.5}\times 10^{-5} \textrm{eV}^2$.
Combined with the KamLAND result, the best-fit oscillation parameters are found to be
$\sin^2\theta_{12} = 0.31 ^{+0.03}_{-0.02} ~(\tan^2\theta_{12} = 0.44^{+0.06}_{-0.04}  )$ and 
$\Delta m^2_{21} = 7.7 \pm 0.3\times 10^{-5} \textrm{eV}^2$. The best fit value of
 $\sin^2 \theta_{13}$ is 0.01, and an upper bound is obtained,
  $\sin^2 \theta_{13} < 0.060$ at the 95\% C.L., for the global solar analysis. 
 Combining with the KamLAND contour, the best fit value of  $\sin^2 \theta _ {13}$ is $0.025^{+0.018}_{-0.016}$ and the 95\% C.L. upper limit of the $\sin^2\theta_{13}$ is found to be 0.059.
  
The flux value of $^8$B neutrinos can be extracted using the oscillation parameters 
 obtained from the fitting of the global solar and KamLAND result.
As in Equation \ref{eq:chi}, $\beta$ is a 
free parameter to minimize the $\chi^2$ and there is no
constraint from the SSM prediction in $\chi^2_{SK+SNO}$.
Table~\ref{tab:b8} summarizes 
the scaled $^8$B flux values by using $\beta_m$ at the best fit point
obtained by the global solar analysis and the global solar + KamLAND analyses
in both two and three flavor analyses.
The size of the error corresponds to the maximum and minimum flux values among the 
1$\sigma$ oscillation parameter region.
As shown in the table, the $^8$B flux agrees well with the latest SSM prediction~\cite{bs09},
and the size of  the uncertainty is $2\sim3\%$ which is consistent with the SNO result~\cite{snoleta}.

\begin{table}
\begin{center}
\begin{tabular}{l c} \hline\hline
                                 & $^8$B flux ($\times 10^{6}$cm$^{-2}$s$^{-1}$) \\ \hline
Global solar (2 flavor)          & 5.3 $\pm 0.2$ \\ 
Global solar + KamLAND (2 flavor)& 5.1 $\pm 0.1$ \\ 
Global solar (3 flavor)          & 5.3 $\pm 0.2$ \\
Global solar + KamLAND (3 flavor)& 5.3 $^{+0.1}_{-0.2}$\\ \hline 
\end{tabular}
\end{center}
\caption{$^8$B neutrino flux obtained from the oscillation parameter fitting. \label{tab:b8}}
\end{table}

\section{Conclusion}
Super-Kamiokande has measured the solar $^8$B flux to be 
$(2.32\pm0.04(\textrm{stat.})\pm 0.05(\textrm{sys.}))\times10^6~\textrm{cm}^{-2}\textrm{sec}^{-1}$ during its third phase; 
 the systematic uncertainty is smaller than for SK-I. 
Combining all solar experiments 
 in a two flavor fit, the best fit is found to favor the LMA region at 
$ \sin^2\theta_{12} = 0.30^{+0.02}_{-0.01}~(\tan^2\theta_{12} = 0.42^{+0.04}_{-0.02})$ and 
$\Delta m^2_{21} = 6.2^{+1.1}_{-1.9}\times 10^{-5} \textrm{eV}^2$.
Combined with the KamLAND result, the best-fit oscillation parameters are found to be
$\sin^2\theta_{12} = 0.31 \pm 0.01~(\tan^2\theta_{12} = 0.44 \pm 0.03 )$ and 
$\Delta m^2_{21} = 7.6 \pm 0.2\times 10^{-5} \textrm{eV}^2$,
 in excellent agreement with previous solar neutrino oscillation measurements. 
 In a three-flavor analysis combining all solar neutrino experiments and the KamLAND result, 
the best fit value of $\sin^2 \theta_{13}$ is found to be  $0.025^{+0.018}_{-0.016}$ and
an upper bound is obtained as $\sin^2\theta_{13} < 0.059$ at 95$\%$
C. L..

\section{Acknowledgments}

The authors gratefully acknowledge the cooperation of the Kamioka Mining and Smelting Company. Super-K has been built and operated from funds provided by the Japanese Ministry of Education, Culture, Sports, Science and Technology, the U.S. Department of Energy, and the U.S. National Science Foundation. This work was partially supported by the  Research Foundation of Korea (BK21 and KNRC), the Korean Ministry of Science and Technology, the National Science Foundation of China, and the Spanish Ministry of Science and Innovation (Grants FPA2009-13697-C04-02 and Consolider-Ingenio-2010/CPAN).

\clearpage
\appendix
\setcounter{figure}{0} \renewcommand{\thefigure}{A.\arabic{figure}} 
\setcounter{table}{0} \renewcommand{\thetable}{A.\arabic{table}} 

\section{Revised SK-III results}

Since the publication of this report\cite{sk3}, two mistakes were found.
One is in how energy-dependent systematic errors are calculated and the
other is related to the flux calculation in SK-III.
The estimates of the energy-correlated uncertainties in the main text
of this report are
based on the Monte Carlo (MC) simulated $^8$B solar neutrino events.
It is found that this evaluation method was not accurate enough.
The statistical error of the MC simulation distorted the shapes of the 
energy-correlated uncertainties systematically.

\begin{figure}[!hbt]
\includegraphics[width=8.0cm,clip]{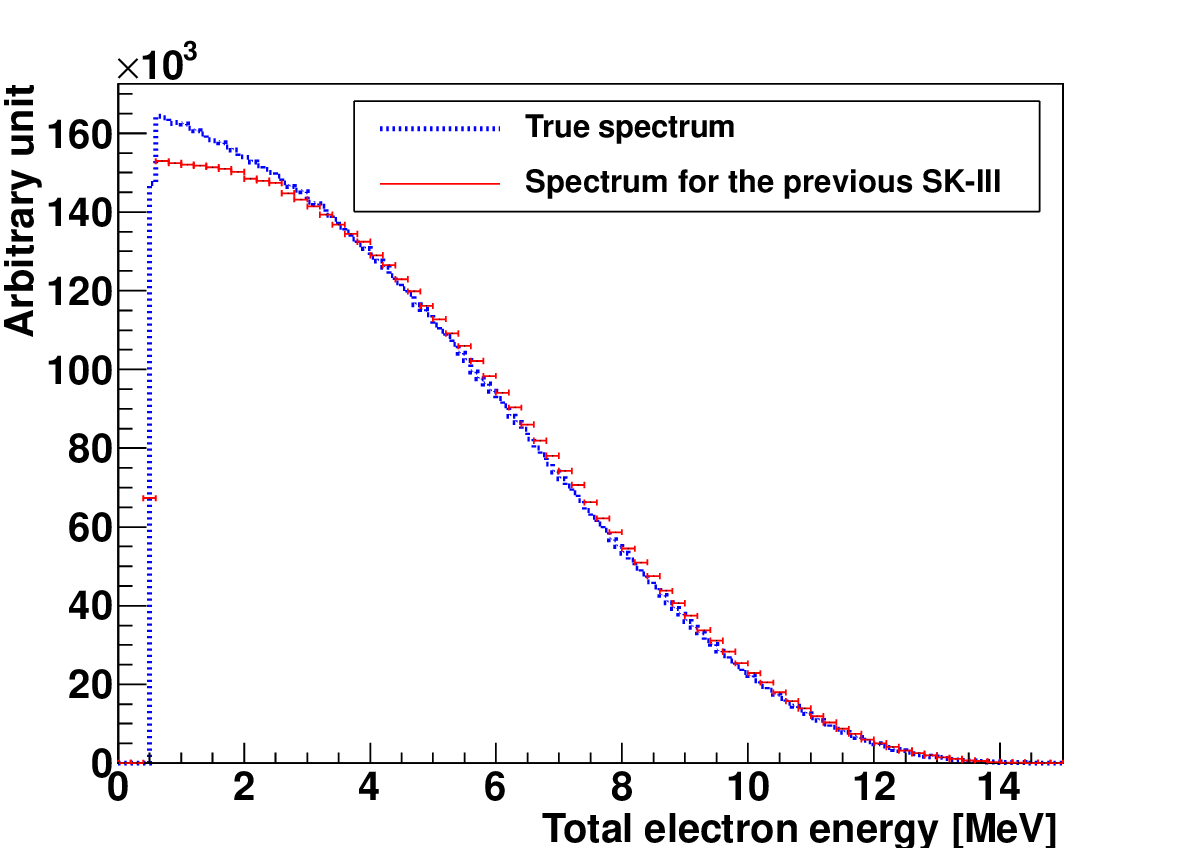}
\caption{Energy spectrum shapes of recoil electrons from $^8$B solar neutrinos
 for SK-III.
   The blue dotted and red solid lines show the true theoretical calculation 
   and wrong spectrum used in the SK-III analysis in the main text of this paper.
   }
   \label{fig:hitmis-sk3}
\end{figure}

The energy dependence of the differential interaction cross-section
between neutrinos and electrons was accidentally eliminated
only for the SK-III flux calculation in the main text.
Figure \ref{fig:hitmis-sk3} shows the energy distributions of recoil
electrons from $^8$B solar neutrinos.
The blue dotted histogram shows the true energy spectrum shape from a
theoretical calculation considering the detector resolutions.
The red solid plot shows the energy spectrum shape used in the SK-III analysis
in the main text of this report.
The expected total flux was normalized correctly, but the expected 
$^8$B energy spectrum shape was wrong in the analysis. 

These mistakes will be fixed in the SK-IV solar neutrino paper\cite{sk4}.
In this appendix, the revised SK-III solar neutrino results are described.
The oscillation results are not revised in this appendix.
The latest oscillation results, including both revised SK-III data 
and SK-IV data, will be reported in the SK-IV paper\cite{sk4}.

\subsection{Systematic uncertainties}

The energy-correlated systematic uncertainties are obtained by counting the number of 
events in the solar neutrino MC simulation with artificially shifted energy scale,
energy resolution and $^8$B solar neutrino energy spectrum.
In the SK-III analysis in the main text, this estimation was done with the generated 
solar neutrino MC events.
However, in the high energy region, not enough MC events were generate
to accurately estimate the small systematic errors.
In the current analysis, this estimation is performed with a theoretical calculation
considering the detector resolutions,
thus eliminating the statistical effects introduced by the small MC
statistics.

\begin{figure}[thb]
\includegraphics[width=7.0cm,clip]{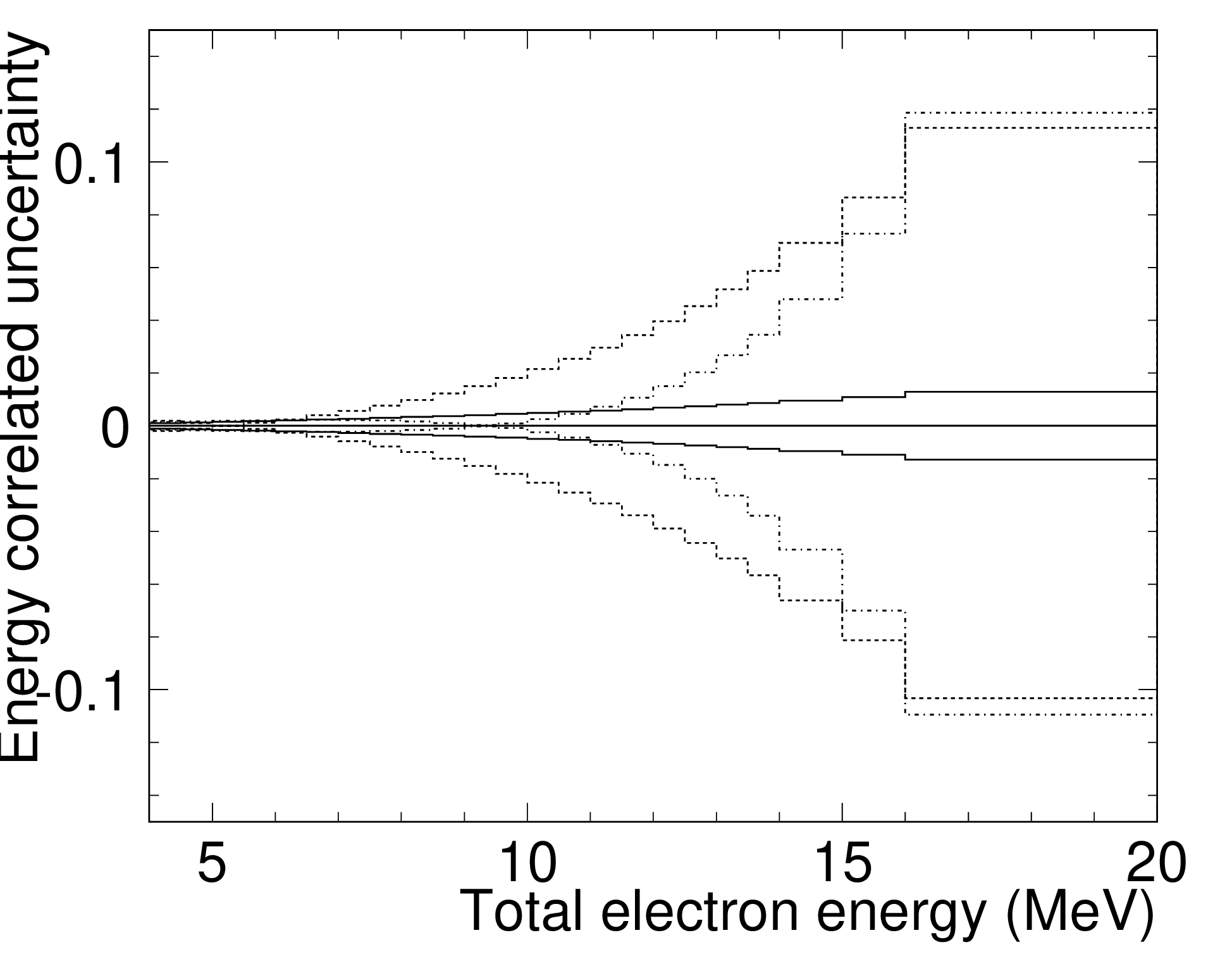}
\caption{Revised energy-correlated systematic uncertainties in SK-III. 
   The solid, dotted, and dashed lines show the uncertainties of the $^8$B
   spectrum, the energy scale, and the energy resolution, respectively.
 This is a revision of Fig.~\ref{fig:ecorr} in the main text.}
   \label{fig:ecorr-sk3}
\end{figure}

The revised results of the energy-correlated systematic uncertainties
are shown in Figure \ref{fig:ecorr-sk3}.
In this update, the uncertainty from $^8$B spectrum shape was improved.

\begin{table}[!hbt]
\label{tab:totalsys-sk3}
\begin{center}
\begin{tabular}{l c  } \hline
Source            &  Total Flux       \\ \hline\hline
Energy scale      &$\pm 1.4   $ \\ 
Energy resolution &$\pm 0.2   $  \\ 
$^8$B spectrum      &$\pm 0.4   $            \\ 
Trigger efficiency        &$\pm 0.5  $   \\ 
Angular resolution   &$\pm 0.67   $ \\ 
Fiducial volume (vertex shift) &$\pm 0.54  $ \\ 
Event quality cuts         &  \\
- Quality cut &$\pm 0.4   $  \\ 
- Hit pattern cut   &$\pm 0.25  $  \\ 
- Second vertex     &$\pm 0.45  $  \\ 
Spallation        &$\pm 0.2   $ \\ 
External event cut     &$\pm 0.25  $ \\
Small cluster hits cut   &$\pm 0.5   $  \\ 
Background shape  &$\pm 0.1   $  \\ 
Signal extraction method&$\pm 0.7  $   \\ 
Livetime          &$\pm 0.1 $ \\
Cross section     &$\pm 0.5  $   \\ \hline
Total             &$\pm 2.2  $\\ \hline
\end{tabular}
\end{center}
\caption{ Revised summary of the systematic uncertainty of the total flux 
 in 5.0--20.0MeV in SK-III.
 This is a revision of Table~\ref{tab:totalsys} in the main text.}
\end{table}

\begin{table*}[!t]
 \label{tab:rates-sk3}
 \begin{center}
  \begin{tabular}{c|c|c|c|c|c}
   \hline
   \hline
   Energy    & \multicolumn{3}{c}{Observed rate} & \multicolumn{2}{c}{Expected rate}\\
   (MeV)     &  ALL & DAY & NIGHT & $^8$B & hep\\
             & $ -1 \leq \cos\theta_{\rm z} \leq 1 $ 
             & $ -1 \leq \cos\theta_{\rm z} \leq 0 $ 
             & $  0 <    \cos\theta_{\rm z} \leq 1 $  & &\\ 
   \hline
  \hline
 $ 5.0- 5.5$ & $ 82.3^{+ 10.3}_{-  9.9}$ & $ 93.4^{+ 15.7}_{- 14.9}$ & $ 72.6^{+ 13.7}_{- 13.0}$ & 189.7 & 0.334 \\
 $ 5.5- 6.0$ & $ 66.4^{+  6.4}_{-  6.1}$ & $ 73.7^{+  9.8}_{-  9.3}$ & $ 59.9^{+  8.4}_{-  7.9}$ & 172.2 & 0.321 \\
 $ 6.0- 6.5$ & $ 62.9^{+  4.9}_{-  4.7}$ & $ 55.3^{+  7.0}_{-  6.5}$ & $ 70.4^{+  7.1}_{-  6.7}$ & 155.2 & 0.310 \\
 $ 6.5- 7.0$ & $ 54.8^{+  2.7}_{-  2.6}$ & $ 50.8^{+  3.8}_{-  3.7}$ & $ 58.7^{+  3.8}_{-  3.7}$ & 134.3 & 0.289 \\
 $ 7.0- 7.5$ & $ 53.8^{+  2.5}_{-  2.4}$ & $ 55.6^{+  3.6}_{-  3.5}$ & $ 52.1^{+  3.5}_{-  3.3}$ & 117.1 & 0.271 \\
 $ 7.5- 8.0$ & $ 40.4^{+  2.2}_{-  2.1}$ & $ 39.6^{+  3.1}_{-  3.0}$ & $ 41.1^{+  3.1}_{-  2.9}$ & 101.2 & 0.257 \\
 $ 8.0- 8.5$ & $ 36.4^{+  1.9}_{-  1.8}$ & $ 37.2^{+  2.7}_{-  2.6}$ & $ 35.7^{+  2.6}_{-  2.5}$ &  85.8 & 0.240 \\
 $ 8.5- 9.0$ & $ 30.5^{+  1.7}_{-  1.6}$ & $ 28.4^{+  2.3}_{-  2.2}$ & $ 32.6^{+  2.4}_{-  2.2}$ &  71.7 & 0.223 \\
 $ 9.0- 9.5$ & $ 22.4^{+  1.4}_{-  1.3}$ & $ 19.8^{+  1.9}_{-  1.8}$ & $ 24.9^{+  2.1}_{-  1.9}$ &  58.5 & 0.205 \\
 $ 9.5-10.0$ & $ 19.1^{+  1.2}_{-  1.2}$ & $ 17.7^{+  1.7}_{-  1.6}$ & $ 20.3^{+  1.8}_{-  1.7}$ &  47.1 & 0.186 \\
 $10.0-10.5$ & $ 14.3^{+  1.0}_{-  1.0}$ & $ 15.0^{+  1.5}_{-  1.4}$ & $ 13.6^{+  1.4}_{-  1.3}$ &  37.0 & 0.169 \\
 $10.5-11.0$ & $ 13.7^{+  1.0}_{-  0.9}$ & $ 14.7^{+  1.4}_{-  1.3}$ & $ 12.9^{+  1.3}_{-  1.2}$ &  28.5 & 0.151 \\
 $11.0-11.5$ & $9.41^{+ 0.79}_{- 0.73}$ & $ 9.36^{+ 1.17}_{- 1.03}$ & $ 9.44^{+ 1.11}_{- 0.98}$ & 21.45 & 0.134 \\
 $11.5-12.0$ & $5.63^{+ 0.64}_{- 0.57}$ & $ 5.24^{+ 0.90}_{- 0.76}$ & $ 6.04^{+ 0.94}_{- 0.81}$ & 15.76 & 0.118 \\
 $12.0-12.5$ & $4.91^{+ 0.57}_{- 0.50}$ & $ 4.08^{+ 0.79}_{- 0.66}$ & $ 5.69^{+ 0.85}_{- 0.73}$ & 11.21 & 0.102 \\
 $12.5-13.0$ & $3.03^{+ 0.44}_{- 0.38}$ & $ 2.67^{+ 0.61}_{- 0.49}$ & $ 3.38^{+ 0.65}_{- 0.53}$ &  7.79 & 0.088 \\
 $13.0-13.5$ & $1.92^{+ 0.35}_{- 0.29}$ & $ 1.59^{+ 0.47}_{- 0.35}$ & $ 2.25^{+ 0.55}_{- 0.43}$ &  5.22 & 0.074 \\
 $13.5-14.0$ & $1.32^{+ 0.29}_{- 0.23}$ & $ 1.13^{+ 0.39}_{- 0.27}$ & $ 1.48^{+ 0.47}_{- 0.35}$ &  3.39 & 0.062 \\
 $14.0-15.0$ & $2.15^{+ 0.36}_{- 0.30}$ & $ 2.00^{+ 0.51}_{- 0.40}$ & $ 2.31^{+ 0.53}_{- 0.42}$ &  3.49 & 0.092 \\
 $15.0-16.0$ & $0.832^{+0.234}_{-0.175}$ & $0.381^{+0.289}_{-0.158}$ & $1.208^{+0.385}_{-0.275}$ & 1.227 & 0.059 \\
 $16.0-20.0$ & $0.112^{+0.130}_{-0.064}$ & $0.244^{+0.238}_{-0.117}$ & $0.000^{+0.123}_{-0.401}$ & 0.513 & 0.068 \\
 \hline
  \hline
 \end{tabular}
  \caption{Revised observed energy spectra expressed in units of event/kton/year in SK-III
  in each recoil electron total energy region.
  The errors in the observed rates are statistical only.  
  The expected rates neglecting oscillation are for the BP2004 SSM flux values.  
  $\theta_{z}$ is the angle between the z-axis of the detector and the
  vector from the Sun to the detector.
  This is a revision of Table VI in the main text.}
 \end{center}
\end{table*}

\begin{figure}[tbh]
 \includegraphics[width=7.0cm,clip]{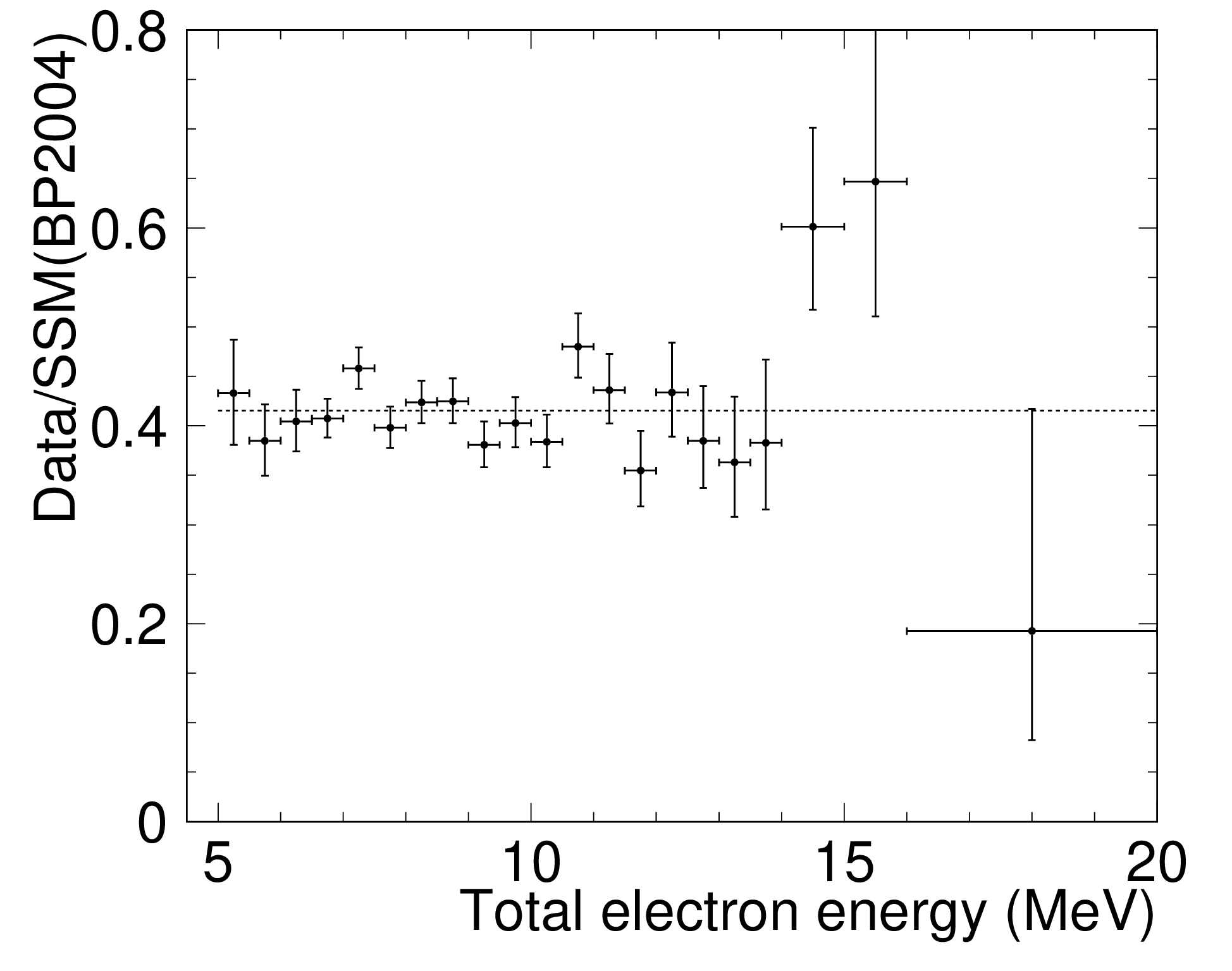}
 \caption{Revised ratio of observed and expected energy spectra in SK-III. 
 The dashed line represents the revised SK-III average.
 This is a revision of Fig.~\ref{fig:energy_spectrum} in the main text.}
 \label{fig:energy_spectrum-sk3}
\end{figure}
The systematic uncertainties on total flux in SK-III are also revised.
The revised uncertainties are summarized in Table~\ref{tab:totalsys-sk3}.
The $^8$B spectrum error was under estimated in the analysis in the main
text.
The revised systematic uncertainty on the total flux in E$_{\rm total}$ = 5.0-20.0 MeV 
in SK-III is estimated to be 2.2$\%$.
\vspace{5mm}

\subsection{$^8$B solar neutrino flux results}

The observed number of solar neutrino events is also updated.
In this analysis, the extracted number of $^8$B solar neutrinos
with the ES reaction in E$_{\rm total}$ = 5.0-20.0 MeV  
for a live time of 548 days of SK-III data
was 8148$^{+133}_{-131}$(stat) $\pm176$(sys).
The corresponding $^8$B flux is obtained to be: 
\begin{eqnarray*}
 (2.40 \pm 0.04 (\textrm{stat.}) \pm 0.05 (\textrm{sys.})) \times 10^ 6~\textrm{cm}^{-2} \textrm{sec}^{-1}.
\end{eqnarray*}
Fixing the cross section problem, a 3.4\% increase was observed.

The observed and expected fluxes are re-estimated in each energy region.
Table~\ref{tab:rates-sk3} shows the revised event rate in each energy region.
Figure~\ref{fig:energy_spectrum-sk3} shows the revised observed energy spectrum 
divided by the BP2004 SSM flux values without oscillation.  



\begin{thebibliography}{99}
\bibitem{sk_detector}
    S.~Fukuda {\it et al.}, Nucl. Instr. Meth. \textbf{A501} (2003) 418.

\bibitem{sk_full_paper}
    J.~Hosaka {\it et al.}, Phys. Rev. \textbf{D73} 112001 (2006).
    
\bibitem{sk_ii_paper}
    J.P.~Cravens {\it et al.}, Phys. Rev. \textbf{D78} 032002 (2008).

\bibitem{SKnim}
        S.~Fukuda {\it et al}. Nucl.Instr.and Meth. A {\bf 51} 418 (2003).

\bibitem{bonsai_smy}
		M.~Smy, Proc. of the 30th ICRC, {\bf 5} 1279 (2007). 
        
\bibitem{icam}
    R.M.Pope and E.S.Fry, Appl. Opt. \textbf{36}, 8710-8723 (1997).

\bibitem{ssm}
    J.N.Bahcall and M.H.Pinsonneault, Phys. Rev. Lett. \textbf{92} 121301 (2004).

\bibitem{ortiz}
    C.E.Ortiz,A.Garcia,R.A.Waltz,M.Bhattacharya,A.K.Komives, Phys. Rev. Lett. \textbf{85} 2909 (2000).

\bibitem{bpspc}
    J.N.Bahcall {\it et al.}, Phys. Rev. \textbf{C54} 411 (1996).

\bibitem{win06}
	W.T.Winter,S.J.Freedman,K.E.Rehm,J.P.Schiffer, Phys. Rev. C, \textbf{73} 025503 (2006).
\bibitem{hep}
     J.N.Bahcall {\it et al.}, Nucl. Phys. B (Proc. Suppl.), 77, 64-72 (1999). 
\bibitem{linac}
    M.Nakahata, {\it et al.} Nucl. Instr. Meth. \textbf{A421} 113 (1999).

\bibitem{dtg}
    E.Blaufuss, {\it et al.} Nucl. Instr. Meth. \textbf{A458} 636 (2001).

\bibitem{pandf}
		R.~M.~Pops {\it et al.} Applied Optics {\bf 35} 33 (1997).
\bibitem{msw}
    S.P.~Mikheyev and A.Y.~Smirnov, Sov. J. Nucl. Phys. \textbf{42} 913 (1985);
    L.~Wolfenstein, Phys. Rev. \textbf{D17} 2369 (1978).

\bibitem{unique_solution}	
	M.B.Smy, arXiv:hep-ex/0202020.
\bibitem{snod2o}
  B. Aharmim {\it et al.}, Phys. Rev. C, \textbf{75} 045502 (2007).
\bibitem{snosalt}
    S.N.Ahmed {\it et al.}, Phys. Rev. C. \textbf{72} 055502 (2005). 
\bibitem{snoncdn}
    B.Aharmim {\it et al.}, Phys. Rev. Lett. \textbf{101} 111301 (2008).
\bibitem{snoleta}
    B. Aharmim {\it et al.}, Phys. Rev. C \textbf{81} 055504 (2010). 
\bibitem{borexino}    
    C.Arpesella {\it et al.}, Phys. Rev. Lett. \textbf{101} 091302 (2008).
\bibitem{cl}
    B.T.Cleveland {\it et al.}, Astrophys. J. \textbf{496} 505 (1998).
\bibitem{gallex_gno}
    M.Altmann {\it et al.}, Phys. Lett. \textbf{B490} 16 (2000);
    also presented at Neutrino 2004 by C.Cattadori.
\bibitem{sage}
    J.N.Abdurashitov {\it et al.}, Phys. Rev. \textbf{C60} 055801 (1999).
\bibitem{fogli}
    G.L.Fogli,E.Lisi,A.Marrone,D.Montanino,A.Palazzo, Phys. Rev. \textbf{D66} 053010 (2002).
\bibitem{kamland}
    S.Abe {\it et al.}, Phys. Rev. Lett. \textbf{100} 221803 (2008).
\bibitem{barger}
    V.Barger,K.Whisnant,S.Pakvasa,R.J.N.Phillips, Phys. Rev. D. \textbf{22} 2718 (1980).
\bibitem{kldb}
    KamLAND official data base wep page:\\   
   {\scriptsize	http://www.awa.tohoku.ac.jp/KamLAND/datarelease/2ndresult.html}
\bibitem{kam2nd}
    T.Araki {\it et al.}, Phys. Rev. Lett. \textbf{94} 081801 (2005)
\bibitem{ichep08kl}
    K.Ichimura for KamLAND collaboration, 34th International Conference on High Energy Physics (ICHEP 2008).
\bibitem{kl3f}
   A.Gando {\it et al.} arXiv:1009.4771
\bibitem{bs09} 
   A.M.Serenelli  {\it et al.} arXiv:0909.2668v2

\bibitem{sk3} K.Abe et al., Phys. Rev. D 83,052010(2011), arXiv:1010.0118v2.
\bibitem{sk4} Super-Kamiokande Collaboration, 
``Solar neutrino measurements in Super--Kamiokande--IV'', {\it to be submitted}.

\end{thebibliography}
\end{document}